\documentclass[manuscript,screen]{acmart}
\AtBeginDocument{%
  }

\usepackage[utf8]{inputenc}
\usepackage{graphicx}
\usepackage[normalem]{ulem}
\usepackage{hyperref}
\usepackage{grffile}
\usepackage{ulem}
\usepackage{tikz}
\usepackage{epsfig}
\usepackage{caption}
\usepackage{algorithm}
\usepackage{hyperref}
\usepackage[noend]{algpseudocode}
\usepackage{subcaption}
\usepackage{sansmath}
\usepackage{physics}
\usepackage{xcolor}
\usepackage{booktabs}
\usepackage{datatool}
\usepackage{multirow}
\usepackage{enumitem}
\usepackage{pifont}%
\usepackage[symbol]{footmisc}

\setlist[itemize]{label=$\bullet$}
\setlist[enumerate]{label=\arabic*.}
\setlist[description]{font=\normalfont\itshape\space}

\newcommand{\cmark}{\ding{51}}%
\newcommand{\xmark}{\ding{55}}%
\newcommand{\circledone}{\ifmmode \text{\ding{182}} \else \ding{182} \fi}%
\newcommand{\circledtwo}{\ifmmode \text{\ding{183}} \else \ding{183} \fi}%
\newcommand{\circledthree}{\ifmmode \text{\ding{184}} \else \ding{184} \fi}%
\newcommand{\ceil}[1]{\left\lceil{#1}\right\rceil}
\newcommand{\floor}[1]{\left\lfloor{#1}\right\rfloor}
\newcommand{\Loopy}{\textsc{Loopy}}
\newcommand{\Fdrake}{\textsc{Firedrake}}
\newcommand{\Fenics}{\textsc{FEniCS}}
\newcommand{\Dune}{\textsc{Dune}}
\newcommand{\coords}{\boldsymbol{x}}
\newcommand{\iq}{i_{\mathrm{q}}}
\newcommand{\icell}{i_{\mathrm{cell}}}

\newcommand{\iodof}{i_{\mathrm{outDoF}}}

\newcommand{\irowtile}{i_{\mathrm{rowtile}}}
\newcommand{\icoltile}{i_{\mathrm{coltile}}}
\newcommand{\Nwi}{N_{\mathrm{WI}}}

\newcommand{\NEval}{N^{\text{deriv}}_{w}}

\newcommand{\etaalias}{\eta_{\mathrm{alias}}}
\newcommand{\etaaliasMin}{\eta^{\min}_{\mathrm{alias}}}
\newcommand{\etasimd}{\eta_{\mathrm{simd}}}
\newcommand{\etasimdMin}{\eta^{\min}_{\mathrm{simd}}}
\newcommand{\NcNwiMax}{(N_c\Nwi)^{\max}}

\newcommand{\oclLid}[1]{\text{\texttt{local\_id}}_{#1}}
\newcommand{\oclGid}[1]{\text{\texttt{group\_id}}_{#1}}
\newcommand{\Nsync}{N_{\mathrm{sync}}}

\newcommand{\etapred}{\eta_{\mathrm{pred}}}
\newcommand{\Nreside}{N_{\mathrm{reside}}}
\newcommand{\NresideEff}{\text{SG}_{\mathrm{reside, eff}}}
\newcommand{\Wmax}{\mathcal{W}_{\max}}
\newcommand{\Lmax}{\mathcal{L}_{\max}}
\newcommand{\FRoofline}{\mathcal{F}_{\mathrm{roofline}}}
\newcommand{\FPeak}{\mathcal{F}_{\mathrm{peak}}}
\newcommand{\bwglobpeak}{\beta_{\mathrm{global}}^{\mathrm{peak}}}
\newcommand{\bwlocpeak}{\beta_{\mathrm{local}}^{\mathrm{peak}}}
\newcommand{\bwtexpeak}{\beta_{\mathrm{LDG}}^{\mathrm{peak}}}
\newcommand{\bwglobmodel}{\beta_{\mathrm{global}}^{\mathrm{model}}}
\newcommand{\bwlocmodel}{\beta_{\mathrm{local}}^{\mathrm{model}}}

\newcommand{\sgSatGlobal}{\mathrm{SG_{\mathrm{sat.,\mathrm{global}}}}}
\newcommand{\sgSatShared}{\mathrm{SG_{\mathrm{sat.,\mathrm{local}}}}}
\newcommand{\algTo}{\mathrm{\textbf{ to }}}
\newcommand{\iquadtile}{i_{\mathrm{quadtile}}}
\newcommand{\quadtiletrunc}{T^{(Q,\mathrm{trunc})}}
\newcommand{\quadtile}{T^Q}
\newcommand{\iquadinner}{i_q'}
\newcommand{\iquadinnerinner}{i_q''}

\newcommand{\iquad}{i_q}
\newcommand{\aiGlobal}{\mathrm{AI}_{\mathrm{global}}}
\newcommand{\aiLocal}{\mathrm{AI}_{\mathrm{local}}}

\newcommand\plusassign{%
  \mathrel{\ooalign{\hss$\leftarrow$\hss\cr%
  \kern0.75ex\raise0.2ex\hbox{\scalebox{0.7}{+}}}}}
\newcommand{\ratioentry}[3]{%
  \DTLgetvalueforkey{\ratioValue}{#3}{#1}{Key}{#2}%
  \DTLround{\ratioRounded}{\ratioValue}{2}%
  \ratioRounded}
\newcommand{\ratiorow}[2]{%
  #1
  & \ratioentry{ratioTitanV}{#2}{Mass}
  & \ratioentry{ratioTitanV}{#2}{Hyperelasticity}
  & \ratioentry{ratioTitanV}{#2}{Elasticity}
  & \ratioentry{ratioH200}{#2}{Mass}
  & \ratioentry{ratioH200}{#2}{Hyperelasticity}
  & \ratioentry{ratioH200}{#2}{Elasticity}
  \\}
\newcommand{\scwpiwgentry}[2]{%
  \DTLgetvalueforkey{\ratioValue}{#2}{scwpiWgStudies}{W}{#1}%
  \DTLround{\ratioRounded}{\ratioValue}{2}%
  \ratioRounded}
\newcommand{\scwpiwgrow}[1]{%
  #1
  & \scwpiwgentry{#1}{titanv_mass_2dp2}
  & \scwpiwgentry{#1}{titanv_hyperelasticity_2dp4}
  & \scwpiwgentry{#1}{titanv_elasticity_3d_p2}
  & \scwpiwgentry{#1}{h200_mass_2dp2}
  & \scwpiwgentry{#1}{h200_hyperelasticity_2dp4}
  & \scwpiwgentry{#1}{h200_elasticity_3d_p2}
  \\}
\newcommand{\hpratioentry}[3]{%
  \DTLgetvalueforkey{\ratioValue}{#3}{#1}{Configuration}{#2}%
  \DTLround{\ratioRounded}{\ratioValue}{2}%
  \ratioRounded}
\newcommand{\hpratiorow}[2]{%
  #1
  & \hpratioentry{hpRatioTitanV}{#2}{mass_p3_2d}
  & \hpratioentry{hpRatioTitanV}{#2}{hyperelasticity_p6_2d}
  & \hpratioentry{hpRatioTitanV}{#2}{elasticity_p4_3d}
  & \hpratioentry{hpRatioH200}{#2}{mass_p3_2d}
  & \hpratioentry{hpRatioH200}{#2}{hyperelasticity_p6_2d}
  & \hpratioentry{hpRatioH200}{#2}{elasticity_p4_3d}
  \\}
\newcommand{\spearmanrow}[2]{%
  #1
  & \ratioentry{spearmanData}{#2}{TitanV}
  & \ratioentry{spearmanData}{#2}{H200}
  & \ratioentry{spearmanData}{#2}{Both}
  \\}
\newcommand{\topkrow}[2]{%
  #1
  & \ratioentry{topkData}{#2}{TitanV1}
  & \ratioentry{topkData}{#2}{TitanV5}
  & \ratioentry{topkData}{#2}{TitanV10}
  & \ratioentry{topkData}{#2}{TitanV20}
  & \ratioentry{topkData}{#2}{H2001}
  & \ratioentry{topkData}{#2}{H2005}
  & \ratioentry{topkData}{#2}{H20010}
  & \ratioentry{topkData}{#2}{H20020}
  & \ratioentry{topkData}{#2}{Both1}
  & \ratioentry{topkData}{#2}{Both5}
  & \ratioentry{topkData}{#2}{Both10}
  & \ratioentry{topkData}{#2}{Both20}
  \\}

\algrenewcommand\alglinenumber[1]{\scriptsize #1}

\errorcontextlines\maxdimen

\RequirePackage{tikz}
\RequirePackage{zref-abspage}
\RequirePackage{zref-user}
\RequirePackage{tikz}
\RequirePackage{atbegshi}
\usetikzlibrary{calc}
\RequirePackage{tikzpagenodes}
\RequirePackage{etoolbox}

\makeatletter
\newcommand*{\algrule}[1][\algorithmicindent]{%
  \hspace*{.2em}%
  \vrule %
  \hspace*{\dimexpr#1-.2em-.4pt}%
}

\newcommand{\StatePar}[1]{%
  \State\parbox[t]{\dimexpr\linewidth-\ALG@thistlm}{\strut #1\strut}%
}
\renewcommand{\ALG@beginalgorithmic}{\offinterlineskip}%

\newcount\ALG@printindent@tempcnta
\def\ALG@printindent{%
  \ifnum \theALG@nested > 0%
    \ifx\ALG@text\ALG@x@notext%
    \else
      \unskip
      \ALG@printindent@tempcnta=1
      \loop
        \algrule[\csname ALG@ind@\the\ALG@printindent@tempcnta\endcsname]%
        \advance \ALG@printindent@tempcnta 1
        \ifnum \ALG@printindent@tempcnta<\numexpr\theALG@nested+1\relax
      \repeat
        \fi
    \fi
}
\patchcmd{\ALG@doentity}{\noindent\hskip\ALG@tlm}{\ALG@printindent}{}{\errmessage{failed to patch}}
\makeatother

\algrenewcommand\algorithmicend{\strut\textbf{end}}
\algrenewcommand\algorithmicdo{\strut\textbf{do}}
\algrenewcommand\algorithmicwhile{\strut\textbf{while}}
\algrenewcommand\algorithmicfor{\strut\textbf{for}}
\algrenewcommand\algorithmicforall{\strut\textbf{for all}}
\algrenewcommand\algorithmicloop{\strut\textbf{loop}}
\algrenewcommand\algorithmicrepeat{\strut\textbf{repeat}}
\algrenewcommand\algorithmicuntil{\strut\textbf{until}}
\algrenewcommand\algorithmicprocedure{\strut\textbf{procedure}}
\algrenewcommand\algorithmicfunction{\strut\textbf{function}}
\algrenewcommand\algorithmicif{\strut\textbf{if}}
\algrenewcommand\algorithmicthen{\strut\textbf{then}}
\algrenewcommand\algorithmicelse{\strut\textbf{else}}

\algrenewcommand\algorithmicrequire{\strut\textbf{Input:}}
\algrenewcommand\algorithmicensure{\strut\textbf{Output:}}

\let\oldState\State
\renewcommand{\State}{\oldState\strut}

\begin{document}
\title{Code Generation for Near-Roofline Finite~Element~Actions on GPUs from Symbolic~Variational~Forms}
\author{Kaushik Kulkarni}
\email{kgk2@illinois.edu}
\orcid{0009-0001-0645-2169}
\affiliation{%
  \institution{Siebel School of Computing and Data Science, University of Illinois at Urbana-Champaign}
  \city{Urbana}
  \state{Illinois}
  \country{USA}
}
\author{Andreas Kl\"ockner}
\email{andreask@illinois.edu}
\orcid{0000-0003-1228-519X}
\affiliation{%
  \institution{Siebel School of Computing and Data Science, University of Illinois at Urbana-Champaign}
  \city{Urbana}
  \state{Illinois}
  \country{USA}
}

\begin{abstract}
We present a novel parallelization strategy for evaluating Finite Element Method
(FEM) variational forms on GPUs, focusing on those that are expressible through
the Unified Form Language (UFL) on simplex meshes.  We base our approach on code
transformations, wherein we construct a space of scheduling candidates and rank
them via a heuristic cost model to effectively handle the large diversity of
computational workloads that can be expressed in this way. We present a design
of a search space to which the cost model is applied, along with an associated
pruning strategy to limit the number of configurations that need to be
empirically evaluated. The goal of our design is to strike a balance between
the device's latency-hiding capabilities and the amount of state space, a key
factor in attaining near-roofline performance.

To make our work widely available, we have prototyped our parallelization
strategy within the \Fdrake{} framework, a UFL-based FEM solver. We evaluate the
performance of our parallelization scheme on three generations of Nvidia GPUs,
specifically the H200, Titan V and Tesla K40c, across a range of operators
commonly used in applications, including fluid dynamics, wave propagation, and
structural mechanics, in 2D and 3D geometries.  Our results demonstrate that our
proposed algorithm achieves more than $50\%$ roofline performance in $60\%$ of
the test cases.
\end{abstract}


\maketitle

\renewcommand{\thefootnote}{\fnsymbol{footnote}}

\section{Introduction}\label{sec:intro}
Domain-specific languages (DSLs) trade generality and ease of maintenance for a
high-level representation meaningful to domain scientists, and, often, superior
performance. This performance gain stems from the utilization of compiler passes
that are specifically tailored to the class of programs captured in such DSLs.
The Unified Form Language (UFL)~\cite{Aln_s_2012} is one such language that
provides near-mathematical notation for variational formulations of Partial
Differential Equations (PDEs). Many state-of-the-art Finite Element Method (FEM)
solvers, like \Fdrake~\cite{Gibson_2019}, \Fenics~\cite{dupont2003fenics},
and, \Dune~\cite{bastian2006distributed} ship a UFL implementation that compiles
variational forms to code accomplishing the \emph{assembly} of FEM operators.

While the desired numerical result is fully specified via UFL's semantics,
evaluation and execution details are implementation-defined. Such common design
choices include selecting the quadrature rule, determining whether the geometry
terms are stored for long-term usage, array of structures versus structure of
arrays encoding of degrees of freedom (DOFs), among others. We focus on one such
implementation choice, specifically, the scheduling of operations in the
assembly stage. This step accounts for a significant fraction of the workload in
the solver's hot loop, and hence has been the subject of active research.

Kirby et al.~\cite{kirby2018solver} present three key reasons why matrix-free
methods are particularly well-suited for devices with high FLOP throughput when
considered relative to their available memory bandwidth. Recent trends suggest
that this includes an increasing number of machines on the market, particularly
GPUs.  First, the memory access patterns in sparse matrix-vector multiplication
(SPMV) and matrix-free action application are comparable, particularly as mesh
refinement increases. Second, matrix-free methods eliminate the computational
cost of matrix assembly and significantly reduce memory usage by avoiding the
storage of the entire matrix, the storage requirements of which exhibit
$O(p^{2d})$ growth with the polynomial approximation order $p$.  Third, the
memory access patterns associated with sparse matrices result in reduced memory
throughput.  Consequently, in this work, \textit{we develop and evaluate code
transformation strategies to accelerate the application of an FEM operator on
GPUs using matrix-free methods}. Importantly, in order to achieve acceptable
convergence in the context of an iterative method, a preconditioner is required
for moderate to large-scale problems. While out of scope for the present
contribution, recent research has shown numerous avenues for progress in this
area, including
$p$-Multigrid~\cite{helenbrook2003analysis,malachi2022chebyshev}, and
hybridization techniques~\cite{brezzi2012mixed}.

GPUs are SIMD (Single Instruction Multiple Data) devices that prioritize
arithmetic throughput over operations like branch prediction and
instruction-level parallelism. Recent advancements in GPU architectures have
further enhanced their suitability for scientific applications. For instance,
the Volta microarchitecture~\cite{nvidia2017v100}, introduced in 2017, can
deliver performance of up to 6 TFLOPS when operating on double-precision
floating-point numbers.  However, achieving performance near the peak
capabilities of a GPU for a specific workload remains challenging. This
complexity arises from the large number of possible implementations for a given
workload, due to the various optimization parameters, such as computational grid
size, local memory usage, and address space selection. These parameters create
a vast search space of potential computational kernels for a particular
workload. Further, the lack of an accurate cost model makes it impractical to
identify the best-performing configuration in the search space.

UFL-based FEM solvers encompass a programming paradigm that allows for the
expression of a wide variety of variational formulations. These options include
specifying arbitrary expressions on meshes, selecting different function spaces,
and choosing quadrature rules, among others. Each of these choices plays a
crucial role in shaping the workload to be executed on the device, as they
directly affect loop counts, loop nestings, and the state-space requirements of
the compute kernel. Consequently, these factors significantly influence
performance and necessitate different code transformation strategies to produce
a high-performant binary for the workload.

We address these challenges by developing a domain-specific transformation for a
subset of integral expressions expressible via UFL. See
Section~\ref{sec:input-space} for a precise description of the variational forms
our proposed algorithm can address. While there are some specific limitations in
the scope of this work compared to the full generality of UFL, none of them are
essential, i.e. they can be removed with additional effort. Where applicable, we
outline approaches towards this goal.

Our algorithm follows a two-step process. First, for a given computational
kernel, we generate a transformed version that accepts parameters such as tile
sizes and computational grid dimensions. Second, we utilize an auto-tuning
strategy centered around a cost model to efficiently explore the parameter space
and identify a configuration that achieves near-roofline performance for the
transformed kernel. This two-step approach is crucial for scaling across a range
of problems independent of the user-provided variational form. To demonstrate
the generality of our method, in Section~\ref{sec:perf_eval}, we evaluate it
across a variety of weak forms commonly used in fluid dynamics, acoustics, and
structural mechanics, considering different function spaces and spatial
dimensions in our benchmark suite.

In this work, we present the following novel contributions:
\begin{itemize}
  \item We have developed a scheme to create a transformation search space
    for action kernels derived from UFL variational forms. This space includes
    configurations with transformed kernels that occupy different design points
    in terms of performance characteristics such as latency-hiding capabilities,
    synchronization requirements, and register pressure. This
    transformation space can be utilized to achieve near-roofline performance
    for a class of UFL-derived kernels that arise from variational forms defined
    over simplex elements.

  \item We have developed an auto-tuning approach that utilizes an analytic
    cost model to explore the aforementioned transformation space. The cost
    model allows us to rank the configurations within the transformation space,
    eliminating the need for an exhaustive search. This approach is beneficial
    in reducing preparation times while considering near-optimal choices in the
    search space.

  \item We have performed a series of experiments comprising FEM operators seen
    in real-world applications. Our test bed consists of 70 variational form evaluation
    scenarios, encompassing different function spaces, spatial dimensions, and
    PDEs being solved. The performance evaluation results indicate that our
    transformation space achieves a minimum of 50\%-roofline performance for
    60\% of the test cases in our suite.
\end{itemize}

\section{Background}\label{sec:background}

\subsection{Finite Element Methods}\label{sec:background_fem}
We will briefly introduce Finite Element Methods (FEM) in this section through a
simple example. For an in-depth review, we refer the reader to the survey by
Logg et al.~\cite{logg2012automated}. Consider the Poisson equation on a
bounded domain $\Omega\subset\mathbb{R}^2$ with Dirichlet boundary conditions:
\begin{equation}
  \label{eq:poissonpde}
  \begin{aligned}
    \laplacian u \pqty{\coords} = -f, \\
    {u}\Bigr{|}_{\partial\Omega} = g,
  \end{aligned}
\end{equation}
where $\boldsymbol{x}\in\Omega$ and $u\in H^1(\Omega)$ is the desired
solution.

To obtain the solution using a Finite Element Method (FEM), we turn the PDE
in~\eqref{eq:poissonpde} into a variational problem by multiplying the equation
with a function $v$. $u$ and $v$ are called the trial and test functions,
respectively. After multiplying~\eqref{eq:poissonpde} by the test function $v$,
integrating it, and applying integration by parts to the operands, we want to
solve for $u \in V$ in
\begin{equation}
  \label{eq:poissonpde_weakform}
  \int\limits_{\Omega} \grad u \cdot \grad v\; \dd x = \int\limits_{\Omega} f v \dd x \qquad \left(v \in W\right),
\end{equation}
where $W$ is a space of test functions and $V$ is a ``trial space'' in which the
solution is sought. We choose the test and trial function spaces as
\begin{equation}
  \begin{split}\label{eq:test_and_trial_function_spaces}
    V &= \left\{u \in H^1\left(\Omega\right):\; u \Bigr{|}_{\partial\Omega} = g\right\}, \\
    W &= \left\{v \in H^1\left(\Omega\right):\; v\Bigr{|}_{\partial\Omega} = 0\right\}.
  \end{split}
\end{equation}
$H^1(\Omega)$ in~\eqref{eq:test_and_trial_function_spaces} is a Sobolev space,
which consists of functions $v$ such that $v^2$ and $|\nabla v|^2$ (evaluated in a weak
sense) have finite integrals, while allowing for discontinuous derivatives. A
key principle in FEM is the construction of finite-dimensional test and trial
function spaces to solve~\eqref{eq:poissonpde_weakform} numerically. To this
end, we introduce the finite-dimensional trial and test function spaces $V_h
\subset V$ and $W_h \subset W$, respectively. Consequently, the numerical
solution $u_h \in V_h$ satisfies
\begin{equation}
  \label{eq:poisson_pde_discrete_weakform}
  \int\limits_{\Omega} \grad u_h \cdot \grad v \;\; \dd x = \int\limits_{\Omega} f
  v\; \dd x \qquad \left(v \in W_h \subset W\right).
\end{equation}

Since $V_h$ is a finite-dimensional space, we can make an ansatz for
$u_h$ as $u_h = \sum_{j=1}^N u_j \phi_j$, where $\{\phi_j\}_{j\in\{1, 2, \ldots,
N\}}$ are the basis functions of $V_h$, and $N$ is the dimensionality of $V_h$.
Similarly, we can require~\eqref{eq:poisson_pde_discrete_weakform} to be true
for all $v\in\{\psi_i\}_{i\in\{1, 2, \ldots, N\}}$ where the $\psi_i$ form the
basis of $W_h$. Consequently, we obtain the linear system for the coefficients
in our ansatz as
\begin{equation}
  \label{eq:poisson_linear_system}
  A \vb u = \vb b,
\end{equation}
where
\begin{equation}
  \label{eq:poisson_pde_matrix_and_vec}
  \begin{split}
    \left(A\right)_{ij} &= \int_{\Omega} \grad \psi_i \cdot \grad \phi_j\; \dd x, \\
    \left(\vb b\right)_{i} &= \int_{\Omega}  f \psi_i\; \dd x.
  \end{split}
\end{equation}
Due to its nature of dependence on the test and trial function, $A$
in~\eqref{eq:poisson_linear_system} is referred to as the \textit{bilinear
form} associated with the Poisson equation.

In this work, we follow UFL and focus on the FEM abstraction as formalized by
Ciarlet~\cite{ciarlet1976numerical}. We begin by decomposing the domain
$\Omega$ into finite, disjoint cells $\mathcal{T}_h = \{T\}$ such
that
\begin{equation}
  \label{eq:union_of_cells}
  \bigcup\limits_{T\in\mathcal{T}_h} T = \Omega.
\end{equation}

We build our finite-dimensional function space by defining local functions on
these cells and use them to define our global function space. To define these
local function spaces, we use Ciarlet's triple $\left(T, \mathcal{V},
\mathcal{L}\right)$ where $T$ is a non-empty, compact subset of $\mathbb{R}^d$,
$\mathcal{V}$ is a $n$-dimensional function space on $T$, $\mathcal{L} =
\{\ell_1, \ell_2, \ldots, \ell_n\}$ is the set of \textit{degrees of freedom},
forming a basis of the dual-space $\mathcal{V}'$.

For each of our cells $T\in\mathcal{T}_h$, we define a \textit{local function
space} based on Ciarlet's triple as $\left\{\left(T, \mathcal{V}_T,
\mathcal{L}_T\right)\right\}_{T\in\mathcal{T}_h}$. Further, we also construct
the mappings $\left\{\iota_T\right\}_{T\in\mathcal{T}_h}$ to map from our local
degrees of freedom to global degrees of freedom, i.e.,
$$\iota_T: \left\{1, 2, \ldots, n\right\} \mapsto \left\{1, 2, \ldots, N\right\}.$$
Recall that $n$ is the dimension of our local function spaces
$\mathcal{V}_T$, and that $N$ is the dimension of the global function space $V_h$.
These $\iota_T$'s establish the mapping between global degrees of freedom, i.e.,
$\mathcal{L} = \{\ell_1, \ell_2,
\ldots, \ell_N\}$ to  $\mathcal{L}_T$'s, as:
\begin{equation*}
\ell^T_{\left(i\right)}(v|_T) = \ell_{\iota_T\left(i\right)}(v),\quad(i\in\left\{1, 2, \ldots, n\right\}).
\end{equation*}

We further simplify our choice of $V_h$ by using a reference Finite Element
$\left(\hat{\mathcal{T}}, \hat{\mathcal{V}}, \hat{\mathcal{L}}\right)$ and a set
of \textit{invertible} mappings $\left\{F_T\right\}_{T\in\mathcal{T}_h}$ such that:
\begin{equation*}
  F_T(\hat{T}) = T\quad(T\in\mathcal{T}_h).
\end{equation*}

Consequently, for each cell the mapping $F_T$ helps us construct the local
function space as:
\begin{equation}
  \mathcal{V}_T = \left\{\hat{v} \circ F_T^{-1}: \hat{v} \in
  \hat{\mathcal{V}}\right\}
\end{equation}

To simplify the construction of the test function space $W_h$ we use the same
framework as above.

\subsection{Towards Discretization}\label{sec:iteratively_solving_numerical_soln}
In Section~\ref{sec:background_fem}, we have discussed the construction of test and
trial function spaces using local function spaces. This process leads to the
formulation of a sparse linear system, as shown
in~\eqref{eq:poisson_linear_system}. These sparse systems are typically solved
using iterative methods, such as Krylov Subspace Methods~\cite{saad1989krylov}.
A significant computational cost in these solvers arises from the evaluation of
the \textit{action} of the FEM operator, i.e., evaluating the integrals
\begin{equation}
  \label{eq:poisson_action_global_no_ciarlet}
  \int_\Omega \grad \psi_i \cdot \grad u_h \; \dd x\qquad \left(i\in\left\{1,\ldots, N\right\}\right),
\end{equation}
where $u_h\in V_h$ is a solution guess, and the $\psi_i$ form the basis of test
function space $W_h$. Accelerating the evaluation of the action operation on a
GPU is the focus of this work. See Section~\ref{sec:input-space} for a precise
description of the bilinear operators whose actions are considered within the scope
of this work.

As discussed in Section~\ref{sec:intro}, to efficiently map the workload to a
GPU, we evaluate the action without explicitly assembling the sparse matrix,
i.e., in a \textit{matrix-free} manner. For our selected test and trial function
spaces in
Section~\ref{sec:background_fem},~\eqref{eq:poisson_action_global_no_ciarlet}
simplifies to:
\begin{equation}
  \label{eq:possion_action_global}
  \int\limits_{\Omega} \grad \psi_i(\vb x) \cdot \grad \left(\sum_{j=1}^{N}\ell_j
  \phi_j(\vb x)\right) \; \dd x \qquad \left(i\in\{1,\ldots, N\}\right),
\end{equation}
where $\{\phi_1, \phi_2, \ldots, \phi_N\}$ form the basis of the trial function
space, and $\{\ell_1, \ell_2, \ldots, \ell_N\}$ are the degrees of freedom for
the guess solution $u_h$.

Our construction of $V_h$, $W_h$ allows us to
express~\eqref{eq:possion_action_global} as a summation of integrals over the
local function spaces for every cell in our mesh:
\begin{equation}
  \label{eq:poisson_action_as_local_sum}
  \sum_{\substack{T\in\mathcal{T}_h\\i\in\operatorname{image}(\iota_T)}} \int\limits_{T} \grad \left(\hat{\psi}_{\iota_T^{-1}(i)} \circ F_T^{-1}\right)(\vb x)  \cdot \grad \left(\sum_{j=1}^{n}\ell_{\iota_T(j)}
  \cdot \left(\hat{\phi}_j \circ F_T^{-1}\right)(\vb x)\right) \; \dd x \qquad \left(i\in\{1,\ldots, N\}\right),
\end{equation}
where $\{\hat{\psi}_i\}_{i\in\{1,2,\ldots,n\}}$ and
$\{\hat{\phi}_i\}_{i\in\{1,2,\ldots,n\}}$ come from our choice of reference
finite element for $V_h$ and $W_h$, respectively.

The cell-local integration operation in~\eqref{eq:poisson_action_as_local_sum} can be
simplified using the substitution $\vb x = F_T(\hat{\vb x})$, leading to:
\begin{equation}
  \label{eq:poisson_action_substituted_integrand}
  \int\limits_{\hat{T}} \left(\mathbf{J}_T^{-T} \cdot \grad_{\hat{\vb x}}
  \hat{\psi}_{i}(\hat{\vb x})\right)  \cdot \left(\mathbf{J}_T^{-T} \cdot \grad_{\hat{\vb x}}
  \left(\sum_{j=1}^{n}\ell_j \cdot \hat{\phi}_j(\hat{\vb x})\right)\right) \;
  \left|\det\left(\mathbf{J}_T\right)\right|\; \dd \hat{x},
\end{equation}
where $\mathbf{J}_T$ is Jacobian matrix for the reference cell to physical cell
mapping $F_T$, and $\hat{T}$ is the reference cell volume.

A key implementation detail in FEM solvers, such as \Fdrake, \Fenics, and \Dune,
involves employing quadrature to automate integral evaluation
in~\eqref{eq:poisson_action_as_local_sum} while being agnostic
of the choice of the basis functions for $\hat{\mathcal{V}}$. Consequently,
the integral in~\eqref{eq:poisson_action_substituted_integrand} can be replaced
with a summation as
\begin{equation}
\label{eq:poisson_action_w_quadrature}
  \sum_{k=1}^{Q} w_k \cdot \left(\mathbf{J}_T^{-T}\left(\mathbf{x}^q_k\right) \cdot \left(\grad
  \hat{\psi}_{i}(\vb x)\right)\Bigr |_{\mathbf{x}^q_k}\right)  \cdot \underbrace{\left(\mathbf{J}_T^{-T}\left(\mathbf{x}^q_k\right) \cdot
  \left(\sum_{j=1}^{n}\ell_j \left(\grad \hat{\phi}_j(\vb x)\right)\Bigr |_{\mathbf{x}^q_k}\right)\right)}_{\text{evaluation}} \;
  \left|\det\left(\mathbf{J}_T\left(\mathbf{x}^q_k\right)\right)\right|,
\end{equation}
where $Q$ is the  number of quadrature nodes being used, $\{\mathbf{x}^q_1,
\ldots, \mathbf{x}^q_Q\}$ are the quadrature nodes and $\{w_1, \ldots, w_Q\}$
are the quadrature weights.~\eqref{eq:poisson_action_w_quadrature} can be
further simplified by tabulating the common subexpressions into matrices as
\begin{equation}
\label{eq:poisson_tabulated_values}
\begin{split}
\Phi_x[i, j] = \left( \frac{\partial \hat{\phi}_i}{\partial x} \right)\Bigr |_{\mathbf{x}^q_j}, \\
\Phi_y[i, j] = \left( \frac{\partial \hat{\phi}_i}{\partial y} \right)\Bigr |_{\mathbf{x}^q_j}, \\
\Psi_x[i, j] = \left( \frac{\partial \hat{\psi}_i}{\partial x} \right)\Bigr |_{\mathbf{x}^q_j}, \\
\Psi_y[i, j] = \left( \frac{\partial \hat{\psi}_i}{\partial y} \right)\Bigr |_{\mathbf{x}^q_j},\\
\end{split}
\end{equation}
where $i\in\left\{1, 2, \ldots, n\right\}$ and $j\in\left\{1, 2, \ldots, Q\right\}$.

We
combine~\eqref{eq:possion_action_global},~\eqref{eq:poisson_action_w_quadrature},
~\eqref{eq:poisson_tabulated_values} in Algorithm~\ref{knl:poisson_action}.
Note that in Algorithm~\ref{knl:poisson_action} we employ an affine
geometry mapping between the physical and reference elements. In this specific example,
we leverage this to compute the Jacobian terms on a per-cell basis rather than on a
per-quadrature-point basis.

\algrenewcommand{\alglinenumber}[1]{\color{magenta}\footnotesize#1}
\begin{algorithm}[H]
\caption{2D Poisson operator action kernel}\label{knl:poisson_action}
\begin{algorithmic}[1]
  \For{$\icell \gets 0 \algTo$ $N_\text{cell}$ }
    \For{$j \gets 0 \algTo n_{V_h}$}
      \State $\ell[j] \gets$ $\ell^{\text{global}}$[$\iota_{V_h}$[$\icell, j$]]
    \EndFor
    \For{$j \gets 0 \algTo n_{P_1}$}
      \For{$d \gets 0 \algTo 2$}
        \State coords$[j,d] \gets$ coords$^{\text{global}}$[$\iota_{P_1}$[$\icell, j$], d]
      \EndFor
    \EndFor
    \State $\vb J \gets$ \Call{Jac}{coords}
    \State det $\gets \left|\Call{Det}{\vb J}\right|$
    \For{$j \gets 0 \algTo n_{W_h}$}
      \State out$[j]\gets 0$
    \EndFor
    \For{$\iq\gets 0$ \textbf{to} $Q$}
      \State $u_x  \gets 0$
      \State $u_y \gets 0$
      \For{$j\gets 0 \algTo N_{V_h}$}
        \State $u_x \gets u_x + \Phi_x[\iq, j]\cdot \ell[j]$
        \State $u_y \gets u_y + \Phi_y[\iq, j]\cdot \ell[j]$
      \EndFor
      \State $\begin{bmatrix}e_1\\ e_2\end{bmatrix} = \vb J^{-1}\vb J^{-T}\begin{bmatrix}u_x\\u_y\end{bmatrix}$
    \algstore{poissonactionapseudocode}
  \end{algorithmic}
\end{algorithm}
\begin{algorithm}[H]
  \begin{algorithmic}[1]
  \algrestore{poissonactionapseudocode}
      \For{$i\gets 0 \algTo n_{W_h}$}
        \State out$[i] \gets $ out$[i]$ + $
        \left(e_1\Psi_x[\iq,i]\ + e_2\Psi_y[\iq, i]\right)\cdot w[\iq]\cdot$ det
      \EndFor
    \EndFor
    \State {}\label{knlline:action.gather_trial_stop}
    \For{$j \gets 0 \algTo n_{W_h}$}
      \State out$^{\text{global}}$[$\iota_{W_h}$[$\icell, j]]$ $\gets$ out$^{\text{global}}$[$\iota_{W_h}$[$\icell, j]]$ + out[$j$]
    \EndFor
  \EndFor
\end{algorithmic}
\end{algorithm}

We now describe the key steps of Algorithm~\ref{knl:poisson_action}. On line 1,
we iterate over all cells in the mesh, corresponding to the summation over all
$T \in \mathcal{T}_h$ in~\eqref{eq:poisson_action_as_local_sum}, where
$N_{\text{cell}}$ denotes the total number of cells, $|\mathcal{T}_h|$. In lines
2--3, the local degrees of freedom (DOFs) of the guess solution are loaded into
the $n_{V_h}$-long array $\ell$, utilizing appropriate local-index to
global-index mappings, i.e., $\iota_{V_h}$, from their global counterparts.
Here, $n_{V_h}$ represents the dimension of the reference element's function
space in $V_h$.

For simplicity in this section, we assume an affine mapping from the reference
element $\hat{T}$ to all physical cells $T$, i.e., the functions $\sum_{T \in
\mathcal{T}_h} F_T$ lie in the degree-1 Lagrange finite element space,
denoted $P_1$. Consequently, the cell-local coordinates are stored in the
variable \texttt{coords} using the indirection $\iota_{P_1}$. These two phases of
loading cell-local DOFs collectively constitute the \textit{gather} phase.
In Section~\ref{sec:extensions_to_non_affine}, we demonstrate that our code
transformation strategy generalizes straightforwardly to the case of non-affine
mapping functions.

On lines 7--8, we compute the Jacobian terms of $F_T$ based
on~\eqref{eq:poisson_action_substituted_integrand}. Then, in line 10, we enter
the loop over quadrature nodes. In lines 11--16, we accumulate the evaluation of
the trial function's derivative at the $Q$ quadrature points, corresponding to
the \textit{evaluation} phase in~\eqref{eq:poisson_action_w_quadrature}.
Notably, these Jacobian terms can be precomputed prior to executing the action
operator, reducing floating-point operations at the cost of increased memory
usage. In this work, we follow the design choices of
TSFC~\cite{homolya2018tsfc} and compute these terms \textit{on the fly}.

On lines 17--18, the evaluated derivatives are used to accumulate the cell-local
action integrals, stored in the $n_{W_h}$-long array `out', where $n_{W_h}$ is
the dimension of the reference element's function space in $W_h$. This step is
referred to as the \textit{quadrature} phase.

Finally, on lines 19--20, the accumulated results are stored back into the
global DOFs of the test functions using indirection through $\iota_{W_h}$, which
corresponds to the local-index to global-index mapping for the test function
space $W_h$. This process of storing the computed integrals via indirection is
known as the \textit{scatter} phase.  In many FEM discretizations, the
$C^0$-continuity requirement causes neighboring cells in the mesh to share DOFs
in $\text{out}^{\text{global}}$, making $ \iota_{W_h}$ a non-injective mapping. As
a result, the scatter operation introduces a loop-carried dependence along
$\icell$.

\subsection{Defining the Input Space of Compute Kernels}\label{sec:input-space}
While Section~\ref{sec:iteratively_solving_numerical_soln} outlines the process
of applying the action operator in a specific simple instance, the focus of this
section is to outline the process at the level of generality supported by our
software system.

The bilinear form associated with a PDE's FEM discretization is a key component
in determining the expression of the action operator. In this work, we focus on
the action operators over the domain $\Omega \subset \mathbb{R}^d$, for a
bilinear form expressed on a mixed trial function space comprising scalar trial
function spaces $\{U_1, \ldots, U_{N_U}\}$, and vector trial function spaces
$\{V_{1}, \ldots, V_{N_V}\}$. The test function space is denoted by $W$, with
the bilinear form expressed as:

\begin{equation}
  \label{eq:generalized_bilinear_form}
  \begin{gathered}
  \sum_{T\in\mathcal{T}_h}\int_{T}\mathcal{I}(\vb{x}) \; \dd x \\
  \mathcal{I}(\vb{x}) = \sum_{j=1}^{\NEval}
  g_j\left(
  \left\{ \frac{\partial^{\eta^{u_i}_{k}} u_i}{\partial\vb{x}^{\eta^{u_i}_{k}}} \right\}_{\begin{array}{l}
  \scriptstyle i = 1, \dotsc, N_U \\[-0.3em]
  \scriptstyle k = 1, \dotsc, N_{u_i}^{\text{deriv}}
  \end{array}},
  \left\{ \frac{\partial^{\eta^{v_i}_{k}} \left(v_i\right)_{C_{i,k}}}{\partial\vb{x}^{\eta^{v_i}_{k}}} \right\}_{\begin{array}{l}
  \scriptstyle i = 1, \dotsc, N_V \\[-0.3em]
  \scriptstyle k = 1, \dotsc, N_{v_i}^{\text{deriv}}
  \end{array}},
  \vb{x}
  \right)
  \cdot \frac{\partial^{\mu_{j}} w}{\partial\vb{x}^{\mu_{j}}},
  \end{gathered}
\end{equation}
where $\{u_i \in U_i\}_{i \in \{1, \ldots, N_U\}}$ are scalar-valued trial
functions, and $\{v_i \in V_i\}_{i \in \{1, \ldots, N_V\}}$ are vector-valued
trial functions. The test function is denoted by $w \in W_h$, and $C_{i,k} \in
\{1, \ldots, d\}$ indicates the subscript used to obtain a component of a vector
function. The mappings $g_{j} :
\mathbb{R}^{\left(\sum_{i}^{N_U}N^{\text{deriv}}_{u_i}+\sum_{i}^{N_V}N^{\text{deriv}}_{v_i}
+ d\right)} \to \mathbb{R}$ for all $j \in \{1, \ldots, \NEval\}$ represent the
trial function derivative mappings. The term
$N^{\text{deriv}}_u$ represents the number of derivative terms of the trial
function $u$ involved in the bilinear form, while $\eta^u_k$ denotes the
multi-index representing the $k$-th derivative term of $u$. Similarly,
$\mu_j$ denotes the multi-index corresponding to the partial derivative of
$w$ involved in the $j$-th contributing derivative term of $w$.

To model the bilinear form of the Poisson equation
using~\eqref{eq:generalized_bilinear_form}, we set the number of scalar trial
function spaces to $N_{U} = 1$, the number of vector trial function spaces to
$N_{V} = 0$, and the number of derivative terms to $N^{\text{deriv}}_{u_1} =
\NEval = 2$. The derivative mappings are defined by $g_{1}(x, y) = x$ and
$g_{2}(x, y) = y$. The derivative multi-indices are defined as
$\eta^{u_1}_{1} = (1,0)$, $\eta^{u_1}_{2} = (0,1)$, $\mu_{1} = (1,0)$, and
$\mu_{2} = (0,1)$.

Applying the process of Section~\ref{sec:background_fem}
to~\eqref{eq:generalized_bilinear_form} results in
Algorithm~\ref{knl:action.vanilla} below.

\begin{algorithm}[H]
  \caption{Action kernel corresponding to the bilinear form in~\eqref{eq:generalized_bilinear_form}.}\label{knl:action.vanilla}
  \begin{algorithmic}[1]

  \For{$\icell \gets 0 \algTo N_{\text{cell}}$}
    \For{$j \gets 0 \algTo n_{P_1}$}
      \For{$i_d \gets 0 \algTo d$}
        \label{knlline:action.gather_coords}
        \State coords$[j,i_d] \gets$ coords$^{\text{global}}$[$\iota_{P_1}$[$\icell, j$], $i_d$]
      \EndFor
    \EndFor

    \For{$j \gets 0 \algTo n_{U_1}$}\label{knlline:action.gather_trial_start}
      \State $\ell^u_1[j] \gets u_1^{\text{global}}[\iota_{U_1}[\icell, j]]$
    \EndFor
    \State $\triangleright$ Similarly gather DOFs for $\{u_2,\ldots,u_{N_{U}}\}$
    \For{$j \gets 0 \algTo n_{V_{1}}$}
      \For{$i_d \gets 0 \algTo d$}
        \State $\ell^v_{1}[j, i_d] \gets v_{1}^{\text{global}}[\iota_{V_1}[\icell, j], i_d$]
      \EndFor
    \EndFor
    \State $\triangleright$ Similarly gather DOFs for $\{v_2,\ldots,v_{N_V}\}$\label{knlline:action.gather_trial_stop}

    \State $J \gets$ \Call{Jac}{coords}\label{knlline:action.jac_start}
    \State $D \gets$ \Call{Det}{$J$}\label{knlline:action.jac_stop}

    \For{$j\gets 0\algTo n_W$}
      \State $\text{out}[j] \gets 0$
    \EndFor

    \For{$i_q\gets 0\algTo Q$}
      \label{knlline:action.eval_start}
      \State $\partial_{1}u_1  \;\;\;\quad\gets 0$
	    \State \vdots
      \State $\partial_{N^{\text{deriv}}_{u_1}}u_1  \gets 0$
    \algstore{actionvanillapseudocode}
  \end{algorithmic}
\end{algorithm}

\begin{algorithm}[H]
  \begin{algorithmic}[1]
  \algrestore{actionvanillapseudocode}

      \For{$i_{U_1}\gets 0 \algTo n_{U_1}$}\label{knlline:action.loop_i1}
        \State $\partial_{1}u_1  \;\;\;\quad\gets \partial_{1}u_1 + \Phi^{U_1}_{1}[i_q,i_1] \cdot \ell^u_1[i_1] $
        \State \vdots
        \State $\partial_{N^{\text{deriv}}_{u_1}}u_1  \gets \partial_{N^{\text{deriv}}_{u_1}}u_1 + \Phi^{U_1}_{N^{\text{deriv}}_{u_1}}[i_q,i_1] \cdot \ell^u_1[i_1]$
      \EndFor
      \State {$\triangleright$ Similar evaluations for all scalar DOFs.}
      \State{}

      \State $\partial_{1}v_1  \;\;\;\quad\gets 0$
	    \State \vdots
      \State $\partial_{N^{\text{deriv}}_{v_1}}v_1  \gets 0$
      \State {}

      \For{$i_{V_1}\gets 0 \algTo n_{V_1}$}\label{knlline:action.loop_i2}
        \State $\partial_{1}v_1  \;\;\;\quad\gets \partial_{1}v_1 + \Phi^{V_1}_{1}[i_q,i_1] \cdot \ell^v_1[i_1, C_{1,1}] $
        \State \vdots
        \State $\partial_{N^{\text{deriv}}_{v_1}}v_1  \gets \partial_{N^{\text{deriv}}_{v_1}}v_1 + \Phi^{V_{1}}_{N^{\text{deriv}}_{v_1}}[i_q,i_1] \cdot \ell^v_1[i_1, C_{1, N^{\text{deriv}}_{v_1}}]$
      \EndFor
      \State {$\triangleright$ Similar evaluations for all vector DOFs.}\label{knlline:action.eval_stop}
      \label{knlline:action.eval_wrap_up_start}
      \State \begin{tikzpicture}
        \node (A) at (0, 0) {};  %
  	    \draw[<-,dotted] (3, 0) -- (3, -0.6);
      \end{tikzpicture}
      \State \hbox{$e_1$, \ldots, $e_{\NEval}$ = Apply $g_j$'s, geometry mappings to $\partial_{j}u_i$, $\partial_{j}v_i$, coords variables.}
      \State \begin{tikzpicture}
        \node (A) at (0, 0) {};  %
	      \draw[->,dotted] (3, 0) -- (3, -0.65);
      \end{tikzpicture}
      \label{knlline:action.eval_wrap_up_stop}

      \label{knlline:action.quad_start}
      \For{$i_W \gets 0 \algTo n_W$}
        \State out$[i_W] \gets$ out$[i_W] + \Psi_1[i_W, i_q] e_1$
        \State \vdots
        \State out$[i_W] \gets$ out$[i_W] + \Psi_{\NEval}[i_W, i_q] \cdot e_{\NEval}$
      \EndFor
    \EndFor
    \label{knlline:action.quad_stop}\label{knlline:action.before_scatter}
    \For{$j \gets 0 \algTo n_W$}
      \State out$^{\text{global}}$[$\iota_{W}$[$\icell, j]]$ $\gets$ out$^{\text{global}}$[$\iota_{W}$[$\icell, j]]$ + out[$j$]\label{knlline:action.scatter}
    \EndFor
  \EndFor\label{knlline:action.complete_stop}
  \end{algorithmic}
\end{algorithm}

Algorithm~\ref{knl:action.vanilla} follows the same four phases as
Algorithm~\ref{knl:poisson_action}. The loop in line 1 corresponds to the
summation of local action operators.
Lines~\ref{knlline:action.gather_coords}--\ref{knlline:action.gather_trial_stop}
correspond to the \textit{gathering} of coordinates and the DOFs of all the
trial functions involved, using their respective indirection maps, $\iota_{U_j}$
and $\iota_{V_j}$ for the scalar and vector trial function spaces involved,
respectively.

Lines~\ref{knlline:action.eval_start}--\ref{knlline:action.eval_stop}
correspond to the derivative \textit{evaluation} phase at the quadrature nodes. The
tabulations $\Phi^{U_i}_{j}$ correspond to the evaluation of the
$\eta^{u_i}_{j}$-th partial derivative of the function space $U_i$'s reference
basis functions at the quadrature points, and $\Phi^{V_i}_{j}$ correspond to
the evaluation of the $\eta^{v_i}_{j}$-th partial derivative of $C_{i,j}$-th component
of $V_i$'s reference basis functions at the quadrature points. These matrices
are computed similarly to $\Phi_x$ and $\Phi_y$
in~\eqref{eq:poisson_tabulated_values}. The variable $\partial_{k}u$ refers to
the evaluation of the partial derivative $\eta^u_k$ of the trial function $u$.

Lines~\ref{knlline:action.quad_start}--\ref{knlline:action.quad_stop} correspond
to the \textit{quadrature} phase, where local integrals are computed using the
derivatives evaluated at the quadrature nodes. These partial derivatives of the
test function, $w$, are used in the construction of the tabulations $\left\{\Psi_1,
\ldots, \Psi_{\NEval}\right\}$, which store the evaluations of the basis
function derivatives of $W$ on the reference element at the quadrature nodes.

Finally, in line~\ref{knlline:action.scatter}, the results are
\textit{scattered} into the global DOFs corresponding to the test function
space, $W$, through the indirection map $\iota_W$.

The lowering of~\eqref{eq:generalized_bilinear_form} to the kernel in
Algorithm~\ref{knl:action.vanilla} has been explored in previous works, such as
those introducing the FFC~\cite{kirby2006compiler} and
TSFC~\cite{homolya2018tsfc} form compilers.  In this work, we develop
domain-specific transformations for the action kernel to obtain high-performance
executables on GPUs. It is important to note that the expression
in~\eqref{eq:generalized_bilinear_form} does not encompass the full range of
variational forms expressible using UFL. Surface integrals and sum
factorization are beyond the scope of this work, as the kernels derived from
them require different loop nesting and memory access patterns, necessitating a
different transformation space.

\subsection{Extensions for non-affine geometry}\label{sec:extensions_to_non_affine}
Algorithm~\ref{knl:poisson_action} describes an action operator with affine
geometry mappings between the physical and reference elements. In this work, we
follow UFL and support non-affine geometry mappings as well. The primary
distinction in Algorithm~\ref{knl:action.vanilla} lies in the treatment of the Jacobian
terms, as the derivatives of the geometry must be computed at every quadrature
point. Furthermore, the coordinates are handled similarly to other vector trial
functions. Specifically, they will involve a loop nest corresponding to the
gathering of the cell's coordinates, and another loop nest within $\iquad$ to
compute the geometry derivative terms. These derivative terms replace the
cell-wide Jacobian term found on lines 15 and 16 of
Algorithm~\ref{knl:poisson_action}. Consequently, in the case of non-affine
geometry mappings, our transformations proposed in
Section~\ref{sec:transform_space} will treat coordinates in the same manner as
other vector trial functions.

\subsection{Machine Model}\label{sec:nomenclature}
Throughout the upcoming sections, we will use a dialect of
\textsc{OpenCL}'s~\cite{munshi2009opencl} kernel language to describe a GPU
kernel in a Single Program Multiple Data (SPMD) sense. The subset of
\textsc{OpenCL} we will utilize allows for a direct mapping to Nvidia's Compute
Unified Device Architecture (CUDA) kernel language. To describe the available
data parallelism, we will adopt the nomenclature of the \textsc{OpenCL} device
model. We will consider the execution device as consisting of multiple levels of
data parallelism, where a device is composed of a set of \textit{compute units},
and each compute unit comprises a collection of \textit{processing elements}.
Additionally, kernels will be launched as a grid of \textit{work-groups}, where
each \textit{work-group} consists of a fixed number of \textit{work-items}. The
computations assigned to a work-item are executed by a processing element, while
the computations associated with a work-group are performed by a compute unit.

We use the term \textit{state-space} to denote memory requirements within a
given computational context. More concretely, a work-item's state-space
refers to the private memory (i.e. registers) allocated to that work-item.
In the context of a work-group, \textit{state-space} refers to the local memory
allocated to the group. When used in relation to a loop, it denotes the private memory whose live
range coincides with the dynamic extent of the loop. Finally, at the level of a
computational kernel, \textit{state-space} encompasses the total private
and local memory allocated within a kernel's invocation.

The table below lists a number of symbols that are used in what follows.
\begin{table}[htbp]
  \centering
\begin{tabular}{lp{70ex}}
\toprule
Symbol & Meaning\\
\midrule
\(\oclLid{i}\) & Equivalent to \texttt{get\_local\_id(i)} in CL kernel language
                for the getting \(i\text{-th}\) axis index of the work-item in a work-group.\\
\(\oclGid{i}\) & Equivalent to \texttt{get\_global\_id(i)} in CL kernel language
                for the getting \(i\text{-th}\) axis index of the work-group in
                an execution grid.\\
\(\Wmax\) & Maximum number of work-groups that can reside on a compute unit.\\
\(\Lmax\) & Maximum local memory pool per compute unit.\\
\(\mathcal{L}\) & Local memory allocated by a work-group in a compute kernel.\\
\(\FPeak\) & Peak Floating-point operations (FLOPS) throughput of a device.\\
\(\bwglobpeak\) & Peak global memory bandwidth of a device. \\
\(\bwlocpeak\) & Peak local memory bandwidth of a device.\\
\(\bwtexpeak\) & Peak on-chip read-only data cache bandwidth of a device.\\
$\aiGlobal$ & Global Arithmetic Intensity. Ratio of the total floating-point
              operations in a kernel to its global memory footprint in bytes\\
$\aiLocal$ & Local Arithmetic Intensity. Ratio of the total floating-point
             operations in a kernel to its total local memory accesses in bytes\\
\bottomrule
\end{tabular}
\caption{Nomenclature}\label{tbl:ufl2gpu_nomeclature}
\end{table}

\section{Related Work}\label{sec:related_work}
The literature on FEM action evaluation on simplex elements on GPUs can be
classified roughly according to their choice of work-division strategy across a
GPU's processing elements.

\emph{Single Work-Item per Output DOF.} Kl\"ockner et
al.~\cite{klockner2009nodal} explored optimizing Discontinuous Galerkin (DG)
solvers on GPUs. They employ a \emph{one thread per output} work-division scheme
in which they utilize one work-item to perform the computation of an output
degree of freedom (DOF). To align with the device's SIMD width, they process
multiple elements in a single work-group. For low orders, they prefetch the
reference element data to local memory within each work-group.  Although the
formulation of their transform space was generic, the parameters of their
parallelization strategy, such as padding offsets and work-group sizes, were
fitted empirically for specific discretizations of Maxwell's equations.
Komatitsch et al.~\cite{komatitsch2010high} explored optimizations to the
entire FEM pipeline for seismic applications and used one work-item per output
DOF for their action kernel. They always load their entire reference cell-data
to local memory. Ljunkvist~\cite{ljungkvist2017matrix} also used a one work-item
per DOF approach for optimizing action kernels for tensor product elements, and
they also chose their parallelization strategy parameters such as work-group
size empirically.

\emph{Single Work-Item per Cell.} Kiss et al.~\cite{kiss2012parallel} use one
work-item for the computation of a cell to parallelize their action operator.
Banas et al.~\cite{banas:2019} use a similar strategy but for performing global
assembly for the sparse matrix corresponding to the operator.  They expressed
their GPU kernels parametrically in terms of the address space of the kernel's
variables and the DOF data-layout pattern. They employed an empirical
auto-tuning strategy by traversing over all possible values of the parameters.
Markall et al.~\cite{markall2013finite} also employ a similar workload
partitioning strategy. They further study the performance effects of element
data layout and access patterns for low order elements. This strategy has also
been employed in the context of using the SIMD floating units in a CPU by
Kronbichler et al.~\cite{kronbichler2017fast} and Sun et
al.~\cite{sun2019study}. Some other approaches include Knepley et
al.'s~\cite{knepley2016finite} \emph{thread transposition} strategy, in which
they aggregate and redistribute the computational workload of multiple cells to
the work-items of a work-group to increase the work done by each work-item
for low-order Finite Elements.

In this work, we extend previously studied transformations by constructing a
transform space that contains various work-division possibilities, including
single work-item per cell, single work-item per output DOF, and configurations
in between, such as multiple output DOFs per work-item. Using an auto-tuning
strategy, we select the best-performing configuration from the parameter space.
Additionally, we also introduce the technique of tiled prefetching of
reference cell data to scratchpad memory. This is to make efficient usage of
local memory even in the case of high polynomial orders. Our auto-tuning
strategy is also responsible for determining suitable tile sizes.

\section{Transformation Space}\label{sec:transform_space}
In this section, we transform Algorithm~\ref{knl:action.vanilla} into a
high-performance GPU implementation. We primarily employ loop transformations to
achieve this. Further techniques such as data-layout reordering are out of scope for
this work. Reordering the global DOF vectors so that accesses via their
indirection maps, namely $\iota_{U_i}$, $\iota_{V_i}$, and $\iota_W$, achieve a
near-coalesced access pattern is generally crucial for performance. However, we
argue that coalescing becomes less important for function spaces with high local
DOF counts, as the resulting kernel's arithmetic intensity makes it increasingly compute
bound rather than memory bound. We refer readers to Sun et
al.~\cite{sun2019study} for a tabulation of compute intensities for weak forms
seen in practice.

Table~\ref{tab:action_param_space} presents the parameters of
Algorithm~\ref{knl:action.vanilla} corresponding to different FEM operators. We
observe that the parameters in the untransformed kernel vary with the changes
in the variational form. Consequently, certain key factors that determine
performance also vary when the variational form changes. These factors include:
(1) the amount of state-space required, which is determined by the size of
temporaries in Algorithm~\ref{knl:action.vanilla}, (2) the SIMD alignment of the
loops, which is determined by the loop length of the kernel, and (3) the
arithmetic intensity of the kernel, which is determined by the ratio of FLOPS to
the global memory accesses in the gather/scatter stages of the kernel.

\begin{table*}[htbp]
\centering
\scriptsize
\begin{tabular}{@{}clcccccccccc@{}}
\toprule
& & \multicolumn{3}{c}{Function spaces}
& \multicolumn{3}{c}{Local DOFs}
& \multicolumn{3}{c}{Derivative terms} & \\
\cmidrule(lr){3-5}\cmidrule(lr){6-8}\cmidrule(lr){9-11}
$d$ & Operator
& $\{U_i\}$ & $\{V_i\}$ & $W$
& $\{n_{U_i}\}$ & $\{n_{V_i}\}$ & $n_W$
& $\{N^{\text{deriv}}_{u_i}\}$ & $\{N^{\text{deriv}}_{v_i}\}$ & $\NEval$
& $Q$\\
\midrule
2 & Mass            & $\{P_3\}$ & $\emptyset$ & $P_3$ & $\{10\}$ & --       & 10 & $\{1\}$ & --       & 1 & 12\\
2 & Mass            & $\{P_4\}$ & $\emptyset$ & $P_4$ & $\{15\}$ & --       & 15 & $\{1\}$ & --       & 1 & 16\\
2 & Helmholtz       & $\{P_4\}$ & $\emptyset$ & $P_4$ & $\{15\}$ & --       & 15 & $\{3\}$ & --       & 3 & 16\\
3 & Helmholtz       & $\{P_4\}$ & $\emptyset$ & $P_4$ & $\{35\}$ & --       & 35 & $\{4\}$ & --       & 4 & 44\\
2 & Hyperelasticity & $\emptyset$ & $\{[P_6]^2\}$ & $[P_6]^2$ & --       & $\{28\}$ & 28 & --       & $\{8\}$ & 4 & 79\\
3 & Elasticity      & $\emptyset$ & $\{[P_4]^3\}$ & $[P_4]^3$ & --       & $\{35\}$ & 35 & --       & $\{9\}$ & 9 & 23\\
\bottomrule
\end{tabular}
\caption{\label{tab:action_param_space}Kernel parameters of
Algorithm~\ref{knl:action.vanilla} for different operators
in~\eqref{eq:generalized_bilinear_form} that are defined on a mesh in
\(\mathbb{R}^d\). The corresponding weak forms are given in
Appendix~\ref{app:fem_forms}. $P_k$ denotes the scalar-value continuous
degree-$k$ Lagrange finite element space on simplex elements.}
\end{table*}

This spread of parameters of the input kernel highlights the need for the
transformed version to adapt its state-space consumption and latency hiding
strategy based on changes in the variational form. For example, when computing the
action of a mass operator using lower-order function spaces, the entire
working set size corresponding to a cell can be stored in the register file.
However, for more complex forms that involve higher-order function spaces,
attempting to accommodate the entire working set in registers would exceed the
limited register space available to a compute unit on current GPU architectures.
This would result in register spilling, which is undesirable for achieving high
performance.

In response to these challenges, we explore two transformation strategies that
address these usage scenarios:
\begin{enumerate}
  \item The \emph{single cell per work-item} strategy utilizes the private memory
    address space for temporary variables encountered in the kernel, making it
    suitable for forms with low state-space consumption. More details of this
    strategy are provided in Section~\ref{sec:scwpi}.
  \item The \emph{parametric multi-level tiling} strategy is more suitable for
    kernels with larger working set sizes in the local assembly. It controls
    the working set size by performing tiled computations and prefetching
    accessed variables into local memory. Further explanation of this strategy
    can be found in Section~\ref{sec:multi_level_tiling}.
\end{enumerate}

The difference in the choice of variable address space between these two
strategies prevents us from unifying them parametrically within a single
transform space.

\subsection{Single Cell per Work-item}\label{sec:scwpi}
In this strategy, the kernel is transformed so that each work-item
is responsible for computing the workload arising from a single cell. All
temporary variables from Algorithm~\ref{knl:action.vanilla} are allocated
in the kernel's private address space. Additionally, to align with a GPU's
sub-group size (warp size), a work-group consisting of 32 work-items is used. A
work-group size smaller than the sub-group size is undesirable, as it results in
masked execution within a sub-group. Larger work-group sizes may be profitable
on architectures with different execution widths, such as some AMD GPUs. On the
NVIDIA GPUs evaluated in this work, however, work-group sizes of 64, 96, and 128
demonstrate no performance improvement over a work-group size of 32 and, in some
cases, degrade performance; see Appendix~\ref{app:scpwi_large_workgroups}.

To preserve memory dependence constraints in the transformed kernel, we employ
atomic addition during the scatter stage, as work-items from different
work-groups may update the same global DOF. The transformed version of the
kernel implementing this strategy is presented in
Algorithm~\ref{knl:action.scpt}.

\begin{algorithm}[H]
  \caption{Action kernel after applying \textit{Single Cell per Work-Item}
  transformation to Algorithm~\ref{knl:action.vanilla}.}\label{knl:action.scpt}
  \begin{algorithmic}[1]

  \State $\icell \gets \pqty{32 \oclGid{0}+\oclLid{0}}$

  \If{$\icell \geq N_{\text{cell}}$ }
    \Return
  \EndIf

  \For{$j \gets 0 \algTo n_{P_1}$}
    \For{$i_d \gets 0 \algTo d$}
      \State coords$[j,i_d] \gets$ coords$^{\text{global}}[\iota_{P_1}[\icell, j], i_d]$
    \EndFor
  \EndFor
  \State \vdots
  \State $\triangleright$ Identical to
    lines~\ref{knlline:action.gather_trial_start}--\ref{knlline:action.before_scatter}
    of Algorithm~\ref{knl:action.vanilla}
  \State \vdots

  \For{$j \gets 0 \algTo n_W$}
    \State \Call{GlobalAtomicAdd}{out$^{\text{global}}$[$\iota_{W}$[$\icell, j]]$,out[$j$]}
  \EndFor
  \end{algorithmic}
\end{algorithm}

\subsection{Parametric Multi-level Tiling}\label{sec:multi_level_tiling}
The single-cell-per-work-item strategy exhibits two key limitations. First, it
uses register space to store all test and trial DOFs of a cell. As the
dimensionality of the test and trial function spaces increases, register
pressure also rises, potentially resulting in register spilling.
Second, the strategy relies on the L1 cache to amortize the cost of repeated
accesses to reference cell data within a work-group. An increase in the number
of DOFs per cell or a higher quadrature degree increases the size of the
reference cell data, leading to higher cache pressure and a higher likelihood
of cache misses.

Our parametric multi-level tiling strategy addresses these challenges by using local
memory for the allocation of reference cell data to manually adapt the caching
strategy on a per-operator basis. We outline this strategy in
three steps. First, in Section~\ref{sec:tile_quadloop}, we introduce the tiling
parameters and corresponding loop-tiling transformations that utilize those
parameters. Then in Section~\ref{sec:prftch_deriv_mtrx}, we describe the prefetching
of variables accessed in the inner tiled loops. Finally, in
Section~\ref{sec:map_to_gpu_axes}, we describe the process of mapping the loops
to the GPU's hardware axes.

Throughout our explanation, we treat the evaluation and quadrature stages of
Section~\ref{sec:input-space} separately. To achieve this, we apply an array
expansion transformation~\cite{feautrier1988array} to the variables $e_1$,
$e_2$, \ldots, $e_{\NEval}$ in the loop $i_q$ of
Algorithm~\ref{knl:action.vanilla}. This transformation is followed by loop
distribution of $i_q$, enabling the independent handling of the evaluation and
quadrature stages.

\subsubsection{Tiling the Quadrature Loop}\label{sec:tile_quadloop}
A consequence of applying array expansion to the variables $e_1$, $e_2$,
\ldots, $e_{\NEval}$ of Algorithm~\ref{knl:action.vanilla} is that the kernel's
state-space size varies linearly with the number of quadrature points ($Q$). As
the value of $Q$ becomes larger, the register and local memory pressure of the kernel
also increases, potentially resulting in additional global memory access. To
adjust the amount of state space required for storing these array-expanded
variables, we introduce a parameter $\iquadtile$ to perform loop-tiling on the
$\iquad$ loop, resulting in two nested loops, $\iquadtile$ and $\iquadinner$,
through the mapping $\iquad \mapsto \quadtile\iquadtile + \iquadinner$.

\subsubsection{Prefetching \({\Phi}\)'s and \({\Psi}\)'s}\label{sec:prftch_deriv_mtrx}

The reference matrices, denoted as ${\Phi}^{U_i}_{j}$, ${\Phi}^{V_i}_{j}$ and
${\Psi}_{j}$, are repeatedly accessed during the computation of each cell in the
mesh. To optimize DRAM-to-register traffic associated with the reading of these
matrices, we employ a prefetching strategy to bring them into local memory.

Although accessing data from local memory offers lower latency and higher
bandwidth compared to global memory accesses, its use entails two
important limitations. First, the amount of local memory available to each
work-group is limited. This restriction prevents us from loading all reference
matrices into local memory, especially when dealing with higher-order function
spaces. Second, the number of work-groups that can reside on a multiprocessor
decreases as the amount of local memory allocated per work-group increases. It
is generally advantageous to have more work-groups residing on a multiprocessor,
as it allows the device's scheduler to better hide the latency of instructions
in progress.

To address these limitations, we tile the loops reading the matrices and then
prefetch their accesses within the inner loops. By reducing the lengths of
the inner loops, we effectively limit the amount of local memory allocated,
mitigating the restrictions imposed by the limited local memory available.

The choice of optimal tile sizes depends on other transform parameters such as
the execution grid size. For these reasons, we introduce the tiling parameters
$T_e^r$, $\Bqty{T_{e,U_1}^c, \ldots, T_{e,U_{N_U}}^c}$, $\Bqty{T_{e,v_1}^c,
\ldots, T_{e,V_{N_V}}^c}$, $T_q^r$ and $T_q^c$. We use these parameters to perform
the following tiling transformations:
\begin{align}
    \iquadinner&\mapsto T_e^r\irowtile+\iquadinnerinner,\\
    i_{U_k}&\mapsto T_{e,U_k}^c\icoltile^{U_k} + i_{U_k}'\qquad (k\in\left\{1,\ldots, N_U\right\}),\\
    i_{V_k}&\mapsto T_{e,V_k}^c\icoltile^{V_k} + i_{V_k}'\qquad (k\in\left\{1,\ldots, N_V\right\}),\\
    i_W&\mapsto T_{q}^r\irowtile + i_W',\\
    \iquadinner&\mapsto T_q^c\icoltile+\iquadinnerinner,
\end{align}
where the loop $\iquadinner$ was introduced in Section~\ref{sec:tile_quadloop},
$i_{U_k}$ denote the loops over the local DOFs of the scalar trial function
spaces, such as the one on line~\ref{knlline:action.loop_i1}, $i_{V_k}$ represent
the loops over the local DOFs of the vector trial function spaces, such as the
one on line~\ref{knlline:action.loop_i2}. $\irowtile$, $\icoltile$,
$\iquadinnerinner$, and $i_k'$ are the induction variables of the new loops introduced
by our tiling transformation.

\subsubsection{Mapping to a GPU-kernel grid}\label{sec:map_to_gpu_axes}
The kernel in Algorithm~\ref{knl:action.vanilla} offers two levels of
concurrency. First, the local operator can be applied independently for each
cell in the mesh, provided the scattering sum-reductions are performed
atomically. Second, in the evaluation stage, the inner products within a cell
can be computed in parallel, independently of each other. Similarly, in the
quadrature stage, the inner products within a cell can also be computed
independently of each other. It is important to note, however, that the inner
products in the quadrature stage should be performed after the inner products
in the evaluation stage, in order to avoid violating the dependencies in the
untransformed kernel.

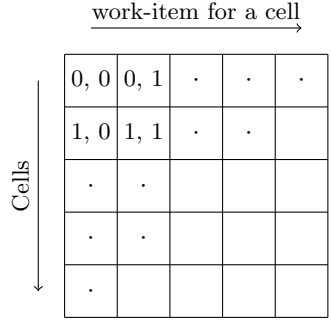
\begin{figure}
  \centering
  \begin{tikzpicture}[scale=0.7]
    \foreach \i in {0, ..., 5} {
      \draw (0, \i) -- (5, \i);
      \draw (\i, 0) -- (\i, 5);
    }

    \node at (0.5, 4.5) {\small 0, 0};
    \node at (1.5, 4.5) {\small 0, 1};

    \node at (2.5, 4.5) {$\cdot$};
    \node at (3.5, 4.5) {$\cdot$};
    \node at (4.5, 4.5) {$\cdot$};
    \node at (2.5, 3.5) {$\cdot$};
    \node at (3.5, 3.5) {$\cdot$};

    \node at (0.5, 2.5) {$\cdot$};
    \node at (0.5, 1.5) {$\cdot$};
    \node at (1.5, 2.5) {$\cdot$};
    \node at (1.5, 1.5) {$\cdot$};
    \node at (0.5, 0.5) {$\cdot$};

    \node at (0.5, 3.5) {\small 1, 0};
    \node at (1.5, 3.5) {\small 1, 1};

    \draw [->] (0.5, 5.5) -- (4.5, 5.5) node[midway, above] {\small work-item for a cell};
    \draw [->] (-0.5, 4.5) -- (-0.5, 0.5) node[midway, above, sloped, rotate=180] {\small Cells};
  \end{tikzpicture}
  \caption{Work-group used in the parametric multi-level tiling. The smallest squares
    represent one work-item annotated with its (\texttt{localid$_0$},
    \texttt{localid$_1$}) index. The shape of the grid is $\pqty{N_c\times\Nwi}$.}
  \label{fig:wg2compute}
\end{figure}

We aim to exploit both levels of concurrency by employing a two-dimensional
work-group structure with dimensions $N_c \times \Nwi$, as depicted in
Fig.~\ref{fig:wg2compute}. Here, $N_c$ represents the number of cells processed
by the work-group, and $\Nwi$ denotes the number of work-items employed for the
computations within each cell, so that $\Nwi$ work-items work in parallel to
compute the inner products within the workload of a cell.

After applying the aforementioned transformations, we present the transformed
kernel in Algorithm~\ref{knl:action.prl}. The early return shown in the
algorithm is for illustrative purposes; in practice, non-barrier statements are
predicated so that all work-items in a work-group reach the barriers.

It is important to highlight two additional consequences of these
transformations. First, the access pattern of the evaluation variables, namely
$e_1$, $e_2$, $\ldots$, $e_{N_{\text{eval}}}$, requires allocating them in the
local memory. This is because the results produced by a work-item during the
evaluation stage are consumed by potentially different work-items during the
quadrature stage.  Second, the live range of the prefetched reference matrices
in the evaluation phase is limited to the computation of the corresponding
derivative terms, and the live range of the prefetched reference matrices in the
quadrature phase is limited to the quadrature phase. Specifically, the following
expressions hold:
\begin{equation}
\begin{aligned}
  \text{LR}\left(\Phi^{U_i,\mathrm{prftch}}_{k}\right) & \cap \text{LR}\left(\Phi^{U_{j},\mathrm{prftch}}_{k}\right) & = \emptyset & \quad \left(j \neq i, k\in \left\{1, \ldots, \min\left({N_{u_i}^{\text{deriv}}, N_{u_{j}}^{\text{deriv}}}\right)\right\}\right), \\
  \text{LR}\left(\Phi^{V_{i'},\mathrm{prftch}}_{k}\right) & \cap \text{LR}\left(\Phi^{V_{j'},\mathrm{prftch}}_{k}\right) & = \emptyset & \quad  \left(j' \neq i', k\in \left\{1, \ldots, \min\left({N_{v_i}^{\text{deriv}}, N_{v_j}^{\text{deriv}}}\right)\right\}\right), \\
  \text{LR}\left(\Phi^{U_i,\mathrm{prftch}}_{k}\right) & \cap \text{LR}\left(\Phi^{V_{i'},\mathrm{prftch}}_{k'}\right) & = \emptyset & \quad \left(k\in\left\{1, \ldots, N_{u_{i}}^\mathrm{deriv}\right\}, k'\in\left\{1, \ldots, N_{v_{i'}}^\mathrm{deriv}\right\}\right), \\
  \text{LR}\left(\Phi^{U_i,\mathrm{prftch}}_{k}\right) & \cap \text{LR}\left(\Psi^{\mathrm{prftch}}_{k'}\right) & = \emptyset & \quad \left(k\in\left\{1, \ldots, N_{u_{i}}^\mathrm{deriv}\right\}, k'\in\left\{1, \ldots, N_{w}^\mathrm{deriv}\right\}\right), \\
  \text{LR}\left(\Phi^{V_{i'},\mathrm{prftch}}_{k'}\right) & \cap \text{LR}\left(\Psi^{\mathrm{prftch}}_{k'}\right) & = \emptyset & \quad \left(k\in\left\{1, \ldots, N_{v_{i}}^\mathrm{deriv}\right\}, k'\in\left\{1, \ldots, N_{w}^\mathrm{deriv}\right\}\right),
\end{aligned}
\end{equation}
where $\text{LR}(\cdot)$ represents the live range of a variable, $i,j \in$
$\left\{1, \ldots, N_U\right\}$, $i',j' \in
\left\{1, \ldots, N_V\right\}$.

We leverage this live-range disjointness to reduce the kernel's local memory
requirements. We do so by allocating a buffer $B$ of size
\begin{equation}
  \max\pqty{\max\limits_{i=1}^{N_{U}}\pqty{N^{\text{deriv}}_{u_i} T_e^r T_{e,U_i}^c}, \max\limits_{i=1}^{N_V}\pqty{N^{\text{deriv}}_{v_i} T_e^r T_{e,V_i}^c}, \NEval T_q^r T_q^c}\label{eq:size_of_buffer_B},
\end{equation}
and, using the following access mapping for the prefetched matrices:
\begin{eqnarray*}
\begin{aligned}
  \Phi^{U_i,\mathrm{prftch}}_{k}[d_0, d_1] \mapsto B[\left(k-1\right)T_{e, U_i}^cT_{e}^r + T_{e, U_i}^c d_0 & + d_1] \\
                                                                                                \; & \left(d_0 \in \left\{1, \ldots, T_{e}^r\right\}, d_1 \in \left\{1, \ldots, T_{e, U_i}^c\right\}\right)\\
                                                                                                \; & \left(i \in \left\{1, \ldots, N_U\right\}, k\in\left\{1, \ldots, N^{\text{deriv}}_{u_i}\right\}\right),
\end{aligned} \\
\begin{aligned}
  \Phi^{V_i,\mathrm{prftch}}_{k}[d_0, d_1] \mapsto B[\left(k-1\right)T_{e, V_i}^cT_{e}^r + T_{e, V_i}^c d_0 & + d_1] \\
                                                                                                \; & \left(d_0 \in \left\{1, \ldots, T_{e}^r\right\}, d_1 \in \left\{1, \ldots, T_{e, V_i}^c\right\}\right)\\
                                                                                                \; & \left(i \in \left\{1, \ldots, N_V\right\}, k\in\left\{1, \ldots, N^{\text{deriv}}_{v_i}\right\}\right),
\end{aligned} \\
\begin{aligned}
  \Psi^{\mathrm{prftch}}_{k}[d_0, d_1] \mapsto B[\left(k-1\right)T_q^cT_q^r + T_q^c d_0 & + d_1] \\
                                                                                                \; & \left(d_0 \in \left\{1, \ldots, T_q^r\right\}, d_1 \in \left\{1, \ldots, T_q^c\right\}\right)\\
                                                                                                \; & \left(k \in \left\{1,\ldots,N_{\NEval}\right\}\right).
\end{aligned}
\end{eqnarray*}

\begin{algorithm}[H]
  \caption{Action kernel after applying parametric multi-level tiling to Algorithm~\ref{knl:action.vanilla}.}
  \label{knl:action.prl}
  \begin{algorithmic}[1]
  \State $\icell \gets\pqty{N_c \oclGid{0}+\oclLid{0}}$
  \If{$\icell \geq N_{\text{cell}}$}
    \Return
  \EndIf

  \For{$j \gets 0 \algTo n_{P_1}$}
    \For{$i_d \gets 0 \algTo d$}
      \State coords$[j,i_d] \gets$ coords$^{\text{global}}[\iota_{P_1}[\icell, j], i_d]$
    \EndFor
  \EndFor

  \State $J \gets$ \Call{Jac}{coords}
  \State $D \gets$ \Call{Det}{$J$}

  \For{$\iquadtile\gets 0 \algTo \ceil{Q/\quadtile}$}
    \State $\quadtiletrunc \gets \quadtile \text{ \bf if } \iquadtile < \floor{Q/\quadtile} \text{ \bf else } \pqty{Q\bmod \quadtile}$
    \For{$\irowtile\gets 0 \algTo \ceil{\quadtiletrunc/T_e^r}$}
      \State $T_e^{(r,\mathrm{trunc})}\gets T_e^r\text{ \bf if } \irowtile < \floor{\quadtiletrunc/T_e^r} \text{ \bf else } \pqty{\quadtiletrunc\bmod T_e^r}$
      \For{$i_q' \gets 0 \algTo \ceil{T_e^{(r,\mathrm{trunc})}/\Nwi}$}
        \State $\partial_{1}u_1[i_q']  \quad\;\;\;\gets 0$
        \State \vdots
        \State $\partial_{N^{\text{deriv}}_{u_1}}u_1[i_q']  \gets 0$
      \EndFor
      \For{$\icoltile\gets 0 \algTo \ceil{n_{U_1}/T_{e,U_1}^c}$}
        \State $T_{e,U_1}^{(c,\mathrm{trunc})}\gets T_{e,U_1}^{c}\text{ \bf if } \icoltile < \floor{n_{U_1}/T_{e,U_1}^c} \text{ \bf else } \pqty{n_{U_1}\bmod T_{e,U_1}^c}$
        \For{$j' \gets 0 \algTo T_{e,U_1}^{(c,\mathrm{trunc})}$}
          \State $j\gets T_{e,U_1}^c\icoltile+j'$
          \State $\ell^u_1[j'] \gets u_1^{\text{global}}[\iota_{U_1}[\icell, j]]$\label{knlline:action.prl.dof_gather}
        \EndFor
  \algstore{actionprlpseudocode1}
  \end{algorithmic}
\end{algorithm}
\begin{algorithm}[H]
\begin{algorithmic}[1]
  \algrestore{actionprlpseudocode1}

        \For{$i \gets 0 \algTo \ceil{\pqty{T_e^{(r,\mathrm{trunc})}T_{e,U_1}^{(c,\mathrm{trunc})}}/(N_c\Nwi)}$}
          \If{$N_c\Nwi i + \Nwi\oclLid{0} + \oclLid{1} \geq T_e^{(r,\mathrm{trunc})}T_{e,U_1}^{(c,\mathrm{trunc})}$}
            \State continue
          \EndIf
          \State $i' \gets \floor{\left(N_c\Nwi i + \Nwi\oclLid{0} + \oclLid{1}\right) / T_{e,U_1}^{(c,\mathrm{trunc})}}$
          \State $j' \gets \left(N_c\Nwi i + \Nwi\oclLid{0} + \oclLid{1}\right) \bmod T_{e,U_1}^{(c,\mathrm{trunc})}$
          \State $I' \gets \iquadtile\quadtile + T_e^r\irowtile+i'$
          \State $J' \gets T_{e,U_1}^c\icoltile + j'$
          \State{}
          \State $\Phi^{U_1,\mathrm{prftch}}_{1}[i', j']\;\;\; \gets \Phi^{U_1}_{1}[I',J']$
          \State \vdots
          \State $\Phi^{U_1,\mathrm{prftch}}_{N^{\text{deriv}}_{u_1}}[i',j'] \gets \Phi^{U_1}_{N^{\text{deriv}}_{u_1}}[I',J']$
        \EndFor

        \State \Call{WorkGroupBarrier}{\null}\label{knlline:action.prl.barrrier_eval1}
        \label{knlline:action.prl.eval_start}

        \For{$i_q^{\text{outer}\prime}\gets 0 \algTo \ceil{T_e^{(r,\mathrm{trunc})}/\Nwi}$}
          \State $i_q' \gets  \Nwi i_q^{\text{outer}\prime} +  \oclLid{1}$

          \If{ $i_q' \geq T_e^{(r,\mathrm{trunc})}$}\label{knlline:action.prl.pred1}
            \State continue
          \EndIf

          \For{$j' \gets 0  \algTo T_{e,U_1}^{(c,\mathrm{trunc})}$}
            \State $\partial_{1}u_1[i_q^{\text{outer}\prime}] \;\;\;\quad\gets \partial_{1}u_1[i_q^{\text{outer}\prime}] + {\Phi^{U_1,\text{prftch}}_{1}}[i_q', j']\cdot \ell^u_1[j']$
            \State \vdots
            \State $\partial_{N^{\text{deriv}}_{u_1}}u_1[i_q^{\text{outer}\prime}] \gets \partial_{N^{\text{deriv}}_{u_1}}u_1[i_q^{\text{outer}\prime}] + {\Phi^{U_1,\text{prftch}}_{N^{\text{deriv}}_{u_1}}}[i_q', j']\cdot \ell^u_1[j']$
          \EndFor %
        \EndFor %

        \State {}
        \State \Call{WorkGroupBarrier}{\null}\label{knlline:action.prl.barrrier_eval2}

      \EndFor %

      \State $\triangleright$ Similar handling of $u_2$, \ldots, $u_{N_U}$, $v_1$, \ldots, $v_{N_V}$.

      \label{knlline:action.prl.eval_end}
      \For{$i_q^{\text{outer}\prime}\gets 0 \algTo \ceil{T_e^{(r,\mathrm{trunc})}/\Nwi}$}
        \State $i_q' \gets  + \Nwi i_q^{\text{outer}\prime} +  \oclLid{1}$
        \If{ $i_q' \geq T_e^{(r,\mathrm{trunc})}$}
          \State continue
        \EndIf

        \State \begin{tikzpicture}
          \node (A) at (0, 0) {};  %
          \draw[<-,dotted] (4, 0) -- (4, -0.2);
        \end{tikzpicture}
        \State \hbox{$e_1[\oclLid{0},T_e^r\irowtile+i_q']$, \ldots,
        $e_{\NEval}[\oclLid{0},T_e^r\irowtile+i_q']$ = \parbox[t]{0.3\textwidth}{Apply
        $g_{i,j}$, geometry mappings to all $\partial_{j}u_i[i_q^{\text{outer}\prime}]$,
        $\partial_{j}v_i[i_q^{\text{outer}\prime}]$ variables.}}
        \State \begin{tikzpicture}
          \node (A) at (0, 0) {};  %
          \draw[->,dotted] (4, 0) -- (4, -0.3);
        \end{tikzpicture}

      \EndFor %
    \EndFor
    \State \Call{WorkGroupBarrier}{\null}\label{knlline:action.prl.barrrier_eval_quad_sync}

  \algstore{actionprlpseudocode2}
  \end{algorithmic}
\end{algorithm}
\begin{algorithm}[H]
\begin{algorithmic}[1]
  \algrestore{actionprlpseudocode2}

    \For{$\irowtile\gets 0 \algTo \ceil{n_W/T_q^r}$}
      \State $T_q^{(r,\mathrm{trunc})} \gets T_q^r \text{\bf if } \irowtile < \floor{n_W/T_q^r} \text{\bf else } n_W\bmod T_q^r$
      \For{$i_W''\gets 0 \algTo \ceil{T_q^{(r,\mathrm{trunc})}/\Nwi}$}
        \State out$[i_W'']\gets 0$
      \EndFor

      \For{$\icoltile\gets 0 \algTo \ceil{\quadtiletrunc/T_q^c}$}
        \State $T_q^{(c,\mathrm{trunc})} \gets T_q^c \text{\bf if } \icoltile < \floor{\quadtiletrunc/T_q^c} \text{\bf else } \quadtiletrunc\bmod T_q^c$

        \For{$i \gets 0 \algTo \ceil{\pqty{T_q^{(r,\mathrm{trunc})}T_q^{(c,\mathrm{trunc})}}/(N_c\Nwi)}$}
          \If{$N_c\Nwi i + \Nwi\oclLid{0} + \oclLid{1} \geq T_q^{(r,\mathrm{trunc})}T_q^{(c,\mathrm{trunc})}$}
            \State continue
          \EndIf
          \State $i' \gets \floor{\left(N_c\Nwi i + \Nwi\oclLid{0} + \oclLid{1}\right) / T_q^{(c,\mathrm{trunc})}}$
          \State $j' \gets \left(N_c\Nwi i + \Nwi\oclLid{0} + \oclLid{1}\right) \bmod T_q^{(c,\mathrm{trunc})}$
          \State $I' \gets T_q^r\irowtile+i'$
          \State $J' \gets \iquadtile\quadtile+T_q^c\icoltile+j'$
          \State $\Psi^{\mathrm{prftch}}_{1}[i', j'] \gets \Psi_1[I',J']$
          \State \vdots
          \State $\Psi^{\mathrm{prftch}}_{\NEval}[i',j'] \gets \Psi_{\NEval}[I',J']$
        \EndFor
        \State \Call{WorkGroupBarrier}{\null}\label{knlline:action.prl.barrrier_quad1}
        \For{$\iodof'' \gets 0 \algTo \ceil{T_q^{(r,\mathrm{trunc})}/\Nwi}$}\label{knlline:action.prl.quad_start}
          \State $j\gets \Nwi\iodof'' + \oclLid{1} $
          \If{$j \geq T_q^{(r,\mathrm{trunc})}$}\label{knlline:action.prl.pred2}
            \State continue
          \EndIf
          \For{$\iq'\gets 0 \algTo T_q^{(c,\mathrm{trunc})}$}
            \State $\iq\gets \iq'+T_q^c\icoltile$
            \State out$[\iodof''] \gets \text{out}[\iodof''] + \Psi^{\text{prftch}}_1[j, \iq'] \cdot e_1[\oclLid{0}, \iq]$
            \State \vdots
            \State out$[\iodof''] \gets \text{out}[\iodof''] + \Psi^{\text{prftch}}_{\NEval}[j, \iq'] \cdot e_{\NEval}[\oclLid{0}, \iq]$
          \EndFor %
        \EndFor\label{knlline:action.prl.quad_end}%
        \State \Call{WorkGroupBarrier}{\null}\label{knlline:action.prl.barrrier_quad2}
      \EndFor %
      \For{$i_W' \gets 0 \algTo \ceil{T_q^{(r,\mathrm{trunc})}/\Nwi}$}
        \State $i_W\gets \Nwi i_W' + \oclLid{1} $
        \If{$i_W \geq T_q^{(r,\mathrm{trunc})}$}
          \State continue
        \EndIf
        \State \Call{GlobalAtomicAdd}{$\text{out}^{\text{global}}[\iota_W[\icell,T_q^r\irowtile+i_W]]$, $\text{out}[i_W']$}\label{knlline:action.prl.dof_scatter}
      \EndFor %
    \EndFor %
    \State {}
    \State \Call{WorkGroupBarrier}{\null}\label{knlline:action.prl.barrrier_quadtile_sync}
  \EndFor %
  \end{algorithmic}
\end{algorithm}

\subsubsection{Quantities of Interest}\label{sec:param_space_qois}

The transformed kernel in Algorithm~\ref{knl:action.prl} is parametric in
$\quadtile$, $T_e^r$, $\left\{T_{e,U_i}^c\right\}_{i\in\left\{1, \ldots,
N_U\right\}}$, $\left\{T_{e,V_i}^c\right\}_{i \in \left\{1, \ldots,
N_V\right\}}$, $T_q^r$, $T_q^c$ $N_c$, and $\Nwi$. These parameters enable
tailoring the implementation to a given variational form via auto-tuning.
To fully define the transformation space, it is necessary to fix the
allowable values for each parameter. This involves establishing bounds for the
parameters based on their impact on a set of metrics that significantly
influence the transformed kernel's performance. The metrics considered in our
analysis are as follows.

\begin{description}
  \item[Number of synchronizations] The number of times work-group wide
  barriers are executed within a work-group from start to end.
\begin{align}
  f(q) &= \underbrace{2\sum\limits_{i=1}^{N_U}\pqty{\ceil{\frac{q}{T_e^r}}\ceil{\frac{n_{U_i}}{T_{e,U_i}^c}}} + 2\sum\limits_{i=1}^{N_V}\pqty{\ceil{\frac{q}{T_e^r}}\ceil{\frac{n_{V_i}}{T_{e,V_i}^c}}}}_{\circledone} + \underbrace{2\ceil{\frac{n_W}{T_q^r}}\ceil{\frac{q}{T_q^c}}}_{\circledtwo} + \underbrace{2\cdot\mathbb{1}_{\{q\neq0\}}}_{\circledthree}\label{eq:nsync_fq}\\
  \Nsync &= \floor{\frac{Q}{\quadtile}}f(\quadtile) + f(Q \bmod \quadtile).
\end{align}
In the above equation, $f(q)$ represents the number of work-group wide barriers
executed in an iteration of $\iquadtile$ that computes the partial sum over $q$
quadrature nodes. As defined in~\eqref{eq:nsync_fq}, $f(q)$ is a sum of 3 terms,
\circledone, \circledtwo and \circledthree. \circledone is the number of
barriers in the evaluation phase, specifically the barriers on
lines~\ref{knlline:action.prl.barrrier_eval1},~\ref{knlline:action.prl.barrrier_eval2}
of Algorithm~\ref{knl:action.prl}. \circledtwo is the number of barriers in the
quadrature phase, i.e. the barriers on
lines~\ref{knlline:action.prl.barrrier_quad1},~\ref{knlline:action.prl.barrrier_quad2}
of Algorithm~\ref{knl:action.prl}. Finally, \circledthree is the number of
barriers needed for synchronization between the evaluation and quadrature phase
(line~\ref{knlline:action.prl.barrrier_eval_quad_sync} of
Algorithm~\ref{knl:action.prl}) and synchronization needed between iterations of
$\iquadtile$-loop (line~\ref{knlline:action.prl.barrrier_quadtile_sync} of
Algorithm~\ref{knl:action.prl}). $\mathbb{1}_{\{q\neq 0\}}$ denotes the
indicator function that equals $1$ when $q \neq 0$ and $0$ otherwise. If all
other kernel features are kept the same, having lower $\Nsync$ favors the
kernel's performance.

\item[Local Memory Allocated] Aggregating all the local memory variables of
  the kernel in Algorithm~\ref{knl:action.prl}, we arrive at:
\begin{align}
\mathcal{L} = \underbrace{\max\pqty{\max\limits_{i=1}^{N_{U}}\pqty{N^{\text{deriv}}_{u_i} T_e^r T_{e,U_i}^c}, \max\limits_{i=1}^{N_V}\pqty{N^{\text{deriv}}_{v_i} T_e^r T_{e,V_i}^c}, \NEval T_q^r T_q^c}}_{\circledone} + \underbrace{N_c\quadtile\NEval}_{\circledtwo}.
\end{align}
In the equation above, $\mathcal{L}$ is the sum of two terms,
\circledone and \circledtwo. \circledone corresponds to the local memory
allocations for the prefetched reference matrices
(see~\eqref{eq:size_of_buffer_B}), while \circledtwo captures the local
memory allocations required to store the results of the evaluation phase, i.e.,
$\left\{e_i\right\}_{i \in \left\{1, \ldots, \NEval\right\}}$. It must be noted
that we count $\mathcal{L}$ in terms of number of floating point words which
depends on the floating point precision employed.

If all other kernel features are kept the same, having lower local memory
usage ($\mathcal{L}$) tends to improve the kernel's performance as discussed in
Section~\ref{sec:prftch_deriv_mtrx}.
\item[SIMD Efficiency ($\etasimd$)] GPUs execute instructions in a SIMD
fashion. This implies that if the work-group size is not a multiple of
device's SIMD width (i.e. 32), an inefficiency arises from masked vector
lanes. We capture this effect using the following quantity which we call
SIMD efficiency:
\begin{align}
\etasimd = \frac{N_c\Nwi}{32\ceil{\pqty{N_c\Nwi}/32}}.
\end{align}
If all other kernel features are kept the same, having higher \(\etasimd\) is
favorable for higher performance.
\item[Predication Efficiency ($\etapred$)] Loop tiling inherently leads to conditionals, like
lines~\ref{knlline:action.prl.pred1},~\ref{knlline:action.prl.pred2} of
Algorithm~\ref{knl:action.prl}, and results in diverged execution within a
work-group. On a GPU's SIMD-like execution model, this leads to a loss of
efficiency due to masked vector lanes. We characterize this inefficiency as the
ratio of usable floating point operations performed to the number of floating
point operations that would have been performed had there been no masking.
Consequently, we define
\begin{align}
  \etapred = \frac{\text{Ops}^{\text{usable\quad\;}}}{\text{Ops}^{\text{performed}}},
\end{align}
where $\text{Ops}^{\text{usable}}$ is the total number of floating-point
operations in the untransformed kernel (Algorithm~\ref{knl:action.vanilla})
particularly in the small matrix-vector multiplies for a single cell.
Aggregating the multiply and add operations from the evaluation and quadrature
phase, we obtain
\begin{equation}\label{eq:ops_usable}
  \text{Ops}^{\text{usable}} = \sum\limits_{i=1}^{N_U}\pqty{2\cdot N^{\text{deriv}}_{u_i}  Q n_{U_i}} + \sum\limits_{i=1}^{N_V}\pqty{2\cdot N^{\text{deriv}}_{v_i}  Q n_{V_i}} + 2\cdot \NEval Q n_W.
\end{equation}
$\text{Ops}^{\text{performed}}$ is the number of corresponding floating point
operations performed for a single cell by $\Nwi$ work-items in
Algorithm~\ref{knl:action.prl}. After accounting for the tiling of the quadrature loop
by $T^Q$, we get
\begin{align}
\text{Ops}^{\text{performed}} = \floor{\frac{Q}{\quadtile}}f_e(\quadtile) + f_e(Q \bmod \quadtile) + \floor{\frac{Q}{\quadtile}}f_q(\quadtile) + f_q(Q \bmod \quadtile),
\end{align}
where $f_e(x)$ is the number of floating point operations performed in the
evaluation phase for $x$ quadrature nodes being evaluated in the
$\iquadtile$-loop, and, $f_q(x)$ is the number of floating point
operations performed in the quadrature phase for $x$ quadrature nodes being
evaluated in the $\iquadtile$-loop. Aggregating the floating point operations
from
lines~\ref{knlline:action.prl.eval_start}--\ref{knlline:action.prl.eval_end} of
Algorithm~\ref{knl:action.prl}, we obtain
\begin{equation}
  f_{e}(x) = \left(\Nwi\floor{\frac{x}{T_e^r}}\ceil{\frac{T_e^r}{\Nwi}} + \Nwi\ceil{\frac{x \bmod T_e^r}{\Nwi}}\right) \left(\sum\limits_{i=1}^{N_U} N^{\text{deriv}}_{u_i}n_{U_i} + \sum\limits_{i=1}^{N_V}N^{\text{deriv}}_{v_i}n_{V_i}\right).\\
\end{equation}
Similarly, aggregating the floating point operations from lines~~\ref{knlline:action.prl.quad_start}--\ref{knlline:action.prl.quad_end} of Algorithm~\ref{knl:action.prl}, we obtain
\begin{align}
  f_{q}(x) = \floor{\frac{n_W}{T_q^r}}\Nwi\ceil{\frac{T_q^r}{\Nwi}}\NEval  T_q^r x + \Nwi\ceil{\frac{n_W \bmod T_q^r}{\Nwi}} \NEval x.
\end{align}

If all other kernel features are invariant, higher $\etapred$ typically
corresponds to a better-performing kernel.
\item \emph{Local Memory Aliasing Efficiency}: From our discussion in
Section~\ref{sec:map_to_gpu_axes}, we use the same local memory buffer for the
tiled prefetches in both the evaluation and quadrature stages. However, the tile
size choices might lead to inefficient use of the allocated buffer space. We
define this efficiency factor as the ratio of the minimum of the allocated local
memory being used for an evaluation phase for the prefetched reference
derivative matrices to the size of the allocated local memory for the prefetched
tiles. Hence we say:
\begin{equation}
  \etaalias = \frac{\min\pqty{\min\limits_{i=1}^{N_U}\pqty{N^{\text{deriv}}_{u_i} T_e^rT_{e,U_i}^c}, \min\limits_{i=1}^{N_V}\pqty{N^{\text{deriv}}_{v_i} T_e^rT_{e,V_i}^c}, \NEval T_q^rT_q^c}}{\max\pqty{\max\limits_{i=1}^{N_U}\pqty{N^{\text{deriv}}_{u_i} T_e^rT_{e,U_i}^c}, \max\limits_{i=1}^{N_V}\pqty{N^{\text{deriv}}_{v_i} T_e^rT_{e,V_i}^c}, \NEval T_q^rT_q^c}}.
\end{equation}

If all other kernel features are kept the same, higher $\etaalias$ is
generally favorable for kernel performance as it allows for better usage of the
allocated local memory which competes with the device's latency hiding
capabilities.
\end{description}

\subsubsection{Parameter Search Space}\label{sec:param_search_space}
In this section, we use the quantities of interest defined in
Section~\ref{sec:param_space_qois} to describe our parameter space.
We limit the size of the search space mainly through the enforcement
of four constraints:

\begin{enumerate}[leftmargin=20ex, label={\it Constraint \arabic*. }]
  \item \begin{align}
  \quadtile & \in \Bqty{Q, \ceil{\frac{Q}{2}}, \ceil{\frac{Q}{3}}, \ldots}, \\
  T_e^r     & \in \Bqty{\quadtile, \ceil{\frac{\quadtile}{2}}, \ceil{\frac{\quadtile}{3}}, \ldots}, \\
  T_{e,U_k}^c & \in \Bqty{n_{U_k}, \ceil{\frac{n_{U_k}}{2}}, \ceil{\frac{n_{U_k}}{3}}, \ldots}, \\
  T_{e,V_k}^c & \in \Bqty{n_{V_k}, \ceil{\frac{n_{V_k}}{2}}, \ceil{\frac{n_{V_k}}{3}}, \ldots}, \\
  T_q^r     & \in \Bqty{n_W, \ceil{\frac{n_W}{2}}, \ceil{\frac{n_W}{3}}, \ldots}, \\
  T_q^c     & \in \Bqty{\quadtile, \ceil{\frac{\quadtile}{2}}, \ceil{\frac{\quadtile}{3}}, \ldots},
\end{align}

\item \(\etaalias \geq \etaaliasMin\),
\item \(\etasimd \geq \etasimdMin\), and,
\item $N_c\Nwi \leq \NcNwiMax$.
\end{enumerate}

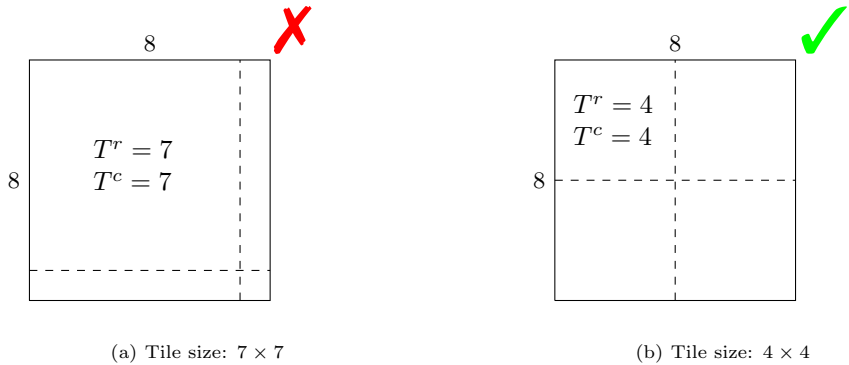
\begin{figure}
  \centering
  \begin{subfigure}{.5\textwidth}
    \begin{center}
    \begin{tikzpicture}[scale=0.4]
      \draw (0, 0) -| (8, 8) -| (0,0) node[pos=0.25,above] {\small $8$} node[pos=0.75,left] {\small $8$};
      \node[text width=1.5cm] at (4, 4.5) {$T^r = 7$\hfill\\$T^c = 7$};
      \draw[dashed] (7, 0) -- (7, 8);
      \draw[dashed] (0, 1) -- (8, 1);
      \node[text width=1.5cm] at (10, 9) {\Huge\color{red}\xmark};
    \end{tikzpicture}
    \end{center}
    \caption{Tile size: $7 \times 7$}
    \label{fig:ex_bad_tile}
  \end{subfigure}%
  \begin{subfigure}{.5\textwidth}
    \begin{center}
    \begin{tikzpicture}[scale=0.4]
      \draw (0, 0) -| (8, 8) -| (0,0) node[pos=0.25,above] {\small $8$} node[pos=0.75,left] {\small $8$};
      \node[text width=1.5cm] at (2.5, 6) {$T^r = 4$\hfill\\$T^c = 4$};
      \draw[dashed] (4, 0) -- (4, 8);
      \draw[dashed] (0, 4) -- (8, 4);
      \node[text width=1.5cm] at (10, 9) {\Huge\color{green}\cmark};
    \end{tikzpicture}
    \end{center}
    \caption{Tile size: $4 \times 4$}
    \label{fig:ex_good_tile}
  \end{subfigure}
  \caption{Both (a) and (b) are ways of tiling an $8\times8$ matrix with a tile shape $T^r\times T^c$.
    Both variants incur $\Nsync = 2$, but variant (a) will use more local memory
    than (b). Configurations like (a) are not considered in our parameter space
    as (b) has lower $\mathcal{L}$ with the same $\Nsync$ which generally
    would correspond to superior performance.}
  \label{fig:tile_constraint_1}
\end{figure}

By applying constraints (1) and (2), we can enumerate all possible tile sizes
within the parameter space. Similarly, constraints (3) and (4) allow for the
enumeration of all valid $N_c$-$\Nwi$ pairs. Constraint (1) has been chosen to
only include tile sizes that allocate the minimum amount of local memory for a
given $\Nsync$. An illustration of constraint (1) is provided in
Fig.~\ref{fig:tile_constraint_1}.

Furthermore, considering our preference for higher values of $\etasimd$ and
$\eta_{\text{alias}}$, we impose lower bounds $\etaaliasMin$ and $\etasimdMin$
on these quantities, along with an upper bound $\NcNwiMax$ on $N_c\Nwi$. These
bounds can be adjusted to either expand or reduce the size of the parameter
space; for the results in Section~\ref{sec:eval_throughput}, we use
$\etaaliasMin = 0.8$, $\etasimdMin = 0.97$, and $\NcNwiMax = 256$. These bounds
were chosen empirically to balance the performance of the obtained
configurations with search space size. We study the sensitivity of the resulting
throughput to $\etaaliasMin$ and $\etasimdMin$ in
Section~\ref{sec:hyperparam_sensitivity_tests}.

\begin{figure}
  \centering
  \begin{subfigure}{.45\textwidth}
    \centering
    \includegraphics[width=\textwidth]{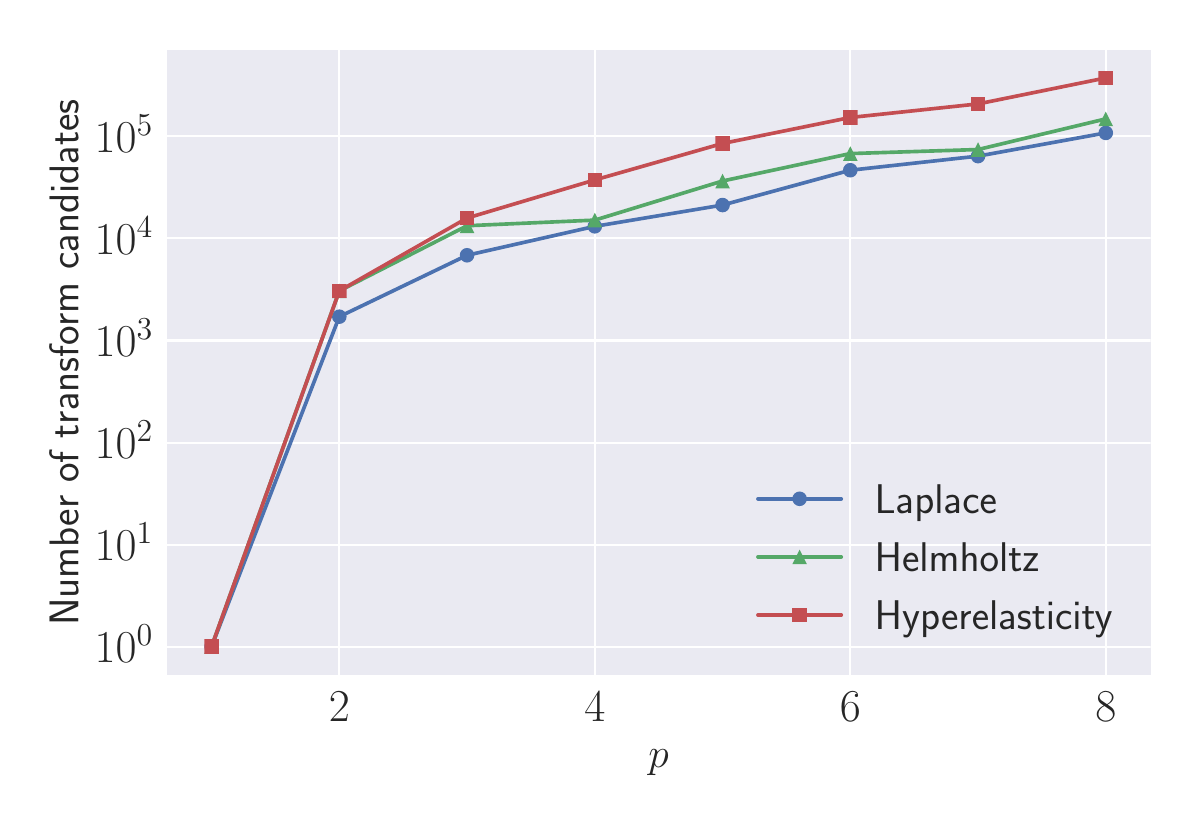}
    \caption{Number of parameters in search space for\\ 2D-FEM variational forms.}
  \end{subfigure}%
  \begin{subfigure}{.45\textwidth}
    \centering
    \includegraphics[width=\textwidth]{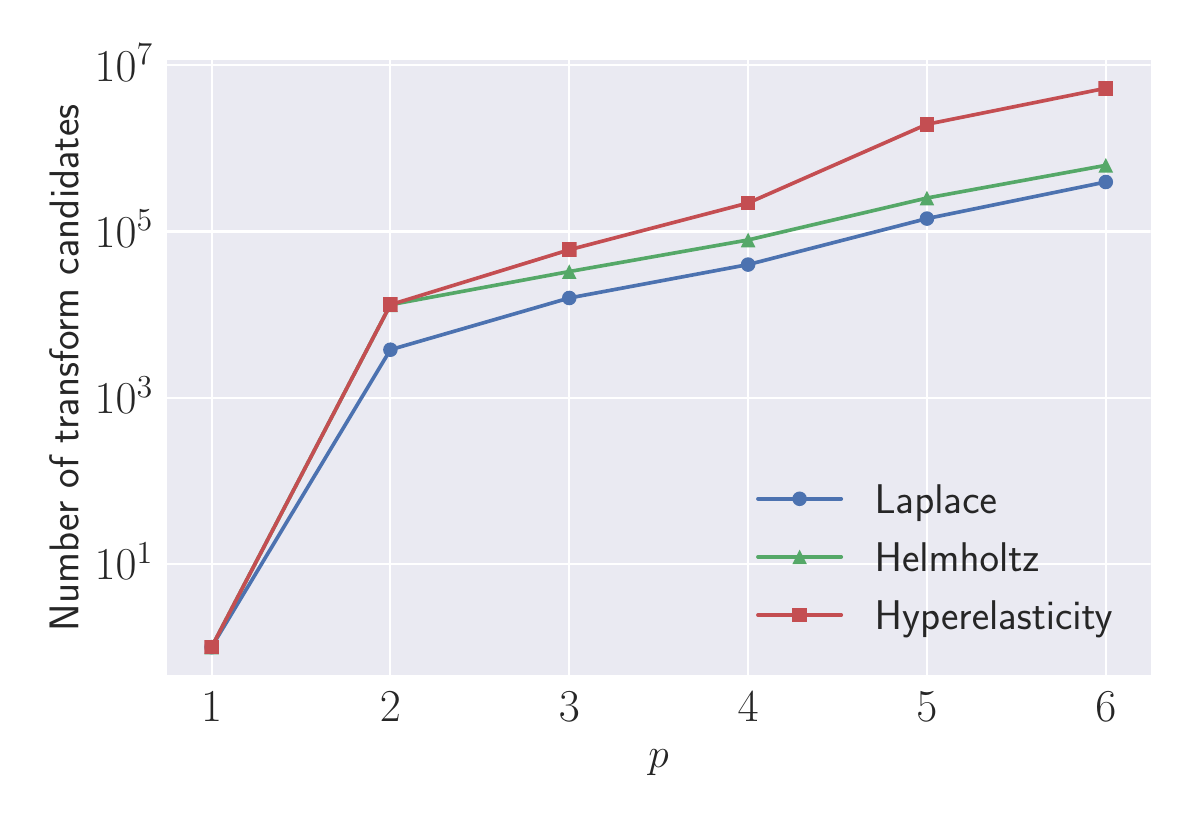}
    \caption{Number of parameters in search space for\\ 3D-FEM variational forms.}
  \end{subfigure}
  \caption{Cardinality of search space as per bounds chosen in Section~\ref{sec:param_search_space}.}\label{fig:search_space_cardinality}
\end{figure}

Section~\ref{sec:perf_eval} shows that
the proposed search space includes candidates achieving near-roofline
throughput. Additionally, Figure~\ref{fig:search_space_cardinality} illustrates
how the size of the search space varies across different variational forms.

\subsubsection{Cost Model}\label{sec:cost_model}
In Section~\ref{sec:param_search_space}, we defined a bounded transformation
space which we enumerate.  However, we see from
Figure~\ref{fig:search_space_cardinality} that the size of this set remains
large enough that exhaustively executing all possible candidates in order to
identify the best-performing variant is impractical. To address this, we propose
ranking the candidates in our search space based on a heuristic cost model
and selecting the $b$ top-ranked kernels to be used with auto-tuning as
described below.

To rank the configurations in our transformation space, we develop a
cost model that approximates the execution time by considering the time
associated with global and local memory accesses. We denote this as
$t_{\text{heur}}$:
\begin{equation}
  \label{eq:ufl2gpu_costmodel}
  t_{\text{heur}} = \frac{\text{Total Global Memory accesses}}{\bwglobmodel} + \frac{\text{Total Local Memory accesses}}{\bwlocmodel}
\end{equation}

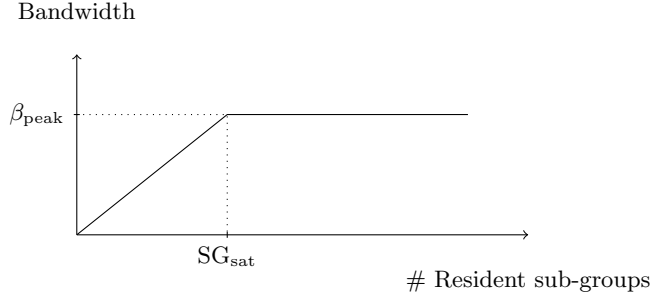
\begin{figure}
\centering
\begin{tikzpicture}[scale=0.4]
  \node[coordinate] (O) at ( 0, 0) {} ;
  \node[coordinate] (X) at (15, 0) {};
  \node[coordinate] (Y) at (0, 6) {};
  \node[coordinate] (flatStart) at (5, 4) {};
  \node[coordinate] (flatEnd) at (13, 4) {};
  \node[coordinate] (satYStart) at (-0.1, 4) {};
  \node[coordinate] (satYEnd)   at ( 0.1, 4) {};
  \node[coordinate] (satXStart) at (5, -0.1) {};
  \node[coordinate] (satXEnd)   at (5, 0.1) {};

  \draw[->] (O) -- node[pos=1,below=10] {\small \# Resident sub-groups} (X);
  \draw[->] (O) -- node[pos=1, above=10] {\small Bandwidth} (Y);
  \draw (O) -- (flatStart) -- (flatEnd);

  \draw[dotted] (0, 4) -- (flatStart) -- (5, 0);

  \draw (satYStart) -- (satYEnd);
  \draw (satXStart) -- (satXEnd);

  \node[below] at (satXStart) {\small $\mathrm{SG}_{\mathrm{sat}}$};
  \node[left] at (satYStart) {\small $\beta_{\mathrm{peak}}$};
\end{tikzpicture}
\caption{Modeled bandwidth}\label{fig:bw_model}
\end{figure}

Further, we approximate the access time to global and local memory using a
bandwidth model, which assumes a linear relationship between bandwidth and the
number of sub-groups resident on the device. The bandwidth model saturates after
reaching the peak value of the device. Figure~\ref{fig:bw_model} illustrates our
bandwidth model. Its shape is motivated by observation of local memory bandwidth
varying with work-group size in SHOC~\cite{danalis2010scalable}. We determine
$\beta^{\text{peak}}_{*}$ and $\text{SG}_{\text{sat},*}$ by fitting to SHOC
measurements. We denote our modeled bandwidths
as $\bwglobmodel$ and $\bwlocmodel$:
\begin{align}
  &\bwglobmodel = \min\pqty{\bwglobpeak, \frac{\bwglobpeak}{\sgSatGlobal} \NresideEff},\\
  &\bwlocmodel = \min\pqty{\bwlocpeak, \frac{\bwlocpeak}{\sgSatShared} \NresideEff}.
\end{align}

We now compute expressions for the number of memory accesses and the effective
number of residing sub-groups i.e. $\NresideEff$. We model the number of global
memory accesses per cell as the sum of those during the \textit{gather} /
\textit{scatter} stages along with the accesses associated with the reference
element data, specifically the derivative matrices and the quadrature weights.
\begin{equation}
  \text{Total Global Memory Accesses} = \left(M^g_{\text{gather}} + M^g_{\text{scatter}} + M^g_{\text{ref}}\right) \cdot s,
\end{equation}
where $s$ is the size of a floating point word for the precision used in the variational form,
\begin{align*}
  M^g_{\text{gather}} &= \pqty{\floor{\frac{Q}{\quadtile}}\ceil{\frac{\quadtile}{T_e^r}} + \ceil{\frac{Q \bmod \quadtile}{T_e^r}}}\pqty{\sum\limits_{i=1}^{N_U}n_{U_i} + d\cdot\sum\limits_{i=1}^{N_V} n_{V_i}} + \underbrace{d(d+1)}_{\text{coords}},
  &\quad \text{(Line~\ref{knlline:action.prl.dof_gather}
  and~\ref{knlline:action.gather_coords}, Alg.~\ref{knl:action.prl})} \\
  M^g_{\text{scatter}} &= \ceil{\frac{Q}{\quadtile}} n_W,
  &\quad \text{(Line~\ref{knlline:action.prl.dof_scatter}, Alg.~\ref{knl:action.prl})}  \\
  M^g_{\text{ref}} &= \frac{1}{N_c} \pqty{\sum\limits_{i=1}^{N_U}n_{U_i} Q + \sum\limits_{i=1}^{N_V}n_{V_i} Q + Q}.
  &\quad \text{(Reference matrices)}
\end{align*}

We model the number of local memory accesses as the sum of reads
to derivative matrices during the evaluation stage,
writes to the evaluation result variables i.e. $e_i$,
and the reads from the evaluation result variables and the derivative matrices
during the quadrature stage.
\begin{equation}
  \text{Total Local Memory Accesses} = \left(M^l_{\text{eval,read}} + M^l_{\text{eval,write}} + M^l_{\text{quad,read}}\right) \cdot s
\end{equation}
where $s$ is the size of a floating point word for the precision used in the variational form,
\begin{align*}
  M^l_{\text{eval,read}} &= \sum\limits_{i=1}^{N_U} \pqty{N^{\text{deriv}}_{u_i}n_{U_i}Q} + \sum\limits_{i=1}^{N_V} \pqty{N^{\text{deriv}}_{v_i}n_{V_i}Q},
  &\quad \text{(Lines~\ref{knlline:action.prl.eval_start}--\ref{knlline:action.prl.eval_end}, Alg.~\ref{knl:action.prl})} \\
  M^l_{\text{eval,write}} &= \NEval Q,
  &\quad \text{(Writes to $\left\{e_1,\ldots,e_{\NEval}\right\}$)}  \\
  M^l_{\text{quad,read}} &= \NEval Q + \NEval Q n_W.
  &\quad \text{(Lines~\ref{knlline:action.prl.quad_start}--\ref{knlline:action.prl.quad_end}, Alg.~\ref{knl:action.prl})}
\end{align*}

The number of residing sub-groups on a multi-processor can be modeled by
assuming that it is limited either by the number of work-groups that can reside
on a multiprocessor or by the total local memory that can be allocated to the
resident work-groups on a multi-processor. And we model an effective number of
resident work-groups, denoted by $\NresideEff$, by accounting for the
efficiency factors $\etasimd$ and $\etapred$:
\begin{align}
\Nreside         = & \; \ceil{\frac{N_c\Nwi}{32}}\min\pqty{\frac{\mathcal{L}_{\max}}{\mathcal{L}}, \mathcal{W}_{\max}} \label{eq:nreside}\\
\NresideEff         = & \; \etapred\etasimd\Nreside \label{eq:eff_nreside}
\end{align}

\subsection{Summary}\label{sec:transform_space_summary}
In Section~\ref{sec:scwpi} and Section~\ref{sec:multi_level_tiling}, we have
defined the\textit{ Single-cell per work-item} and the \textit{Multi-level
tiling} transform strategies respectively. In order to obtain the final
transformed kernel we execute the $b$-best\footnote{For the experiments in
Section~\ref{sec:perf_eval} we chose $b=9$.} ranked variants of the
\textit{Multi-level tiling} and the \textit{Single-cell per work-item} variant
with dummy data, time them and choose the fastest timed kernel variant.

\section{Implementation Details}\label{sec:impl_details}
We implement our approach within the \Fdrake~\cite{Gibson_2019} Finite Element
framework. Firedrake's software design makes it suitable for domain specialists
to work on disparate parts of the code, maintaining separations of concerns.
Within our modified version of \Fdrake{}, a UFL form is lowered to a GPU kernel
as:
\begin{enumerate}
  \item User expresses the weak form as a UFL expression.
  \item The \texttt{ufl}\footnote{\href{https://pypi.org/project/UFL}{\tt https://pypi.org/project/UFL}} package extracts the
    integral expressions in the variational form
    and returns its tensorial sum representation of the integral via
    integration by quadrature.
  \item The Two Stage Form Compiler~\cite{homolya2018tsfc} (TSFC) module then
    converts this representation into \texttt{GEM} IR. TSFC also
    performs algebraic optimizations on the GEM IR.
  \item The local assembly operator and the mesh information is then passed to
    PyOP2~\cite{rathgeber2012pyop2} which generates a \Loopy~\cite{klockner2014loo} kernel that applies the local operator to the
    to the corresponding entities of the mesh as required by the variational
    form.
  \item Performance optimizations, such as mapping operations to SIMD units as in
    \cite{sun2019study} or implementing our algorithm of
    Section~\ref{sec:transform_space}, are then applied to the \Loopy{} kernel.
  \item The transformed \Loopy{} kernel is then translated to the intended target
    such as C/\textsc{OpenCL}/CUDA.
\end{enumerate}

We have implemented our parallelization strategy as \Loopy{} transformations as
a part of Stage (5) of the described pipeline. \Loopy{} is a loop
transformation engine and a source-to-source translator written in Python.
\Loopy{} uses the Integer Set Library~\cite{verdoolaege2010isl} to represent
loop domains in its IR. All the loop and data transformations required in
Section~\ref{sec:transform_space} are available as primitive operations in \Loopy{}.

\section{Performance Evaluation}\label{sec:perf_eval}
We evaluate the effectiveness of our parallelization strategy by comparing the
FLOP rate of a kernel with the maximum achievable FLOP rate predicted by a
roofline model. In this work, we focus on measuring the FLOP rate as it
directly represents the computational throughput. The details of our roofline
model are explained in Section~\ref{sec:roofline_model}.

\subsection{Roofline Model}\label{sec:roofline_model}
A roofline model is a performance model with the intention to capture the
maximum performance that can be obtained for a workload. We develop such a
model for the FEM action kernel in Algorithm~\ref{knl:action.vanilla}.

In our model, we simplify the operations performed by the transformed kernel
and classify them into one of the following three categories:
\begin{enumerate}
  \item Global data accesses in the gather/scatter part of the compute
    kernel.\label{item:stmntcat1}
  \item Local memory accesses during the evaluation/quadrature stages of the
    action kernel. We assume that, for maximum performance the reference
    matrices are cached either in read-only data cache or in the local
    memory.
    \label{item:stmntcat2}
  \item Floating point operations.\label{item:stmntcat3}
\end{enumerate}

Each of the three categories of operations make use of different functional
units on a GPU.  We approximate the roofline by assuming that one of them is the
performance bottleneck and that all latency is hidden. Correspondingly, we obtain:
\begin{equation}
  \label{eq:multilevel_roofline}
\FRoofline = \min\pqty{\aiGlobal\bwglobpeak,\; \aiLocal \cdot \max\left(\bwlocpeak,\bwtexpeak\right),\; \FPeak}.
\end{equation}
Here, we use $\max\left(\bwlocpeak,\bwtexpeak\right)$ as an upper bound of read
throughput for the reference matrices to account for all possible address space
assignment strategies. For $\aiGlobal$, we aggregate the memory footprint associated
with the DOF vectors, coordinate data, and the local-to-global maps. Note that
although a DOF is potentially read from multiple cells in the gather phase, in
our roofline it counts as a single read, and similarly, writes to DOFs from
multiple cells in the scatter phase are counted as a single write. This is a
deliberately conservative choice, made in order to use an absolute lower bound
of global memory traffic when computing the roofline.

\subsection{Experiments}\label{sec:experiments}
We evaluate the proposed transformation space and tuning scheme by addressing
four questions. First, how closely does the generated code approach the roofline
across the full benchmark suite? Second, how sensitive is the resulting
throughput to each parameter of the transformation space? Third, how effectively
does our heuristic cost model rank empirically high-performing candidates?
Finally, how sensitive are the results to the search-space pruning thresholds
and to the number of model-ranked candidates that are measured empirically? We
perform a series of experiments in
Sections~\ref{sec:eval_throughput}--\ref{sec:hyperparam_sensitivity_tests}
address these questions.

\subsubsection{Experimental Setup}\label{sec:experimental_setup}
Our benchmark suite comprises weak forms with varying test and trial function
spaces. We consider the Mass, Poisson, Helmholtz, Elasticity, and
Hyperelasticity operators defined in Appendix~\ref{app:fem_forms}. These forms
arise in applications including fluid dynamics, wave propagation, and structural
mechanics. We evaluate each operator on both 2D and 3D geometries.  The 2D
domain is the unit square, while the 3D domain is the unit cube.  We choose
meshes with high element counts to eliminate performance decreases resulting
from idle compute units. We tabulate the problem sizes of weak forms in our
suite in Table~\ref{tab:cell_count_in_suite}.

\begin{table}[htbp]
  \centering
  \begin{tabular}{lccc}\toprule
    & Tesla K40 & Titan V & H200 NVL \\
    \midrule
    \# Triangles in 2D domain  & $5\times 10^5$           &  $5\times 10^5$      &   $3.2\times 10^6$        \\
    \# Tetrahedra in 3D domain & $2\times 10^5$           &  $2\times 10^5$      &   $1 \times 10^6$       \\
    \bottomrule
  \end{tabular}
  \caption{Cell count for actions operators in the benchmark suite.}\label{tab:cell_count_in_suite}
\end{table}

Our test bed comprises three Nvidia GPUs with distinct micro-architectures:
Tesla K40 (Kepler architecture), Titan V (Volta architecture) and H200 NVL
(Hopper architecture). The specifications for these devices have been tabulated
in Table~\ref{tab:device_specs} as aggregated
from~\cite{wiki:nvgpulist,jia2018dissecting,luo2025dissectinghopper}. For all
three GPUs, we perform the experiments by maintaining the default on-chip memory
carveout between the L1 data cache and CUDA's shared memory.

\begin{table}[htbp]
\centering
\begin{tabular}{lrrr}
\toprule
 & Tesla K40 & Titan V & H200 NVL\\
\midrule
$\FPeak$ (in GFLOPS) & 1430 & 6144 & 30000\\
$\bwglobpeak$ (in GB/s) & 288 & 653 & 4800\\
$\bwlocpeak$ (in GB/s) & 1000\footnotemark[6] & 13800 & 29700\\
$\bwtexpeak$ (in GB/s) & 840\footnotemark[7] & 12600 & 29240\\
Maximum 32-bit registers per work-item & 255 & 255 & 255\\
$\Lmax$ (in KB) & 48 & 96 & 228\\
$\Wmax$ & 32 & 32 & 32\\
$\sgSatGlobal$ & 8 & 1 & 1 \\
$\sgSatShared$ & 10 & 12 & 1 \\
\bottomrule
\end{tabular}
\caption{\label{tab:device_specs}Nvidia Microarchitecture specifications}
\end{table}
\footnotetext[6]{Measured experimentally using {\tt readLocalMemory}
benchmark of the SHOC suite.}
\footnotetext[7]{Tesla K40c contains 15 SMs each with 16 texture units. Each
texture unit can perform 32-bit fetch in a cycle. At its boost clock of 875\,MHz,
this evaluates to $15\times 16\times 4 \times 0.875$ GB/s.}
\addtocounter{footnote}{1}

The transformed \textsc{Loopy} kernel is lowered to a CUDA kernel
(Section~\ref{sec:impl_details}), which is then compiled with \texttt{nvcc} from
\textsc{CUDA} toolkit version 12.0.0 for the H200 and Titan V, and version
11.4.0 for the Tesla K40. We use \textsc{PyCUDA}~\cite{klockner2012pycuda} to
invoke the \textsc{CUDA} driver API for offloading the generated kernels.  We
have compiled the \textsc{Firedrake} toolkit in double-precision mode,
consequently, the DOF values in Algorithm~\ref{knl:action.vanilla} are encoded
using 64-bit floating-point numbers. To obtain the kernel execution time for a
test case, we perform five warm-up runs and then record wall-clock times until
either at least 15 timed runs have been collected or their cumulative runtime
reaches 0.2 seconds. We use the arithmetic mean of these measurements as the
kernel execution time.

\subsubsection{Throughput Relative to the Roofline}\label{sec:eval_throughput}
For each test case, we evaluate the FLOP rate of the compute kernel by
calculating the ratio of the number of FLOPS in the untransformed
Algorithm~\ref{knl:action.vanilla} to the measured execution time. We
specifically use the untransformed kernel to compute the FLOPS. This choice is
made because transformations that utilize multiple work-items for the
computation within a cell involve recomputation of geometric factors,
leading to additional FLOPS that are not indicative of the net DOF-throughput,
which is a key factor in determining the computational efficiency for the
computational scientist.

\begin{figure}
\centering
\begin{subfigure}{.5\textwidth}
  \centering
  \includegraphics[width=\textwidth, trim=0 90 0 90, clip]{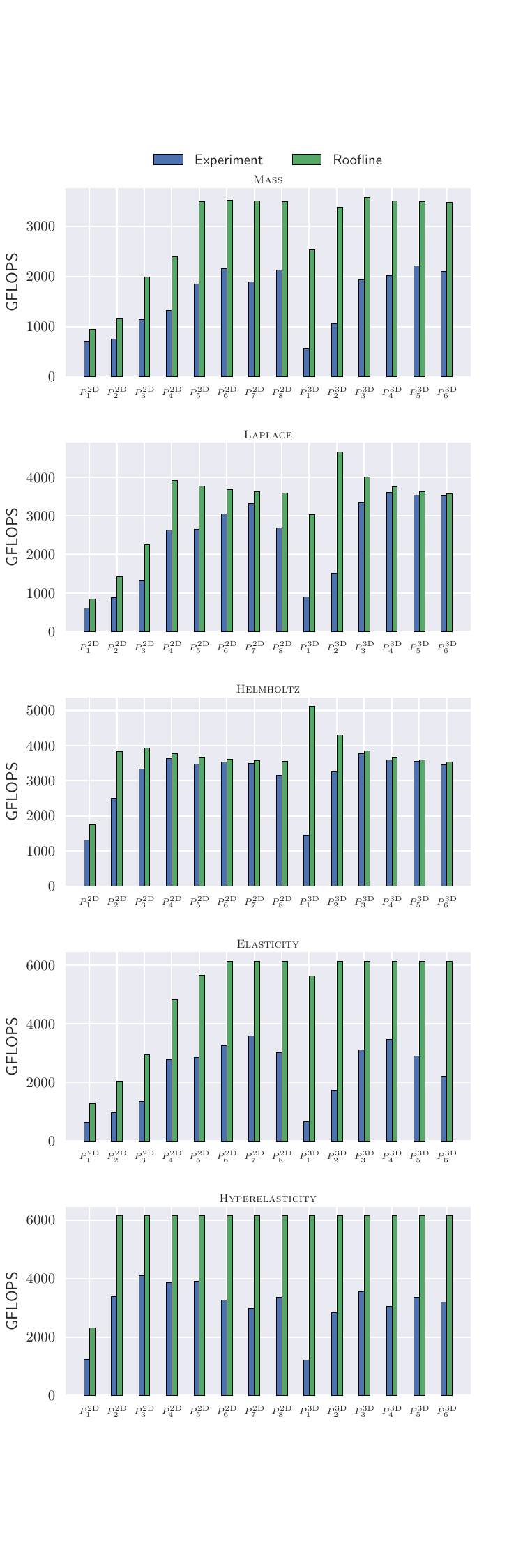}
  \caption{Experimental and roofline performances\\ on a Titan V.}\label{subfig:perf_eval_v100}
\end{subfigure}%
\begin{subfigure}{.5\textwidth}
  \centering
  \includegraphics[width=\textwidth, trim=0 99 0 99, clip]{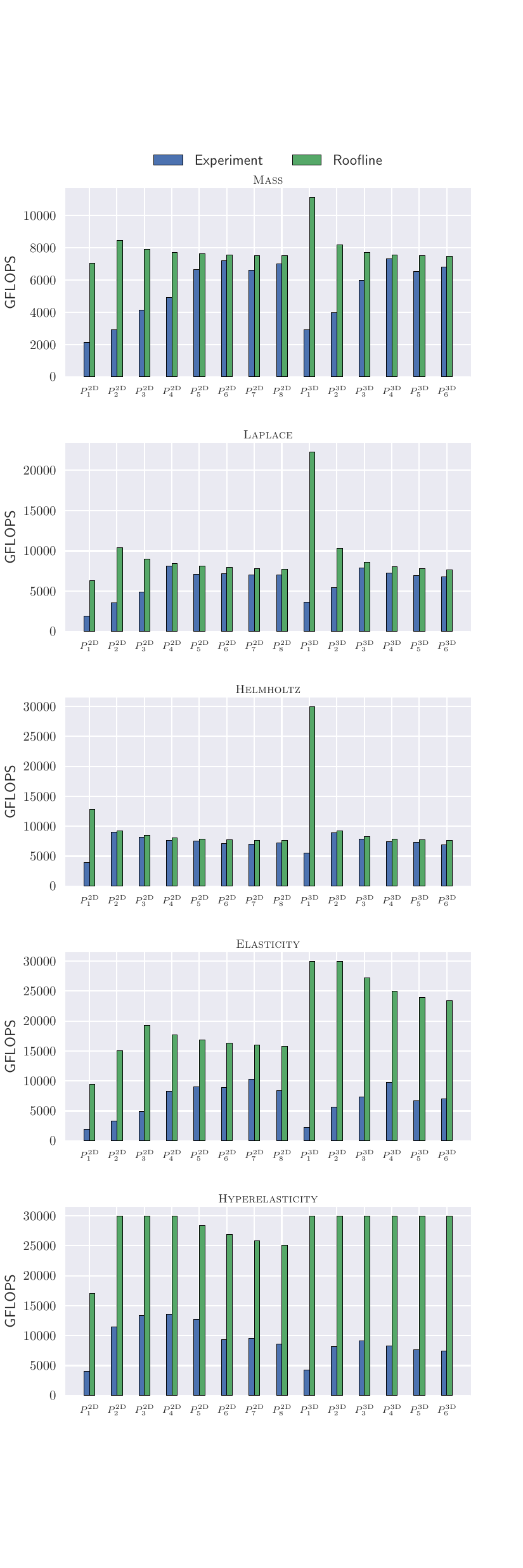}
  \caption{Experimental and roofline performances\\ on a H200 NVL.}\label{subfig:perf_eval_h200}
\end{subfigure}
\caption{Evaluated performance versus the roofline for Nvidia Titan V (left) and Nvidia H200 NVL (right). The bars denote the performance in GFLOPS for an FEM operator.}\label{fig:perf_eval_all}
\end{figure}

In Figure~\ref{fig:perf_eval_all}, we present our experimental results for the
Titan V and H200 NVL, showcasing the empirically measured FLOP throughput along
with the roofline performance according to our model for all the operators in
our benchmark suite. Results for the older Tesla K40c are reported separately in
Appendix~\ref{app:teslak40_throughput}, as its performance characteristics are
less representative of current GPU microarchitectures. Each test case in our suite
is labeled using the naming convention $P^{nD}_{p}$, where $n$ denotes the
spatial dimension, and $p$ signifies the degree of the function space chosen.
We summarize the key observations from these results in
Section~\ref{sec:results_discussion}.

\subsubsection{Sensitivity to Transformation Parameters}\label{sec:param_ablation_tests}
In the previous section, we have presented the throughput of our transformation
strategy of Section~\ref{sec:transform_space}, however, it does not establish
whether each parameter involved in our multi-level tiling scheme materially
affects performance. In this section, we study the sensitivity of the throughput
to $N_c$, $\Nwi$, $\quadtile$, $T_e^r$, $\{T_{e,U_i}^c\}_{i=1}^{N_U}$,
$\{T_{e,V_i}^c\}_{i=1}^{N_V}$, $T_q^r$, and $T_q^c$.

We consider three representative weak forms: Mass ($P_3$, 2D), Hyperelasticity
($P_6$, 2D), and Elasticity ($P_4$, 3D), on an Nvidia Titan V and an Nvidia H200
NVL. In Table~\ref{tab:action_param_space}, we saw that these cases include
scalar- and vector-valued function spaces and span substantially different test
and trial function spaces, derivative counts, and quadrature sizes. Furthermore,
from Appendix~\ref{app:winning_confs}, we know that for these weak forms
multi-level tiling outperforms the single-cell-per-work-item strategy, making
these workloads appropriate for assessing the parameters of the multi-level
transformation.

Let $\Theta$ denote the pruned transformation space defined in
Section~\ref{sec:param_search_space}, and let $P(\boldsymbol{\theta})$ denote the
measured FLOP throughput of a candidate $\boldsymbol{\theta}\in \Theta$. For any subset
$\Theta'\subseteq\Theta$, we apply the tuning procedure from
Section~\ref{sec:cost_model} and define its selected throughput as
\begin{equation}
  P_{\Theta'} =
  \max_{\boldsymbol{\theta}\in\operatorname{Top}_b(\Theta')}
  P(\boldsymbol{\theta}),
\end{equation}
where $\operatorname{Top}_b(\Theta')$ contains the $b$ candidates in
$\Theta'$ ranked most favorably by our cost
model~\eqref{eq:ufl2gpu_costmodel}. Notably, $P_\Theta$ is the throughput selected
by the unconstrained tuning procedure and is the quantity reported in
Section~\ref{sec:eval_throughput}. 
In this section, we consider a sequence of
subsets $\Theta'$ of $\Theta$, each obtained by constraining certain
parameters of the transformation space. Subsequently, we measure the throughput
$P_{\Theta'}$ for each subset to study the throughput's sensitivity to individual
parameters.

We perform three groups of experiments. First, we constrain all tile sizes using
one of three policies: full extents, a uniform tile size of four, or a uniform
tile size of eight. The full-extent policy sets $\quadtile=T_e^r=T_q^c=Q$,
$T_{e,U_i}^c=n_{U_i}$, $T_{e,V_i}^c=n_{V_i}$, and $T_q^r=n_W$. Second, we clamp
several $(N_c,\Nwi)$ pairs. Third, we choose the outer quadrature tile to be
either $\quadtile=Q$ or $\quadtile=\ceil{Q/2}$. In every experiment, parameters
not specified by the policy remain unconstrained and correspondingly selected by
our tuning scheme of Section~\ref{sec:cost_model}. For each resulting subset
$\Theta'\subseteq\Theta$, Table~\ref{tab:param_sensitivity} reports
$P_{\Theta'}/P_\Theta$. Ratios below one indicate that the performance
degraded under the corresponding constraint, while a value above one indicates
that the restricted search allowed the heuristic to select a candidate faster
than the one chosen from $\Theta$. Ideally we would expect all ratios to be
less than one, however, due to a non-exhaustive search of the transform candidates
along with an approximate cost model, we notice some deviations in practice.

From Table~\ref{tab:param_sensitivity}, we see that varying every dimension of
our parameter space has a significant performance impact, with individual cases
performing between $0.22\times$ to $1.42\times$ the performance reported in
Section~\ref{sec:eval_throughput}. This justifies our choice to include each
parameter in the tuning stage of our transformation. The influence of these
parameters is less pronounced for weak forms with smaller working sets, like
Mass ($P_3$, 2D), as the performance of the SCPWI scheme is close to that of
multi-level tiling, thereby, exhibiting a relatively higher lower bound
compared to the other weak forms in consideration.

\begin{filecontents*}{ratio_titanv.csv}
Key,Mass,Hyperelasticity,Elasticity
full_extents,1.0003452442603142,0.5694195938098378,0.7835844822058352
uniform_4,0.9830830312446056,0.35376840254889036,0.5572298813722347
uniform_8,0.9839461418953911,0.6808550710989736,0.7614620070535428
nc32_nwi1,0.9830830312446056,0.2987412499607622,0.21843539596024367
nc16_nwi4,0.9839461418953911,1.0489374391813417,0.8285668483488298
nc8_nwi8,0.9830830312446056,1.126596980255517,0.9175056107726834
nc16_nwi8,0.9830830312446056,1.4175848322189786,1.110612375761462
nc16_nwi16,0.9839461418953911,1.1118435508679412,0.86437960884899
tq_q,1.0607629898152944,1.1128794299526008,0.999647322859891
tq_half,0.9848092525461765,0.9999686097247074,0.6640910548252645
\end{filecontents*}
\DTLloaddb{ratioTitanV}{ratio_titanv.csv}
\begin{filecontents*}{ratio_h200.csv}
Key,Mass,Hyperelasticity,Elasticity
full_extents,0.9926201923076924,0.7315828638248113,0.9258992805755396
uniform_4,0.9926201923076924,0.6920378441586053,0.4603288797533402
uniform_8,0.9926201923076924,0.9447007547570957,0.9367934224049332
nc32_nwi1,0.9926201923076924,0.5711704050175401,0.6595066803699897
nc16_nwi4,0.9926201923076924,0.5637291378760497,0.9735868448098663
nc8_nwi8,0.9925480769230769,1.1169341979377059,1.0225077081192189
nc16_nwi8,0.9925480769230769,1.0025512915913681,1.051901336073998
nc16_nwi16,0.9925480769230769,0.9068778569150633,0.7359712230215827
tq_q,0.9925480769230769,0.9071967683639842,1.0
tq_half,0.9925480769230769,0.9537578399064527,0.6471736896197328
\end{filecontents*}
\DTLloaddb{ratioH200}{ratio_h200.csv}

\begin{table*}[htbp]
  \centering
  \scriptsize
  \begin{tabular}{@{}lcccccc@{}}
    \toprule
    & \multicolumn{3}{c}{Titan V}
    & \multicolumn{3}{c}{H200 NVL} \\
    \cmidrule(lr){2-4}\cmidrule(l){5-7}
    Constraint defining $\Theta'$
    & Mass ($P_3$, 2D) & Hyperelasticity ($P_6$, 2D) & Elasticity ($P_4$, 3D)
    & Mass ($P_3$, 2D) & Hyperelasticity ($P_6$, 2D) & Elasticity ($P_4$, 3D) \\
    \midrule
    \multicolumn{7}{@{}l}{\textit{Tile-size policy}} \\
    \ratiorow{Full extents}{full_extents}
    \ratiorow{Uniform tile size 4}{uniform_4}
    \ratiorow{Uniform tile size 8}{uniform_8}
    \midrule
    \multicolumn{7}{@{}l}{\textit{Work division}} \\
    \ratiorow{$(N_c,\Nwi)=(32,1)$}{nc32_nwi1}
    \ratiorow{$(N_c,\Nwi)=(16,4)$}{nc16_nwi4}
    \ratiorow{$(N_c,\Nwi)=(8,8)$}{nc8_nwi8}
    \ratiorow{$(N_c,\Nwi)=(16,8)$}{nc16_nwi8}
    \ratiorow{$(N_c,\Nwi)=(16,16)$}{nc16_nwi16}
    \midrule
    \multicolumn{7}{@{}l}{\textit{Outer quadrature tile}} \\
    \ratiorow{$\quadtile=Q$}{tq_q}
    \ratiorow{$\quadtile=\ceil{Q/2}$}{tq_half}
    \bottomrule
  \end{tabular}
  \caption{Throughput sensitivity to constraints on the transformation space.
  Each entry reports $P_{\Theta'}/P_\Theta$, where parameters not fixed by the
  listed policy are selected using the same tuning procedure as for the full
  space $\Theta$.}
  \label{tab:param_sensitivity}
\end{table*}

\subsubsection{Quality of Cost Model Rankings}\label{sec:cost_model_ranking_tests}
In Section~\ref{sec:eval_throughput}, we saw the performance of our transform
strategy of Section~\ref{sec:transform_space}. We now study the
effectiveness of the heuristic cost model of Section~\ref{sec:cost_model}. Since
the model is only used to rank candidates of our transform space, we evaluate the
correctness of its ordering rather than its absolute runtime prediction
accuracy.

We first evaluate the Spearman's rank-correlation coefficient,
$\rho$~\cite{spermann1904association}, between the modeled and empirical
rankings. A correlation of $\rho=1$ indicates ideal preservation of the ranking
by the cost model, while $\rho=0$ indicates no monotonic association
between the empirical and predicted rankings.

For each of the 70 variational forms used for reporting throughputs in
Section~\ref{sec:eval_throughput}, we then randomly select, without replacement,
20 transform candidates from the search space described in
Section~\ref{sec:param_search_space}, such that each candidate has an equal
probability of being selected and utilize this sample for computing $\rho$ as
follows.
\begin{equation}
\label{eq:spearmann_rho}
\rho = 1 - \frac{6\sum_{i=1}^{n}\left(X_i - Y_i\right)^2}{n(n^2-1)},
\end{equation}
where $n = 20$ is the sample size, and $X_i$ and $Y_i$ correspond to the
empirical rank and the rank predicted by our cost
model~\eqref{eq:ufl2gpu_costmodel}, respectively, for the $i$-th transform
candidate in our sample, $i \in \{1, 2, \ldots, n\}$.

In Table~\ref{tab:cost_model_spearman}, we report the arithmetic mean of $\rho$
values computed using~\eqref{eq:spearmann_rho} for the specified subset of test
cases.  Throughout this section, we classify the test cases whose function
spaces lie in $\{P_1^{2D},\ldots, P_4^{2D}, P_1^{3D}, P_2^{3D}, P_3^{3D} \}$ as
low order, and refer to the remaining test cases as high order.  We observe a
moderate to strong correlation~\cite{schober2018correlation} across all the test
cases, and stronger correlation for test cases employing low order function
spaces and simpler weak forms.

\begin{filecontents*}{spearman.csv}
Key,TitanV,H200,Both
all,0.70,0.66,0.68
twod,0.69,0.71,0.70
threed,0.71,0.59,0.65
loworder,0.78,0.63,0.70
highorder,0.63,0.68,0.66
mass,0.71,0.70,0.71
laplace,0.74,0.72,0.73
helmholtz,0.68,0.74,0.71
elasticity,0.73,0.71,0.72
hyperelasticity,0.64,0.42,0.53
\end{filecontents*}
\DTLloaddb{spearmanData}{spearman.csv}
\begin{filecontents*}{topk_regret.csv}
Key,TitanV1,TitanV5,TitanV10,TitanV20,H2001,H2005,H20010,H20020,Both1,Both5,Both10,Both20
all,0.88,0.94,0.96,0.97,0.83,0.94,0.95,0.96,0.85,0.94,0.95,0.96
twod,0.86,0.93,0.95,0.96,0.83,0.93,0.95,0.96,0.85,0.93,0.95,0.96
threed,0.91,0.96,0.97,0.97,0.83,0.94,0.95,0.96,0.87,0.95,0.96,0.97
loworder,0.91,0.95,0.96,0.96,0.87,0.97,0.97,0.98,0.89,0.96,0.96,0.97
highorder,0.85,0.94,0.96,0.97,0.79,0.91,0.93,0.95,0.82,0.92,0.94,0.96
mass,0.86,0.97,0.98,0.99,0.85,0.98,0.99,0.99,0.85,0.98,0.98,0.99
laplace,0.92,0.96,0.98,0.98,0.90,0.99,0.99,1.00,0.91,0.97,0.99,0.99
helmholtz,0.93,0.97,0.97,0.98,0.95,0.99,1.00,1.00,0.94,0.98,0.98,0.99
elasticity,0.86,0.92,0.94,0.94,0.86,0.92,0.93,0.96,0.86,0.92,0.93,0.95
hyperelasticity,0.84,0.89,0.91,0.93,0.63,0.80,0.84,0.87,0.73,0.85,0.88,0.90
\end{filecontents*}
\DTLloaddb{topkData}{topk_regret.csv}

\begin{table}[htbp]
  \centering
  \begin{tabular}{@{}lccc@{}}
    \toprule
    Benchmark subset & Titan V & H200 NVL & Both \\
    \midrule
    \spearmanrow{All cases}{all}
    \midrule
    \spearmanrow{2D}{twod}
    \spearmanrow{3D}{threed}
    \spearmanrow{Low order}{loworder}
    \spearmanrow{High order}{highorder}
    \midrule
    \spearmanrow{Mass}{mass}
    \spearmanrow{Laplace}{laplace}
    \spearmanrow{Helmholtz}{helmholtz}
    \spearmanrow{Elasticity}{elasticity}
    \spearmanrow{Hyperelasticity}{hyperelasticity}
    \bottomrule
  \end{tabular}
  \caption{Spearman rank correlation $\rho$ between the cost model and
  empirical candidate rankings. ``Both'' aggregates results from the two
  devices.}
  \label{tab:cost_model_spearman}
\end{table}

Spearman's rank correlation, $\rho$, is stricter than our use case of the model
since we only need our tuning scheme to pick a high-performing transform
candidate in the top-$b$ candidates as ranked by the cost model
of~\eqref{eq:ufl2gpu_costmodel}. For this reason, we also assess the quality of
our top-$b$ transform candidates using the regret metric, $Q_k$:
\begin{equation}
  Q_k =
  \frac{
    \max_{\boldsymbol{\theta}\in\operatorname{Top}_k(\Theta)}
    P(\boldsymbol{\theta})
  }{
    \max_{\boldsymbol{\theta}\in\Theta}
    P(\boldsymbol{\theta})
  },
\end{equation}
where $P(\boldsymbol{\theta})$ is the empirical throughput for a search-space
candidate, $\boldsymbol{\theta}$, $\Theta$ is the search-space of transform
candidate coming solely from the multi-level tiling scheme,
$\operatorname{Top}_k(\Theta)$ are $k$ candidates of $\Theta$ ranked most
favorable as per our cost model of~\eqref{eq:ufl2gpu_costmodel}.

In Figure~\ref{fig:search_space_cardinality}, we saw that the size of our search
space is too prohibitively large to measure
$\max_{\boldsymbol{\theta}\in\Theta}P(\boldsymbol{\theta})$ exhaustively.
We, therefore, report
\begin{equation}
  \tilde{Q_k} =
  \frac{
    \max_{\boldsymbol{\theta}\in\operatorname{Top}_k(\Theta)}
    P(\boldsymbol{\theta})
  }{
    \max_{\boldsymbol{\theta}\in\operatorname{Top}_{70}(\Theta)}
    P(\boldsymbol{\theta})
  }.\label{eq:define_tilde_qk}
\end{equation}
We use $\tilde{Q_k}$ as a proxy to study $Q_k$, by using the
approximation $\tilde{Q_k} \approx Q_k$ on the basis that our cost model
achieves a moderate-to-strong Spearman's correlation coefficient of 0.68 across
several randomly selected samples of our search space, $\Theta$. While it does
not guarantee that the optimal configurations lies within
$\operatorname{Top}_{70}(\Theta)$, we argue that the quality of our cost model's
global ranking is sufficient to capture the true optimum within the top
predictions.

Table~\ref{tab:cost_model_topk} reports the geometric mean of $\tilde{Q_k}$ measured
over several subsets of our benchmark suite of
Section~\ref{sec:eval_throughput}, grouped by dimension, polynomial order, and
physics. Values close to one indicate that measuring only the first $k$
model-ranked candidates recovers throughput close to the empirical optimum.

\begin{table*}[htbp]
  \centering
  \scriptsize
  \begin{tabular}{@{}lcccccccccccc@{}}
    \toprule
    & \multicolumn{4}{c}{Titan V}
    & \multicolumn{4}{c}{H200 NVL}
    & \multicolumn{4}{c}{Both} \\
    \cmidrule(lr){2-5}\cmidrule(lr){6-9}\cmidrule(l){10-13}
    Benchmark subset
    & $k=1$ & $k=5$ & $k=9$ & $k=20$
    & $k=1$ & $k=5$ & $k=9$ & $k=20$
    & $k=1$ & $k=5$ & $k=9$ & $k=20$ \\
    \midrule
    \topkrow{All cases}{all}
    \midrule
    \topkrow{2D}{twod}
    \topkrow{3D}{threed}
    \topkrow{Low order}{loworder}
    \topkrow{High order}{highorder}
    \midrule
    \topkrow{Mass}{mass}
    \topkrow{Laplace}{laplace}
    \topkrow{Helmholtz}{helmholtz}
    \topkrow{Elasticity}{elasticity}
    \topkrow{Hyperelasticity}{hyperelasticity}
    \bottomrule
  \end{tabular}
  \caption{Mean fraction $\tilde{Q_k}$~\eqref{eq:define_tilde_qk} for various
  subsets of our benchmark suite. ``Both'' aggregates results from the two
  devices.}
  \label{tab:cost_model_topk}
\end{table*}
A key takeaway from Table~\ref{tab:cost_model_topk} is that we recover
95\% of the throughput by considering the subset
$\operatorname{Top}_{9}(\Theta)$ (see $k=9$ in the table), and, reassuringly,
that throughput exhibits strongly diminishing returns as $k$ increases.

\subsubsection{Sensitivity to Pruning Thresholds and Tuning Budget}\label{sec:hyperparam_sensitivity_tests}
Our search space pruning strategy of Section~\ref{sec:param_search_space}
characterizes parameters such as $\etaaliasMin$, $\etasimdMin$,
$\NcNwiMax$, and $b$. In Section~\ref{sec:eval_throughput}, we saw the
performance of our transformation strategy obtained for $\etaaliasMin = 0.8$,
$\etasimdMin = 0.97$, $\NcNwiMax=256$ and, $b=9$. In this section, we study the
sensitivity of the observed throughputs to these parameters.
Similar to Section~\ref{sec:param_ablation_tests}, we consider three
representative weak forms: Mass ($P_3$, 2D), Hyperelasticity ($P_6$, 2D), and
Elasticity ($P_4$, 3D) on Nvidia Titan V and Nvidia H200 NVL.

Let $\mathbf{h} = (\etaaliasMin, \etasimdMin, \NcNwiMax, b)$ denote the vector of
pruning and tuning-budget parameters characterizing the search space of
Section~\ref{sec:param_search_space}, and let $\Theta(\mathbf{h})$ denote the
resulting pruned transformation space. Reusing the throughput notation of
Section~\ref{sec:param_ablation_tests}, we write throughput as a function
of the parameters as
\begin{equation}
  P(\mathbf{h}) =
  \max_{\boldsymbol{\theta}\in\operatorname{Top}_b(\Theta(\mathbf{h}))}
  P(\boldsymbol{\theta}).
\end{equation}

Consequently, we had reported $P(\mathbf{h}^*)$ in
Section~\ref{sec:eval_throughput} with $\mathbf{h}^*=(0.8, 0.97, 256, 9)$.  To
study the sensitivity of performance to each parameter individually, we
vary one component of $\mathbf{h}^*$ at a time while clamping the rest to their
values in $\mathbf{h}^*$.  Specifically, we measure $P(\mathbf{h})$ for several
$h$-values that exhibit perturbation from $\mathbf{h}^*$ along exactly one
dimension. We report these measurements as normalized throughput, i.e.
$P(\mathbf{h})/P(\mathbf{h}^*)$, for each such perturbation.

\begin{filecontents*}{hyperparam_ratio_titanv.csv}
Configuration,mass_p3_2d,hyperelasticity_p6_2d,elasticity_p4_3d
ETA_ALIAS_MIN=0,1.032168968482021,1.012886452460517,0.9445547461813972
ETA_ALIAS_MIN=0.5,1.0368347247433662,0.9917813895811654,0.9488743078842342
ETA_ALIAS_MIN=0.6,1.0599042973689061,0.9916248238328023,0.9492872071646525
ETA_ALIAS_MIN=0.7,0.9974350052031183,1.0138258469506958,0.9747916781012563
ETA_ALIAS_MIN=0.9,1.0008047180585342,1.0343046468365948,0.9930545308889863
ETA_SIMD_MIN=0,1.0294904787764338,0.9813227975905074,1.0046157107406972
ETA_SIMD_MIN=0.5,1.026552780389661,0.9793187560114591,1.0156687068626626
ETA_SIMD_MIN=0.7,1.0226646501718735,0.9801642110526201,1.0152240460991353
ETA_SIMD_MIN=0.9,1.024997528302546,0.9797884532565486,1.0041710499771699
ETA_SIMD_MIN=0.99,1.0014095383146344,0.9969167461274764,1.0824948673241999
WG_MAX=64,0.9849929885061979,1.0284490878478134,0.8085203140251402
WG_MAX=128,0.9895723418738144,1.2704057953682144,1.0999954445173117
WG_MAX=512,0.9908683852797435,0.9657288490535383,0.7917184894604287
b=1,0.7532424202090292,0.992762521222014,0.8542814421426139
b=3,0.999558690378095,0.992762521222014,0.9546710405948835
b=5,1.0,0.992762521222014,0.9546710405948835
b=20,1.0623827792685903,1.0384280448706584,1.0
b=50,1.0623827792685903,1.0752597593488058,1.1060355572383518
b=70,1.0623827792685903,1.0752597593488058,1.1060355572383518
\end{filecontents*}
\DTLloaddb{hpRatioTitanV}{hyperparam_ratio_titanv.csv}
\begin{filecontents*}{hyperparam_ratio_h200.csv}
Configuration,mass_p3_2d,hyperelasticity_p6_2d,elasticity_p4_3d
ETA_ALIAS_MIN=0,0.9895198178127571,0.9984920226635524,1.000272951156366
ETA_ALIAS_MIN=0.5,0.9889926617687846,0.9866813990111111,1.0016300899121748
ETA_ALIAS_MIN=0.6,0.9868600759545331,0.9865750927676148,1.0006122358453182
ETA_ALIAS_MIN=0.7,0.9887290837467986,0.9534394366698111,1.0003140765732086
ETA_ALIAS_MIN=0.9,0.9860453802502124,0.9536307879081046,1.0011057408474306
ETA_SIMD_MIN=0,0.9892802014291333,0.9567242995938476,1.0009618018884812
ETA_SIMD_MIN=0.5,0.9894958561743946,0.9570751101973854,1.000303795218998
ETA_SIMD_MIN=0.7,0.9893041630674956,0.9568624977103928,1.001085178139009
ETA_SIMD_MIN=0.9,0.9891603932373214,0.9566392545990505,1.0008795510547956
ETA_SIMD_MIN=0.99,0.987554963467042,0.9537052022785519,0.8388762527572229
WG_MAX=64,0.9892562397907709,0.5515486831319666,1.0576634703603423
WG_MAX=128,0.9884655057248124,1.0000440938185988,1.0710189494800062
WG_MAX=512,0.989136431598959,0.9239713459726309,1.0003140765732086
b=1,0.975232518130706,0.7938375437993678,1.0
b=3,0.975232518130706,0.9521979240038948,1.0
b=5,0.975232518130706,0.9521979240038948,1.0
b=20,1.0,1.256934998298833,1.0
b=50,1.0,1.3339416031771874,1.0706953648897002
b=70,1.0,1.3698170832762924,1.0706953648897002
\end{filecontents*}
\DTLloaddb{hpRatioH200}{hyperparam_ratio_h200.csv}

\begin{table*}[htbp]
  \centering
  \scriptsize
  \begin{tabular}{@{}lcccccc@{}}
    \toprule
    & \multicolumn{3}{c}{Titan V}
    & \multicolumn{3}{c}{H200 NVL} \\
    \cmidrule(lr){2-4}\cmidrule(l){5-7}
    & Mass ($P_3$, 2D) & Hyperelasticity ($P_6$, 2D) & Elasticity ($P_4$, 3D)
    & Mass ($P_3$, 2D) & Hyperelasticity ($P_6$, 2D) & Elasticity ($P_4$, 3D) \\
    \midrule
    \hpratiorow{$\etaaliasMin=0$}{ETA_ALIAS_MIN=0}
    \hpratiorow{$\etaaliasMin=0.5$}{ETA_ALIAS_MIN=0.5}
    \hpratiorow{$\etaaliasMin=0.6$}{ETA_ALIAS_MIN=0.6}
    \hpratiorow{$\etaaliasMin=0.7$}{ETA_ALIAS_MIN=0.7}
    \hpratiorow{$\etaaliasMin=0.9$}{ETA_ALIAS_MIN=0.9}
    \midrule
    \hpratiorow{$\etasimdMin=0$}{ETA_SIMD_MIN=0}
    \hpratiorow{$\etasimdMin=0.5$}{ETA_SIMD_MIN=0.5}
    \hpratiorow{$\etasimdMin=0.7$}{ETA_SIMD_MIN=0.7}
    \hpratiorow{$\etasimdMin=0.9$}{ETA_SIMD_MIN=0.9}
    \hpratiorow{$\etasimdMin=0.99$}{ETA_SIMD_MIN=0.99}
    \midrule
    \hpratiorow{$\NcNwiMax=64$}{WG_MAX=64}
    \hpratiorow{$\NcNwiMax=128$}{WG_MAX=128}
    \hpratiorow{$\NcNwiMax=512$}{WG_MAX=512}
    \midrule
    \hpratiorow{$b=1$}{b=1}
    \hpratiorow{$b=3$}{b=3}
    \hpratiorow{$b=5$}{b=5}
    \hpratiorow{$b=20$}{b=20}
    \hpratiorow{$b=50$}{b=50}
    \hpratiorow{$b=70$}{b=70}
    \bottomrule
  \end{tabular}
  \caption{Empirical measurements to study the sensitivity of the transformation
  scheme's throughput to the model's parameters.  Each entry reports the
  normalized throughput, $P(\mathbf{h})/P(\mathbf{h}^*)$, where, $P(\mathbf{h})$
  is the measured throughput for parameters $\mathbf{h}$ and
  $\mathbf{h}^*=(0.8, 0.97, 256, 9)$ correspond to our prescribed defaults.
  Each row corresponds to performance of our transformation strategy with
  exactly one parameter perturbed while the rest clamped to the defaults in
  $\mathbf{h}^*$.}
  \label{tab:hyperparam_sensitivity}
\end{table*}

We see from Table~\ref{tab:hyperparam_sensitivity} that throughput is
generally unchanged as we vary $\etasimdMin$ and $\etaaliasMin$. In such cases, a
higher value is generally favorable because it limits the number of candidates
in the search space and reduces time spent in the form compiler.  In
contrast, $\NcNwiMax$ and $b$ exhibit a relatively higher impact on the
throughput.  For example, aggressively pruning the search space to only consider
candidates with $N_cN_{\text{WI}}\leq 64$ leads to a 45\% slowdown for
Hyperelasticity ($P_6$, 2D). Our choice of $\NcNwiMax=256$ is justified as we do
not observe any improvement as we loosen the pruning criterion to
$\NcNwiMax=512$. Similarly, we observe that relaxing $b$ to higher values does
not result in substantial gains. For Hyperelasticity ($P_6$, 2D), we see a
speedup of 26\% over our default of $b=9$, however, we argue that this is an
outlier across the entire suite, based on cost model evaluation in
Section~\ref{sec:cost_model_ranking_tests}.

Although performance generally improves as a pruning threshold is relaxed, the
selected throughput need not be monotone: with a fixed tuning budget, newly
admitted candidates can change the model-ranked top-$b$ subset and displace a
faster candidate. Table~\ref{tab:hyperparam_sensitivity} shows this broad trend,
although there are outliers. For example, tuning Hyperelasticity ($P_6$,
2D) with $\NcNwiMax=128$ yields $1.27\times$ the performance relative to
$\NcNwiMax=256$. This is a limitation that arises from the approximate nature of
our cost model in~\eqref{eq:ufl2gpu_costmodel}.

\subsection{Discussion}\label{sec:results_discussion}
\begin{figure}[h]
  \centering
  \includegraphics[width=0.6\textwidth]{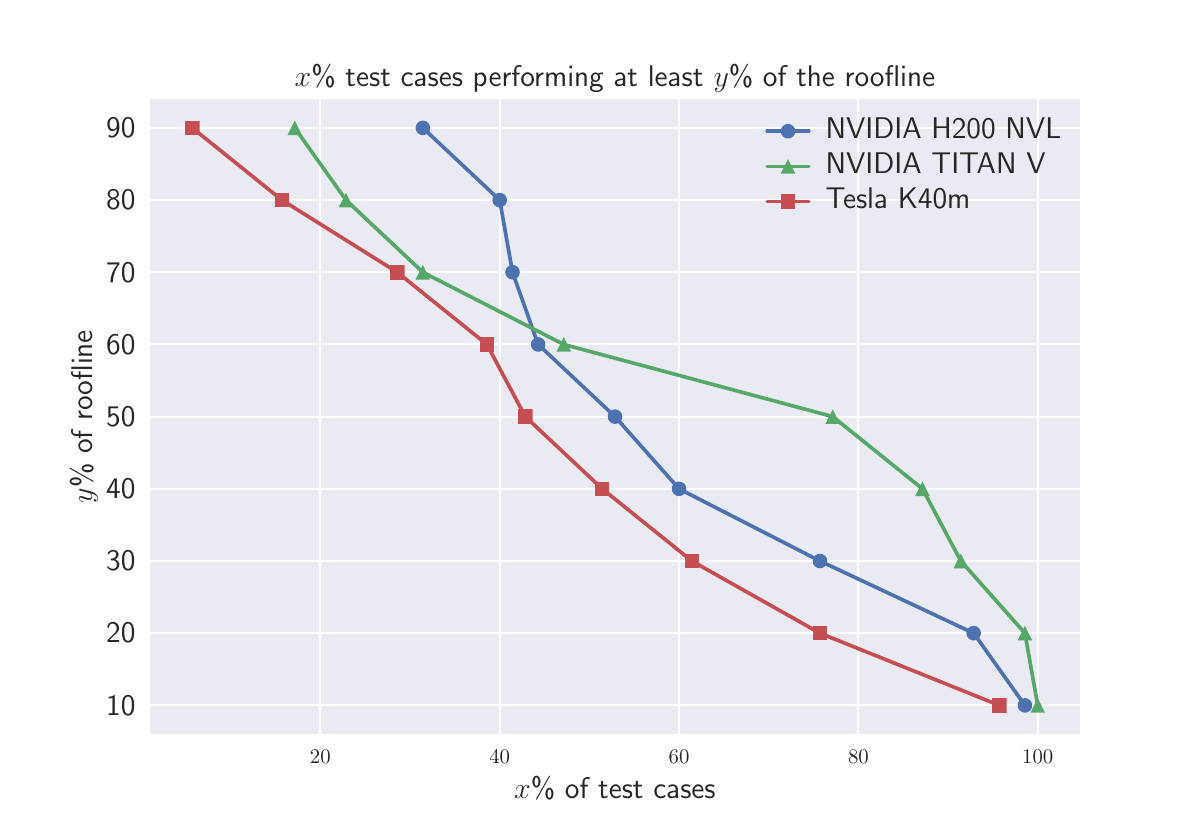}
  \caption{Comparison of \% of cases achieving a \% roofline on the Tesla K40, Titan V and H200 GPUs}\label{fig:roofline_comp}
\end{figure}

The throughputs measured in Section~\ref{sec:eval_throughput} can be seen in an
aggregated way in Figure~\ref{fig:roofline_comp}. From the figure, we see that,
across all three devices, our method generates kernels that attain at least
50\% of the roofline performance for approximately $60$\% of the test cases.
This highlights the consistency of our transform strategy across
micro-architectures and various finite element operators.

\subsubsection{Behavior at low orders}\label{sec:discussion_low_order_limiters}
At low orders, repeated and indirect DOF accesses widen the gap
between attained throughput and the footprint-based roofline.
In  Figure~\ref{fig:perf_eval_all}, we observe that operators based on the $P_1$
function space significantly lag the roofline performance.  In particular, on
the H200 GPU, the Elasticity ($P_1$, 3D) kernel achieves only $7.7\%$ of its
roofline performance. To understand this discrepancy, we use this kernel as a
representative case and collect hardware performance counters with the NVIDIA
Nsight Compute tool. From Appendix~\ref{app:winning_confs}, we know that SCPWI
is the selected transformation for this operator, as it is for the other $P_1$
operators.

Recall that global bandwidth term in roofline model
in~\eqref{eq:ufl2gpu_costmodel} is conservative and chooses memory footprint
that counts each unique DOF exactly once. In a $P_1$ function space, neighboring
cells share DOFs, however, our transformed kernel in
Algorithm~\ref{knl:action.scpt} gathers the local DOFs of each cell
independently. This leads to repeated loads issued for a DOF by all the cells
that contain it. Furthermore, these accesses are performed indirectly through
the local-to-global maps and are not necessarily coalesced across neighboring
work-items.

In Table~\ref{tab:elast_3dp1_profile}, we report the memory traffic at each level of
the cache hierarchy. The L1 entry was obtained from the profile counters
\texttt{l1tex\_\_t\_sectors\_pipe\_lsu\_mem\_global\_op\_ld} and
\texttt{l1tex\_\_t\_sectors\_pipe\_lsu\_mem\_global\_op\_red}; the L2
entry from \texttt{lts\_\_t\_sectors\_srcunit\_tex\_op\_read} and
\texttt{lts\_\_t\_sectors\_srcunit\_tex\_op\_red}; and the DRAM entry
from \texttt{dram\_\_bytes\_read} and
\texttt{dram\_\_bytes\_write}. The footprint used to compute $\aiGlobal$
for this operator is 34.6\,MB.

\begin{table}[htbp]
  \centering
  \begin{tabular}{lr}
    \toprule
    Memory-hierarchy interface & Traffic volume \\
    \midrule
    Global loads and reductions at L1           & 758.6\,MB \\
    Global reads and reductions presented to L2 & 446.7\,MB \\
    Reads and writes transferred at DRAM        & 33.1\,MB \\
    \bottomrule
  \end{tabular}
  \caption{Memory-hierarchy traffic for the Elasticity ($P_1$, 3D) kernel on
  the H200 GPU.}
  \label{tab:elast_3dp1_profile}
\end{table}

We see that the measured DRAM traffic is close to the footprint used by the
roofline model.  However, the repeated and indirect DOF accesses generate
traffic corresponding to $22\times$ and $13\times$ the footprint at the L1 and
L2 caches, respectively. The cache hit rate is high and most of the memory
traffic does not reach DRAM, but it still exerts pressure on the memory hierarchy
at earlier stages. Since the footprint-based roofline does not account for this
additional on-chip traffic, it substantially overestimates the attainable
performance of this memory-bound kernel.

A DOF-layout order that prioritizes proximity of cells sharing DOFs could lower
the number of requests made to the caches. Additionally, transforming the gather loops
(lines~\ref{knlline:action.gather_trial_start}--\ref{knlline:action.gather_trial_stop}
of Algorithm~\ref{knl:action.vanilla}) to perform a collective gather of all the
DOFs of the cells being processed by a work-group could further reduce the
pressure on the L1 cache. Such layout reordering and aggregation strategies are
outside the scope of this work.

\subsubsection{Behavior at high orders}\label{sec:discussion_high_order_limiters}
At high orders, the performance is primarily determined by large state-space requirements.
Performance is sometimes limited because of the increased per-cell
data requirements of the finite element operators, specifically the reference
matrices \(\vb{\Phi}\) and \(\vb{\Psi}\). These matrices have larger size in
these cases,
leading to increased utilization of local memory and more frequent
synchronizations. Both of these factors impede the latency hiding capabilities
of the hardware, resulting in lower throughputs.

\subsubsection{On Single-Cell per Work-item}\label{sec:scwpi_kepler_v_volta} SCPWI performs better on Volta and Hopper
than Kepler micro-architectures.
From the results presented in
Tables~\ref{tbl:perf_eval_winner_2d_k40}--\ref{tbl:perf_eval_winner_3d_h200nvl},
of Appendix~\ref{app:winning_confs}, we observe that on the Tesla K40c GPU, 79\%
of the configurations that performed the best were obtained from the multi-level
tiling algorithm. On the other hand, for the Titan V and H200 NVL, only 58\% and
61\% of the best performing configurations, respectively, were achieved using
the multi-level tiling algorithm.

We attribute this behavior primarily to SCPWI's high reliance on
hardware-managed caching for reducing repeated accesses to the reference
matrices. Inspection of the SASS assembly confirms these global loads appear as
\texttt{LDG.E.64} on the Tesla K40c and \texttt{LDG.E.64.CONSTANT.SYS}
instructions on the Volta and Hopper. On Kepler, these instructions use the
48\,KB read-only/texture cache per compute unit.  In contrast, Volta and Hopper
utilize the read-only and texture-cache which is integrated into unified 128\,KB
and 256\,KB L1/texture-cache, shared memory, respectively. Since SCPWI does not
allocate any variables in the local-memory address space, the device can
potentially make this entire resource available for L1/texture caching at
runtime. This larger cache capacity enables Volta and Hopper to better mitigate
the latency associated with global memory reads of the derivative matrices in
the kernel transformed using SCPWI.

\subsection{Code Availability}\label{sec:code_avail}
To ensure reproducibility, we provide citations for the archived versions of the
software used to generate the results presented in this article. The primary
components of \textsc{Firedrake} have been archived on
Zenodo~\cite{zenodo/Firedrake-20241102.1} at the versions studied here. Similarly, the software utilized to
obtain the empirical data in Section~\ref{sec:experiments} has been archived on
Zenodo~\cite{kulkarni_2026_22003825}. Detailed installation instructions are
provided in the README file included in the archived records.

\section{Conclusion}\label{sec:conclusion}
In this work, we have studied optimizing one of the major bottlenecks of an FEM
solver on a GPU, that is evaluating an FEM operator's action. We have introduced a
sequence of loop/data transformations to define a schedule space. We have then
performed auto-tuning over the schedule space via a heuristic cost model.
Through performance evaluation, we have observed that our method delivers
consistent near-roofline performance for different FEM operators and on
different micro-architectures. Our auto-tuning approach is able to approach
near-roofline configurations in our transformation space with just 10 benchmark
runs, which makes it highly attractive for practical use.

This leads to some interesting questions for further study.

\textit{Can we deliver comparable near-roofline performances if action kernels
sharing trial functions were fused?} Fusing multiple action kernels, e.g. computing both the
velocity and pressure terms of a Navier Stokes solver in a single kernel, would reduce the number
of memory accesses. However, the fusion will also increase the kernel's state
space requirements resulting in a larger search space and more complex
trade-offs.

\textit{How can the cost model be adapted to meshes with low element count?} Operators
with lower element counts offer limited outer-loop concurrency. In order to
ensure work-groups being scheduled to all the multi-processors, the model would need to
prefer configurations in the schedule space that target higher amounts of inner loop
parallelization. Cantwell et al.~\cite{cantwell2011h}
study and tabulate the performance of some kernel variants while varying the $h-p$
parameters of an FEM solver. However, an effective cost model remains to be found.

\textit{Can GPU vector shuffles accelerate action kernels in a performance
and platform portable manner?} Nvidia GPUs provide
shuffle functions that allow intra sub-group communication between the
private variables of work-items. This would decrease the amount of local
memory allocated per work-group. However, doing so would put further restrictions
on the kernel's grid size as now the work-items performing the computation
would have to exactly align with the device's sub-group length.

\begin{acks}
The authors would like to thank Tianjiao Sun, David Ham, and Lawrence Mitchell
for their valuable discussions and insights.

The authors' research was funded by the US National Science Foundation under
awards SHF-1911019 and OAC-1931577, by the US Department of Energy under award
number DE-NA0003963, as well as by the Siebel School of Computing and Data
Science at the University of Illinois at Urbana-Champaign. Any opinions,
findings, and conclusions, or recommendations  expressed in this article are
those of the authors and do not necessarily reflect the views of the sponsors;
the sponsors have not approved or endorsed its content.

The authors would also like to thank the anonymous reviewers for their
careful, constructive feedback, which helped substantially strengthen this paper.
\end{acks}

\bibliographystyle{ACM-Reference-Format}
\bibliography{refs}

\appendix
\section{Variational forms of FEM operators}\label{app:fem_forms}
Here we describe the operators used as the test cases for experimental evaluation. They are defined as bilinear forms, and we take their \texttt{action} in UFL to obtain the corresponding linear forms.

\begin{description}
  \item[Mass] Here $u$ and $v$ are scalar-valued trial and test functions.
  \begin{equation}
    a = \int uv\ \dd x
  \end{equation}

   \item[Helmholtz] Here $u$ and $v$ are scalar-valued trial and test
     functions.
     \begin{equation}
       a = \int (\grad u \cdot \grad v + uv)\ \dd x
     \end{equation}

  \item[Laplacian] Here $u$ and $v$ are scalar-valued trial
    and test functions.
    \begin{equation}
      a = \int\grad u \cdot \grad v\ \dd x
    \end{equation}

  \item[Elasticity] The linear elasticity model solves for a displacement
    vector field. Here $\mathbf{u}$ and $\mathbf{v}$ are vector-valued trial and test
    functions, $\epsilon$ is the infinitesimal strain tensor, and $\sigma$ is the
    Cauchy stress tensor. The bilinear form is defined as:
    \begin{equation}
      \begin{aligned}
        \epsilon(\vb u) &= \frac{1}{2}\bqty{\grad\vb u + (\grad\vb u)^T}, \\
        \sigma(\vb u) &= \lambda\operatorname{tr}(\epsilon(\vb u))\vb I
          + 2\mu\epsilon(\vb u), \\
        a &= \int \sigma(\vb u):\epsilon(\vb v) \ \dd x.
      \end{aligned}
    \end{equation}
    We use the Lam\'{e} modulus $\lambda=0$ and shear modulus $\mu=1/2$, for
    which $\sigma(\vb u)=\epsilon(\vb u)$.

  \item[Hyperelasticity] In this simple hyperelastic model, we define the strain
  energy function $\Psi$ over vector field $\vb u$:
    \begin{equation}
      \begin{aligned}
      \vb{F} &= \vb{I} + \grad \vb{u}\\
      \vb{C} &= \vb{F}^T\vb{F}\\
      \vb{E} &=(\vb{C} - \vb{I})/2,\\
      \Psi &= \frac{\lambda}{2}\bqty{\textsf{tr}(\vb{E})}^2+\mu\textsf{tr}(\vb{E}^2)
      \end{aligned}
    \end{equation}
    where $\vb{I}$ is the identity matrix, $\lambda$ and $\mu$ are the Lam\'{e}
    parameters of the material, $\vb{F}$ is the deformation gradient, $\vb{C}$
    is the right Cauchy-Green tensor, $\vb{E}$ is the Lagrangian Green strain tensor.
    We define the first and second Piola-Kirchhoff stress tensors as:
    \begin{equation}
      \begin{aligned}
        \vb S &= \pdv{\Psi}{\vb E}\\
        \vb P &= \vb F\vb S.
      \end{aligned}
    \end{equation}
    Finally, we arrive at the residual form of this nonlinear problem:
    \begin{equation}
    r=\int \vb P:\grad \vb v - \vb b\cdot\vb v \ \dd x
    \end{equation}
    where $\vb b$ is the external forcing. To solve this nonlinear problem, we need
    to linearize the residual form at an approximate solution $\mathbf{u}$, this gives
    us the bilinear form $a$:
    \begin{equation}
      a=\lim_{\epsilon \to 0}\frac{r(\vb u+\epsilon\delta\vb u)-r(\vb u)}{\epsilon},
    \end{equation}
    where the trial function is $\delta\mathbf{u}$, the test function is $\mathbf{v}$,
    and $\mathbf{u}$ is a coefficient of the operator. We use the automatic
    differentiation of UFL to compute the operator symbolically.

\end{description}

\section{Exploring different work-group sizes for Single cell per
Work-item}~\label{app:scpwi_large_workgroups}
In Section~\ref{sec:scwpi}, the SCPWI scheme employed a work-division with a
work-group size of 32. In this section, we explore the impact on the kernel's
performance with a different work-group size. Similar to our parameter ablation
studies (Section~\ref{sec:param_ablation_tests}), we choose three representative
weak forms: Mass ($P_2$, 2D), Hyperelasticity ($P_4$, 2D) and Elasticity ($P_2$,
3D) and measure the throughput on the Titan V and H200 NVL GPUs. Notably, for
these test cases, the SCPWI scheme outperforms multi-level tiling on both the
devices -- highlighting the significance of our study for these test cases.

Let $P_W$ denote the throughput of the SCPWI scheme using a work-group size of
$W$. Correspondingly, we had considered $P_{32}$ in the tuning phase of our
transform strategy while reporting the throughputs in
Section~\ref{sec:eval_throughput}. In Table~\ref{tab:scwpi_wg_studies}, we
report the quantities $P_W / P_{32}$ for different values of $W$.

\begin{filecontents*}{scwpi_wg_studies.csv}
W, titanv_mass_2dp2, titanv_hyperelasticity_2dp4, titanv_elasticity_3d_p2, h200_mass_2dp2, h200_hyperelasticity_2dp4, h200_elasticity_3d_p2
64, 0.99, 0.99, 1.03, 1.00, 0.99, 1.03
96, 0.98, 0.93, 1.01, 1.00, 0.86, 1.00
128, 0.98, 0.99, 1.01, 1.00, 0.99, 1.03
\end{filecontents*}
\DTLloaddb{scwpiWgStudies}{scwpi_wg_studies.csv}

\begin{table*}[htbp]
  \centering
  \scriptsize
  \begin{tabular}{@{}lcccccc@{}}
    \toprule
    & \multicolumn{3}{c}{Titan V}
    & \multicolumn{3}{c}{H200 NVL} \\
    \cmidrule(lr){2-4}\cmidrule(l){5-7}
    Work-group size ($W$)
    & Mass ($P_2$, 2D) & Hyperelasticity ($P_4$, 2D) & Elasticity ($P_2$, 3D)
    & Mass ($P_2$, 2D) & Hyperelasticity ($P_4$, 2D) & Elasticity ($P_2$, 3D) \\
    \midrule
    \scwpiwgrow{64}
    \scwpiwgrow{96}
    \scwpiwgrow{128}
    \bottomrule
  \end{tabular}
  \caption{Sensitivity of SCPWI's throughput to the chosen work-group size for work-division.
  Each entry denotes the normalized throughput, $P_W / P_{32}$, for the specified work-group
  size, $W$.}
  \label{tab:scwpi_wg_studies}
\end{table*}

From Table~\ref{tab:scwpi_wg_studies}, we observe that increasing the work-group
size beyond 32 leads to either no change in the scheme's performance or degrades
the performance. This substantiates our claims from Section~\ref{sec:scwpi},
that, either the latency hiding was already limited by microarchitecture's
maximum number of work-groups in flight, ($\Wmax$ of
Table~\ref{tab:device_specs}), or would lead to performance degradation due to
increased register space usage by a work-group.

\section{Detailed throughput evaluation on Tesla K40}\label{app:teslak40_throughput}
\begin{figure}
  \centering
  \includegraphics[width=0.5\textwidth, trim=0 90 0 90, clip]{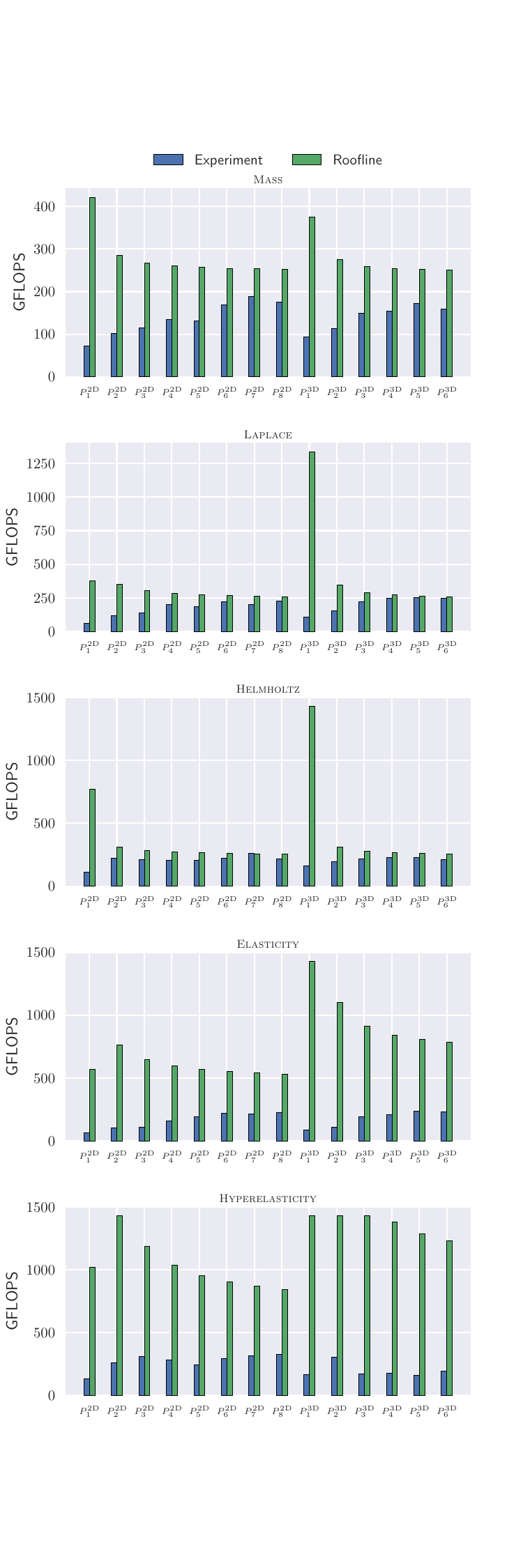}
  \caption{Evaluated performance versus the roofline for Nvidia Tesla K40c. The bars denote the performance in GFLOPS for an FEM operator.}\label{fig:perf_eval_teslak40m_app}
\end{figure}

To keep the discussion in Section~\ref{sec:eval_throughput} concise, we present
detailed performance results only for the more recent GPUs, namely the Titan V
and H200 NVL. Figure~\ref{fig:perf_eval_teslak40m_app} reports the fine-grained
throughput measurements for the Tesla K40c. See
Section~\ref{sec:experimental_setup} for details on the experimental
methodology.

\section{Best Performing Candidates in the Transformation Space}\label{app:winning_confs}

In
Tables~\ref{tbl:perf_eval_winner_2d_k40},~\ref{tbl:perf_eval_winner_2d_v100},~\ref{tbl:perf_eval_winner_3d_k40},
and~\ref{tbl:perf_eval_winner_3d_v100}, we present the configurations from the
transformation space that were selected by our auto-tuning approach
of Section~\ref{sec:cost_model} to report the FLOP throughput
for each test case shown in Figure~\ref{fig:perf_eval_all} of
Section~\ref{sec:experiments}.

\begin{table}
\begin{center}
\begin{tabular}{ccccccccccc}
\toprule

\multicolumn{2}{c}{} & \multicolumn{1}{c}{SCPT} & \multicolumn{1}{c}{Multi-level tiling} & \multicolumn{1}{c}{$N_c$} & \multicolumn{1}{c}{$N_{WI}$} & \multicolumn{1}{c}{$\quadtile$} & \multicolumn{1}{c}{$T_e^r$} & \multicolumn{1}{c}{$T_e^c$} & \multicolumn{1}{c}{$T_q^r$} & \multicolumn{1}{c}{$T_q^c$}\\

\midrule

\multirow{8}{*}{Mass}  &         
\multirow{1}{*}{$P_1$}  &     \cmark  &    &    &    &    &    &    &    &  \\

    &         
\multirow{1}{*}{$P_2$}  &       &  \cmark  &  64  &  2  &  6  &  6  &  6  &  6  &  6\\

    &         
\multirow{1}{*}{$P_3$}  &     \cmark  &    &    &    &    &    &    &    &  \\

    &         
\multirow{1}{*}{$P_4$}  &     \cmark  &    &    &    &    &    &    &    &  \\

    &         
\multirow{1}{*}{$P_5$}  &       &  \cmark  &  64  &  4  &  36  &  36  &  7  &  21  &  12\\

    &         
\multirow{1}{*}{$P_6$}  &       &  \cmark  &  51  &  5  &  49  &  49  &  7  &  28  &  13\\

    &         
\multirow{1}{*}{$P_7$}  &       &  \cmark  &  37  &  6  &  64  &  64  &  9  &  36  &  16\\

    &         
\multirow{1}{*}{$P_8$}  &       &  \cmark  &  64  &  4  &  41  &  41  &  9  &  45  &  9\\
\midrule[0.1pt]
        
\multirow{8}{*}{Laplace}  &         
\multirow{1}{*}{$P_1$}  &     \cmark  &    &    &    &    &    &    &    &  \\

    &         
\multirow{1}{*}{$P_2$}  &       &  \cmark  &  64  &  1  &  3  &  3  &  6  &  6  &  3\\

    &         
\multirow{1}{*}{$P_3$}  &       &  \cmark  &  64  &  2  &  6  &  6  &  10  &  10  &  6\\

    &         
\multirow{1}{*}{$P_4$}  &       &  \cmark  &  64  &  4  &  12  &  12  &  15  &  15  &  12\\

    &         
\multirow{1}{*}{$P_5$}  &       &  \cmark  &  51  &  5  &  25  &  25  &  7  &  21  &  9\\

    &         
\multirow{1}{*}{$P_6$}  &       &  \cmark  &  64  &  4  &  18  &  18  &  14  &  28  &  9\\

    &         
\multirow{1}{*}{$P_7$}  &       &  \cmark  &  51  &  5  &  25  &  25  &  9  &  18  &  13\\

    &         
\multirow{1}{*}{$P_8$}  &       &  \cmark  &  51  &  5  &  22  &  22  &  9  &  45  &  5\\
\midrule[0.1pt]
        
\multirow{8}{*}{Helmholtz}  &         
\multirow{1}{*}{$P_1$}  &     \cmark  &    &    &    &    &    &    &    &  \\

    &         
\multirow{1}{*}{$P_2$}  &     \cmark  &    &    &    &    &    &    &    &  \\

    &         
\multirow{1}{*}{$P_3$}  &     \cmark  &    &    &    &    &    &    &    &  \\

    &         
\multirow{1}{*}{$P_4$}  &       &  \cmark  &  63  &  4  &  13  &  13  &  15  &  15  &  13\\

    &         
\multirow{1}{*}{$P_5$}  &     \cmark  &    &    &    &    &    &    &    &  \\

    &         
\multirow{1}{*}{$P_6$}  &       &  \cmark  &  51  &  5  &  17  &  17  &  7  &  28  &  5\\

    &         
\multirow{1}{*}{$P_7$}  &       &  \cmark  &  51  &  5  &  16  &  16  &  9  &  36  &  4\\

    &         
\multirow{1}{*}{$P_8$}  &       &  \cmark  &  51  &  5  &  17  &  17  &  8  &  15  &  9\\
\midrule[0.1pt]
        
\multirow{8}{*}{Elasticity}  &         
\multirow{1}{*}{$P_1$}  &     \cmark  &    &    &    &    &    &    &    &  \\

    &         
\multirow{1}{*}{$P_2$}  &       &  \cmark  &  64  &  3  &  3  &  3  &  6  &  6  &  3\\

    &         
\multirow{1}{*}{$P_3$}  &       &  \cmark  &  64  &  3  &  6  &  6  &  10  &  10  &  6\\

    &         
\multirow{1}{*}{$P_4$}  &       &  \cmark  &  56  &  4  &  12  &  12  &  15  &  15  &  12\\

    &         
\multirow{1}{*}{$P_5$}  &       &  \cmark  &  25  &  10  &  25  &  25  &  7  &  21  &  9\\

    &         
\multirow{1}{*}{$P_6$}  &       &  \cmark  &  36  &  7  &  18  &  18  &  7  &  14  &  9\\

    &         
\multirow{1}{*}{$P_7$}  &       &  \cmark  &  25  &  10  &  25  &  25  &  9  &  36  &  5\\

    &         
\multirow{1}{*}{$P_8$}  &       &  \cmark  &  28  &  9  &  22  &  22  &  9  &  45  &  5\\
\midrule[0.1pt]
        
\multirow{8}{*}{Hyperelasticity}  &         
\multirow{1}{*}{$P_1$}  &     \cmark  &    &    &    &    &    &    &    &  \\

    &         
\multirow{1}{*}{$P_2$}  &       &  \cmark  &  53  &  3  &  6  &  6  &  6  &  6  &  6\\

    &         
\multirow{1}{*}{$P_3$}  &       &  \cmark  &  25  &  10  &  25  &  25  &  5  &  10  &  13\\

    &         
\multirow{1}{*}{$P_4$}  &     \cmark  &    &    &    &    &    &    &    &  \\

    &         
\multirow{1}{*}{$P_5$}  &       &  \cmark  &  23  &  11  &  27  &  27  &  7  &  21  &  9\\

    &         
\multirow{1}{*}{$P_6$}  &       &  \cmark  &  21  &  12  &  31  &  31  &  7  &  28  &  7\\

    &         
\multirow{1}{*}{$P_7$}  &       &  \cmark  &  21  &  12  &  29  &  29  &  6  &  36  &  6\\

    &         
\multirow{1}{*}{$P_8$}  &       &  \cmark  &  28  &  9  &  23  &  23  &  9  &  45  &  5\\

\bottomrule
\end{tabular}

\end{center}
\caption{Best performing configurations for 2D cases on an Nvidia Tesla K40c.}
\label{tbl:perf_eval_winner_2d_k40}
\end{table}

\begin{table}
\begin{center}
\begin{tabular}{ccccccccccc}
\toprule

\multicolumn{2}{c}{} & \multicolumn{1}{c}{SCPT} & \multicolumn{1}{c}{Multi-level tiling} & \multicolumn{1}{c}{$N_c$} & \multicolumn{1}{c}{$N_{WI}$} & \multicolumn{1}{c}{$\quadtile$} & \multicolumn{1}{c}{$T_e^r$} & \multicolumn{1}{c}{$T_e^c$} & \multicolumn{1}{c}{$T_q^r$} & \multicolumn{1}{c}{$T_q^c$}\\

\midrule

\multirow{8}{*}{Mass}  &         
\multirow{1}{*}{$P_1$}  &     \cmark  &    &    &    &    &    &    &    &  \\

    &         
\multirow{1}{*}{$P_2$}  &     \cmark  &    &    &    &    &    &    &    &  \\

    &         
\multirow{1}{*}{$P_3$}  &       &  \cmark  &  64  &  3  &  12  &  12  &  5  &  5  &  12\\

    &         
\multirow{1}{*}{$P_4$}  &     \cmark  &    &    &    &    &    &    &    &  \\

    &         
\multirow{1}{*}{$P_5$}  &     \cmark  &    &    &    &    &    &    &    &  \\

    &         
\multirow{1}{*}{$P_6$}  &     \cmark  &    &    &    &    &    &    &    &  \\

    &         
\multirow{1}{*}{$P_7$}  &     \cmark  &    &    &    &    &    &    &    &  \\

    &         
\multirow{1}{*}{$P_8$}  &       &  \cmark  &  64  &  4  &  81  &  81  &  9  &  45  &  14\\
\midrule[0.1pt]
        
\multirow{8}{*}{Laplace}  &         
\multirow{1}{*}{$P_1$}  &     \cmark  &    &    &    &    &    &    &    &  \\

    &         
\multirow{1}{*}{$P_2$}  &       &  \cmark  &  64  &  2  &  3  &  3  &  2  &  2  &  3\\

    &         
\multirow{1}{*}{$P_3$}  &       &  \cmark  &  64  &  3  &  6  &  6  &  5  &  10  &  3\\

    &         
\multirow{1}{*}{$P_4$}  &     \cmark  &    &    &    &    &    &    &    &  \\

    &         
\multirow{1}{*}{$P_5$}  &     \cmark  &    &    &    &    &    &    &    &  \\

    &         
\multirow{1}{*}{$P_6$}  &     \cmark  &    &    &    &    &    &    &    &  \\

    &         
\multirow{1}{*}{$P_7$}  &     \cmark  &    &    &    &    &    &    &    &  \\

    &         
\multirow{1}{*}{$P_8$}  &       &  \cmark  &  64  &  4  &  32  &  32  &  15  &  45  &  11\\
\midrule[0.1pt]
        
\multirow{8}{*}{Helmholtz}  &         
\multirow{1}{*}{$P_1$}  &     \cmark  &    &    &    &    &    &    &    &  \\

    &         
\multirow{1}{*}{$P_2$}  &     \cmark  &    &    &    &    &    &    &    &  \\

    &         
\multirow{1}{*}{$P_3$}  &     \cmark  &    &    &    &    &    &    &    &  \\

    &         
\multirow{1}{*}{$P_4$}  &     \cmark  &    &    &    &    &    &    &    &  \\

    &         
\multirow{1}{*}{$P_5$}  &     \cmark  &    &    &    &    &    &    &    &  \\

    &         
\multirow{1}{*}{$P_6$}  &       &  \cmark  &  64  &  4  &  25  &  25  &  7  &  28  &  5\\

    &         
\multirow{1}{*}{$P_7$}  &       &  \cmark  &  55  &  4  &  32  &  32  &  6  &  18  &  11\\

    &         
\multirow{1}{*}{$P_8$}  &       &  \cmark  &  64  &  4  &  27  &  27  &  9  &  45  &  5\\
\midrule[0.1pt]
        
\multirow{8}{*}{Elasticity}  &         
\multirow{1}{*}{$P_1$}  &     \cmark  &    &    &    &    &    &    &    &  \\

    &         
\multirow{1}{*}{$P_2$}  &       &  \cmark  &  64  &  2  &  3  &  3  &  2  &  2  &  3\\

    &         
\multirow{1}{*}{$P_3$}  &       &  \cmark  &  64  &  4  &  6  &  6  &  5  &  5  &  6\\

    &         
\multirow{1}{*}{$P_4$}  &       &  \cmark  &  56  &  4  &  12  &  12  &  15  &  15  &  12\\

    &         
\multirow{1}{*}{$P_5$}  &       &  \cmark  &  56  &  4  &  25  &  25  &  7  &  21  &  7\\

    &         
\multirow{1}{*}{$P_6$}  &       &  \cmark  &  36  &  7  &  36  &  36  &  7  &  28  &  8\\

    &         
\multirow{1}{*}{$P_7$}  &       &  \cmark  &  56  &  4  &  25  &  25  &  9  &  8  &  25\\

    &         
\multirow{1}{*}{$P_8$}  &       &  \cmark  &  42  &  6  &  32  &  32  &  7  &  23  &  8\\
\midrule[0.1pt]
        
\multirow{8}{*}{Hyperelasticity}  &         
\multirow{1}{*}{$P_1$}  &     \cmark  &    &    &    &    &    &    &    &  \\

    &         
\multirow{1}{*}{$P_2$}  &     \cmark  &    &    &    &    &    &    &    &  \\

    &         
\multirow{1}{*}{$P_3$}  &     \cmark  &    &    &    &    &    &    &    &  \\

    &         
\multirow{1}{*}{$P_4$}  &     \cmark  &    &    &    &    &    &    &    &  \\

    &         
\multirow{1}{*}{$P_5$}  &       &  \cmark  &  32  &  8  &  41  &  41  &  7  &  21  &  11\\

    &         
\multirow{1}{*}{$P_6$}  &       &  \cmark  &  32  &  8  &  41  &  41  &  7  &  28  &  9\\

    &         
\multirow{1}{*}{$P_7$}  &       &  \cmark  &  32  &  8  &  43  &  43  &  6  &  36  &  8\\

    &         
\multirow{1}{*}{$P_8$}  &       &  \cmark  &  28  &  9  &  45  &  45  &  9  &  45  &  9\\

\bottomrule
\end{tabular}

\end{center}
\caption{Best performing configurations for 2D cases on an Nvidia Titan V.}
\label{tbl:perf_eval_winner_2d_v100}
\end{table}

\begin{table}
\begin{center}
\begin{tabular}{ccccccccccc}
\toprule

\multicolumn{2}{c}{} & \multicolumn{1}{c}{SCPT} & \multicolumn{1}{c}{Multi-level tiling} & \multicolumn{1}{c}{$N_c$} & \multicolumn{1}{c}{$N_{WI}$} & \multicolumn{1}{c}{$\quadtile$} & \multicolumn{1}{c}{$T_e^r$} & \multicolumn{1}{c}{$T_e^c$} & \multicolumn{1}{c}{$T_q^r$} & \multicolumn{1}{c}{$T_q^c$}\\

\midrule

\multirow{8}{*}{Mass}  &         
\multirow{1}{*}{$P_1$}  &     \cmark  &    &    &    &    &    &    &    &  \\

    &         
\multirow{1}{*}{$P_2$}  &     \cmark  &    &    &    &    &    &    &    &  \\

    &         
\multirow{1}{*}{$P_3$}  &       &  \cmark  &  64  &  2  &  12  &  12  &  5  &  10  &  6\\

    &         
\multirow{1}{*}{$P_4$}  &       &  \cmark  &  64  &  2  &  16  &  16  &  15  &  15  &  16\\

    &         
\multirow{1}{*}{$P_5$}  &     \cmark  &    &    &    &    &    &    &    &  \\

    &         
\multirow{1}{*}{$P_6$}  &     \cmark  &    &    &    &    &    &    &    &  \\

    &         
\multirow{1}{*}{$P_7$}  &       &  \cmark  &  64  &  2  &  42  &  42  &  36  &  36  &  42\\

    &         
\multirow{1}{*}{$P_8$}  &       &  \cmark  &  64  &  2  &  55  &  55  &  45  &  45  &  55\\
\midrule[0.1pt]
        
\multirow{8}{*}{Laplace}  &         
\multirow{1}{*}{$P_1$}  &     \cmark  &    &    &    &    &    &    &    &  \\

    &         
\multirow{1}{*}{$P_2$}  &     \cmark  &    &    &    &    &    &    &    &  \\

    &         
\multirow{1}{*}{$P_3$}  &       &  \cmark  &  64  &  2  &  6  &  6  &  10  &  10  &  6\\

    &         
\multirow{1}{*}{$P_4$}  &       &  \cmark  &  64  &  2  &  12  &  12  &  15  &  15  &  12\\

    &         
\multirow{1}{*}{$P_5$}  &       &  \cmark  &  64  &  2  &  16  &  16  &  11  &  11  &  16\\

    &         
\multirow{1}{*}{$P_6$}  &       &  \cmark  &  64  &  2  &  25  &  25  &  14  &  28  &  13\\

    &         
\multirow{1}{*}{$P_7$}  &       &  \cmark  &  64  &  2  &  33  &  33  &  18  &  18  &  33\\

    &         
\multirow{1}{*}{$P_8$}  &       &  \cmark  &  64  &  2  &  42  &  42  &   8  &  45  &   7\\
\midrule[0.1pt]
        
\multirow{8}{*}{Helmholtz}  &         
\multirow{1}{*}{$P_1$}  &     \cmark  &    &    &    &    &    &    &    &  \\

    &         
\multirow{1}{*}{$P_2$}  &     \cmark  &    &    &    &    &    &    &    &  \\

    &         
\multirow{1}{*}{$P_3$}  &       &  \cmark  &  64  &  2  &  12  &  12  &  10  &  10  &  12\\

    &         
\multirow{1}{*}{$P_4$}  &       &  \cmark  &  64  &  2  &  16  &  16  &  15  &  15  &  16\\

    &         
\multirow{1}{*}{$P_5$}  &       &  \cmark  &  64  &  2  &  25  &  25  &  11  &  21  &  13\\

    &         
\multirow{1}{*}{$P_6$}  &     \cmark  &    &    &    &    &    &    &    &  \\

    &         
\multirow{1}{*}{$P_7$}  &       &  \cmark  &  64  &  2  &  21  &  21  &  18  &  18  &  21\\

    &         
\multirow{1}{*}{$P_8$}  &       &  \cmark  &  64  &  2  &  28  &  28  &  8  &  45  &  5\\
\midrule[0.1pt]
        
\multirow{8}{*}{Elasticity}  &         
\multirow{1}{*}{$P_1$}  &     \cmark  &    &    &    &    &    &    &    &  \\

    &         
\multirow{1}{*}{$P_2$}  &       &  \cmark  &  64  &  2  &  3  &  3  &  2  &  2  &  3\\

    &         
\multirow{1}{*}{$P_3$}  &       &  \cmark  &  64  &  3  &  6  &  6  &  10  &  10  &  6\\

    &         
\multirow{1}{*}{$P_4$}  &     \cmark  &    &    &    &    &    &    &    &  \\

    &         
\multirow{1}{*}{$P_5$}  &       &  \cmark  &  64  &  4  &  16  &  16  &  21  &  21  &  16\\

    &         
\multirow{1}{*}{$P_6$}  &       &  \cmark  &  56  &  4  &  25  &  25  &  7  &  28  &  5\\

    &         
\multirow{1}{*}{$P_7$}  &       &  \cmark  &  42  &  3  &  33  &  33  &  6  &  36  &  6\\

    &         
\multirow{1}{*}{$P_8$}  &       &  \cmark  &  32  &  2  &  42  &  42  &  8  &  8  &  42\\
\midrule[0.1pt]
        
\multirow{8}{*}{Hyperelasticity}  &         
\multirow{1}{*}{$P_1$}  &     \cmark  &    &    &    &    &    &    &    &  \\

    &         
\multirow{1}{*}{$P_2$}  &     \cmark  &    &    &    &    &    &    &    &  \\

    &         
\multirow{1}{*}{$P_3$}  &     \cmark  &    &    &    &    &    &    &    &  \\

    &         
\multirow{1}{*}{$P_4$}  &     \cmark  &    &    &    &    &    &    &    &  \\

    &         
\multirow{1}{*}{$P_5$}  &     \cmark  &    &    &    &    &    &    &    &  \\

    &         
\multirow{1}{*}{$P_6$}  &       &  \cmark  &  32  &  4  &  40  &  40  &  10  &  28  &  14\\

    &         
\multirow{1}{*}{$P_7$}  &       &  \cmark  &  36  &  7  &  38  &  38  &  6  &  36  &  7\\

    &         
\multirow{1}{*}{$P_8$}  &       &  \cmark  &  36  &  7  &  38  &  38  &  7  &  45  &  7\\

\bottomrule
\end{tabular}

\end{center}
\caption{Best performing configurations for 2D cases on an Nvidia H200 NVL.}
\label{tbl:perf_eval_winner_2d_h200nvl}
\end{table}

\begin{table}
\begin{center}
\begin{tabular}{ccccccccccc}
\toprule

\multicolumn{2}{c}{} & \multicolumn{1}{c}{SCPT} & \multicolumn{1}{c}{Multi-level tiling} & \multicolumn{1}{c}{$N_c$} & \multicolumn{1}{c}{$N_{WI}$} & \multicolumn{1}{c}{$\quadtile$} & \multicolumn{1}{c}{$T_e^r$} & \multicolumn{1}{c}{$T_e^c$} & \multicolumn{1}{c}{$T_q^r$} & \multicolumn{1}{c}{$T_q^c$}\\

\midrule

\multirow{6}{*}{Mass}  &         
\multirow{1}{*}{$P_1$}  &     \cmark  &    &    &    &    &    &    &    &  \\

    &         
\multirow{1}{*}{$P_2$}  &       &  \cmark  &  64  &  2  &  14  &  14  &  5  &  10  &  7\\

    &         
\multirow{1}{*}{$P_3$}  &       &  \cmark  &  64  &  4  &  24  &  24  &  10  &  20  &  12\\

    &         
\multirow{1}{*}{$P_4$}  &       &  \cmark  &  64  &  4  &  42  &  42  &  7  &  35  &  7\\

    &         
\multirow{1}{*}{$P_5$}  &       &  \cmark  &  56  &  4  &  44  &  44  &  8  &  28  &  15\\

    &         
\multirow{1}{*}{$P_6$}  &       &  \cmark  &  51  &  5  &  49  &  49  &  10  &  42  &  10\\
\midrule[0.1pt]
        
\multirow{6}{*}{Laplace}  &         
\multirow{1}{*}{$P_1$}  &     \cmark  &    &    &    &    &    &    &    &  \\

    &         
\multirow{1}{*}{$P_2$}  &       &  \cmark  &  64  &  2  &  4  &  4  &  5  &  10  &  2\\

    &         
\multirow{1}{*}{$P_3$}  &       &  \cmark  &  64  &  4  &  14  &  14  &  7  &  20  &  5\\

    &         
\multirow{1}{*}{$P_4$}  &       &  \cmark  &  63  &  4  &  12  &  12  &  18  &  35  &  6\\

    &         
\multirow{1}{*}{$P_5$}  &       &  \cmark  &  64  &  4  &  14  &  14  &  7  &  28  &  4\\

    &         
\multirow{1}{*}{$P_6$}  &       &  \cmark  &  42  &  6  &  18  &  18  &  14  &  42  &  6\\
\midrule[0.1pt]
        
\multirow{6}{*}{Helmholtz}  &         
\multirow{1}{*}{$P_1$}  &     \cmark  &    &    &    &    &    &    &    &  \\

    &         
\multirow{1}{*}{$P_2$}  &     \cmark  &    &    &    &    &    &    &    &  \\

    &         
\multirow{1}{*}{$P_3$}  &       &  \cmark  &  56  &  4  &  12  &  12  &  7  &  20  &  4\\

    &         
\multirow{1}{*}{$P_4$}  &       &  \cmark  &  51  &  5  &  13  &  13  &  7  &  18  &  5\\

    &         
\multirow{1}{*}{$P_5$}  &       &  \cmark  &  51  &  5  &  13  &  13  &  7  &  14  &  7\\

    &         
\multirow{1}{*}{$P_6$}  &       &  \cmark  &  51  &  5  &  25  &  25  &  10  &  42  &  5\\
\midrule[0.1pt]
        
\multirow{6}{*}{Elasticity}  &         
\multirow{1}{*}{$P_1$}  &     \cmark  &    &    &    &    &    &    &    &  \\

    &         
\multirow{1}{*}{$P_2$}  &       &  \cmark  &  64  &  4  &  4  &  4  &  10  &  10  &  4\\

    &         
\multirow{1}{*}{$P_3$}  &       &  \cmark  &  22  &  10  &  14  &  14  &  5  &  20  &  4\\

    &         
\multirow{1}{*}{$P_4$}  &       &  \cmark  &  21  &  12  &  12  &  12  &  7  &  35  &  3\\

    &         
\multirow{1}{*}{$P_5$}  &       &  \cmark  &  18  &  14  &  14  &  14  &  14  &  56  &  4\\

    &         
\multirow{1}{*}{$P_6$}  &       &  \cmark  &  17  &  15  &  15  &  15  &  14  &  14  &  15\\
\midrule[0.1pt]
        
\multirow{6}{*}{Hyperelasticity}  &         
\multirow{1}{*}{$P_1$}  &     \cmark  &    &    &    &    &    &    &    &  \\

    &         
\multirow{1}{*}{$P_2$}  &       &  \cmark  &  18  &  14  &  14  &  14  &  10  &  10  &  14\\

    &         
\multirow{1}{*}{$P_3$}  &       &  \cmark  &  17  &  13  &  18  &  18  &  5  &  20  &  5\\

    &         
\multirow{1}{*}{$P_4$}  &       &  \cmark  &  14  &  18  &  18  &  18  &  9  &  18  &  9\\

    &         
\multirow{1}{*}{$P_5$}  &       &  \cmark  &  17  &  15  &  30  &  30  &  14  &  56  &  8\\

    &         
\multirow{1}{*}{$P_6$}  &       &  \cmark  &  13  &  17  &  46  &  16  &  12  &  17  &  12\\

\bottomrule
\end{tabular}

\end{center}
\caption{Best performing configurations for 3D cases on an Nvidia Tesla K40c.}
\label{tbl:perf_eval_winner_3d_k40}
\end{table}

\begin{table}
\begin{center}
\begin{tabular}{ccccccccccc}
\toprule

\multicolumn{2}{c}{} & \multicolumn{1}{c}{SCPT} & \multicolumn{1}{c}{Multi-level tiling} & \multicolumn{1}{c}{$N_c$} & \multicolumn{1}{c}{$N_{WI}$} & \multicolumn{1}{c}{$\quadtile$} & \multicolumn{1}{c}{$T_e^r$} & \multicolumn{1}{c}{$T_e^c$} & \multicolumn{1}{c}{$T_q^r$} & \multicolumn{1}{c}{$T_q^c$}\\

\midrule

\multirow{6}{*}{Mass}  &         
\multirow{1}{*}{$P_1$}  &     \cmark  &    &    &    &    &    &    &    &  \\

    &         
\multirow{1}{*}{$P_2$}  &     \cmark  &    &    &    &    &    &    &    &  \\

    &         
\multirow{1}{*}{$P_3$}  &     \cmark  &    &    &    &    &    &    &    &  \\

    &         
\multirow{1}{*}{$P_4$}  &     \cmark  &    &    &    &    &    &    &    &  \\

    &         
\multirow{1}{*}{$P_5$}  &       &  \cmark  &  64  &  4  &  72  &  72  &  14  &  12  &  72\\

    &         
\multirow{1}{*}{$P_6$}  &       &  \cmark  &  64  &  4  &  69  &  69  &  14  &  84  &  12\\
\midrule[0.1pt]
        
\multirow{6}{*}{Laplace}  &         
\multirow{1}{*}{$P_1$}  &     \cmark  &    &    &    &    &    &    &    &  \\

    &         
\multirow{1}{*}{$P_2$}  &     \cmark  &    &    &    &    &    &    &    &  \\

    &         
\multirow{1}{*}{$P_3$}  &     \cmark  &    &    &    &    &    &    &    &  \\

    &         
\multirow{1}{*}{$P_4$}  &       &  \cmark  &  64  &  4  &  24  &  24  &  7  &  35  &  5\\

    &         
\multirow{1}{*}{$P_5$}  &       &  \cmark  &  64  &  4  &  25  &  25  &  10  &  56  &  5\\

    &         
\multirow{1}{*}{$P_6$}  &       &  \cmark  &  64  &  4  &  27  &  27  &  9  &  28  &  7\\
\midrule[0.1pt]
        
\multirow{6}{*}{Helmholtz}  &         
\multirow{1}{*}{$P_1$}  &     \cmark  &    &    &    &    &    &    &    &  \\

    &         
\multirow{1}{*}{$P_2$}  &     \cmark  &    &    &    &    &    &    &    &  \\

    &         
\multirow{1}{*}{$P_3$}  &     \cmark  &    &    &    &    &    &    &    &  \\

    &         
\multirow{1}{*}{$P_4$}  &       &  \cmark  &  64  &  4  &  21  &  21  &  7  &  12  &  11\\

    &         
\multirow{1}{*}{$P_5$}  &       &  \cmark  &  64  &  4  &  20  &  20  &  10  &  28  &  7\\

    &         
\multirow{1}{*}{$P_6$}  &       &  \cmark  &  42  &  6  &  29  &  29  &  9  &  42  &  6\\
\midrule[0.1pt]
        
\multirow{6}{*}{Elasticity}  &         
\multirow{1}{*}{$P_1$}  &     \cmark  &    &    &    &    &    &    &    &  \\

    &         
\multirow{1}{*}{$P_2$}  &     \cmark  &    &    &    &    &    &    &    &  \\

    &         
\multirow{1}{*}{$P_3$}  &       &  \cmark  &  44  &  5  &  14  &  14  &  5  &  20  &  4\\

    &         
\multirow{1}{*}{$P_4$}  &       &  \cmark  &  25  &  10  &  24  &  24  &  9  &  35  &  6\\

    &         
\multirow{1}{*}{$P_5$}  &       &  \cmark  &  28  &  8  &  21  &  21  &  7  &  8  &  21\\

    &         
\multirow{1}{*}{$P_6$}  &       &  \cmark  &  21  &  12  &  27  &  27  &  12  &  12  &  27\\
\midrule[0.1pt]
        
\multirow{6}{*}{Hyperelasticity}  &         
\multirow{1}{*}{$P_1$}  &     \cmark  &    &    &    &    &    &    &    &  \\

    &         
\multirow{1}{*}{$P_2$}  &       &  \cmark  &  44  &  5  &  14  &  14  &  10  &  10  &  14\\

    &         
\multirow{1}{*}{$P_3$}  &       &  \cmark  &  25  &  10  &  25  &  25  &  5  &  20  &  7\\

    &         
\multirow{1}{*}{$P_4$}  &       &  \cmark  &  21  &  12  &  29  &  29  &  7  &  35  &  6\\

    &         
\multirow{1}{*}{$P_5$}  &       &  \cmark  &  22  &  10  &  27  &  27  &  7  &  28  &  7\\

    &         
\multirow{1}{*}{$P_6$}  &       &  \cmark  &  12  &  16  &  45  &  45  &  9  &  42  &  9\\

\bottomrule
\end{tabular}

\end{center}
\caption{Best performing configurations for 3D cases on an Nvidia Titan V.}
\label{tbl:perf_eval_winner_3d_v100}
\end{table}

\begin{table}
\begin{center}
\begin{tabular}{ccccccccccc}
\toprule

\multicolumn{2}{c}{} & \multicolumn{1}{c}{SCPT} & \multicolumn{1}{c}{Multi-level tiling} & \multicolumn{1}{c}{$N_c$} & \multicolumn{1}{c}{$N_{WI}$} & \multicolumn{1}{c}{$\quadtile$} & \multicolumn{1}{c}{$T_e^r$} & \multicolumn{1}{c}{$T_e^c$} & \multicolumn{1}{c}{$T_q^r$} & \multicolumn{1}{c}{$T_q^c$}\\

\midrule

\multirow{6}{*}{Mass}  &         
\multirow{1}{*}{$P_1$}  &     \cmark  &    &    &    &    &    &    &    &  \\

    &         
\multirow{1}{*}{$P_2$}  &     \cmark  &    &    &    &    &    &    &    &  \\

    &         
\multirow{1}{*}{$P_3$}  &     \cmark  &    &    &    &    &    &    &    &  \\

    &         
\multirow{1}{*}{$P_4$}  &       &  \cmark  &  64  &  2  &  44  &  44  &  35  &  35  &  44\\

    &         
\multirow{1}{*}{$P_5$}  &       &  \cmark  &  64  &  2  &  74  &  74  &  14  &  14  &  74\\

    &         
\multirow{1}{*}{$P_6$}  &       &  \cmark  &  64  &  2  &  61  &  61  &  28  &  28  &  61\\
\midrule[0.1pt]
        
\multirow{6}{*}{Laplace}  &         
\multirow{1}{*}{$P_1$}  &     \cmark  &    &    &    &    &    &    &    &  \\

    &         
\multirow{1}{*}{$P_2$}  &     \cmark  &    &    &    &    &    &    &    &  \\

    &         
\multirow{1}{*}{$P_3$}  &       &  \cmark  &  64  &  2  &  11  &  11  &  20  &  20  &  11\\

    &         
\multirow{1}{*}{$P_4$}  &       &  \cmark  &  64  &  2  &  23  &  23  &  18  &  18  &  23\\

    &         
\multirow{1}{*}{$P_5$}  &       &  \cmark  &  64  &  4  &  22  &  22  &  28  &  28  &  22\\

    &         
\multirow{1}{*}{$P_6$}  &       &  \cmark  &  64  &  4  &  25  &  25  &  17  &  17  &  25\\
\midrule[0.1pt]
        
\multirow{6}{*}{Helmholtz}  &         
\multirow{1}{*}{$P_1$}  &     \cmark  &    &    &    &    &    &    &    &  \\

    &         
\multirow{1}{*}{$P_2$}  &       &  \cmark  &  64  &  2  &  11  &  11  &  10  &  10  &  11\\

    &         
\multirow{1}{*}{$P_3$}  &     \cmark  &    &    &    &    &    &    &    &  \\

    &         
\multirow{1}{*}{$P_4$}  &       &  \cmark  &  63  &  2  &  22  &  22  &  6  &  18  &  8\\

    &         
\multirow{1}{*}{$P_5$}  &       &  \cmark  &  53  &  3  &  25  &  25  &  7  &  28  &  7\\

    &         
\multirow{1}{*}{$P_6$}  &       &  \cmark  &  63  &  3  &  21  &  21  &  9  &  17  &  11\\
\midrule[0.1pt]
        
\multirow{6}{*}{Elasticity}  &         
\multirow{1}{*}{$P_1$}  &     \cmark  &    &    &    &    &    &    &    &  \\

    &         
\multirow{1}{*}{$P_2$}  &     \cmark  &    &    &    &    &    &    &    &  \\

    &         
\multirow{1}{*}{$P_3$}  &       &  \cmark  &  56  &  4  &  11  &  11  &  10  &  10  &  11\\

    &         
\multirow{1}{*}{$P_4$}  &       &  \cmark  &  27  &  7  &  23  &  23  &  7  &  35  &  5\\

    &         
\multirow{1}{*}{$P_5$}  &       &  \cmark  &  28  &  8  &  22  &  22  &  7  &  8  &  22\\

    &         
\multirow{1}{*}{$P_6$}  &       &  \cmark  &  24  &  4  &  25  &  25  &  9  &  28  &  7\\
\midrule[0.1pt]
        
\multirow{6}{*}{Hyperelasticity}  &         
\multirow{1}{*}{$P_1$}  &     \cmark  &    &    &    &    &    &    &    &  \\

    &         
\multirow{1}{*}{$P_2$}  &     \cmark  &    &    &    &    &    &    &    &  \\

    &         
\multirow{1}{*}{$P_3$}  &       &  \cmark  &  28  &  8  &  22  &  22  &  7  &  20  &  8\\

    &         
\multirow{1}{*}{$P_4$}  &       &  \cmark  &  25  &  10  &  25  &  25  &  6  &  18  &  9\\

    &         
\multirow{1}{*}{$P_5$}  &       &  \cmark  &  9  &  14  &  57  &  57  &  8  &  28  &  15\\

    &         
\multirow{1}{*}{$P_6$}  &       &  \cmark  &  5  &  19  &  103  &  52  &  9  &  42  &  11\\

\bottomrule
\end{tabular}

\end{center}
\caption{Best performing configurations for 3D cases on an Nvidia H200 NVL.}
\label{tbl:perf_eval_winner_3d_h200nvl}
\end{table}

\end{document}